\documentclass[a4paper,11pt]{article}
\pdfoutput=1 % if your are submitting a pdflatex (i.e. if you have
\usepackage{jheppub} % for details on the use of the package, please
\usepackage[T1]{fontenc} % if needed

\usepackage{amsmath}
\usepackage{graphicx,slashed,xcolor,multirow,bbold,mathtools,sidecap,tikz,bm,enumitem,booktabs,array}

\usepackage[normalem]{ulem}

\usepackage{tikz}
\usetikzlibrary{positioning,arrows}
\usetikzlibrary{decorations.pathmorphing}
\usetikzlibrary{decorations.markings}

\usepackage{subcaption}

\usepackage[compat=1.1.0]{tikz-feynman}
\newcommand{\beq}{\begin{equation}}
\newcommand{\eeq}{\end{equation}}
\newcommand{\bea}{\begin{eqnarray}}
\newcommand{\eea}{\end{eqnarray}}
\newcommand{\barr}{\begin{array}}
\newcommand{\earr}{\end{array}}

\usepackage{color}
\usepackage{colortbl}

\long\def\/*#1*/{}
\usepackage{color}
 \usepackage[normalem]{ulem}
\definecolor{darkgreen}{cmyk}{1,0,1,0.4}
\definecolor{darkred}{cmyk}{0,1,1,0.4}

\title{\boldmath Revisiting Universal Extra-Dimension Model with Gravity Mediated Decays}

\author[a]{Kirtiman Ghosh,}
\author[b]{Katri Huitu,}
\author[c]{Rameswar Sahu}

\affiliation[a,c]{\footnotesize Institute of Physics, Bhubaneswar, Sachivalaya Marg, Sainik School Post, Bhubaneswar 751005, India}
\affiliation[a,c]{\footnotesize Homi Bhabha National Institute, Training School Complex, Anushakti Nagar, Mumbai 400094, India}
\affiliation[b]{\footnotesize Department of Physics, and Helsinki Institute of Physics, University of Helsinki, Finland 00014}
\emailAdd{kirti.gh@gmail.com}
\emailAdd{katri.huitu@helsinki.fi}
\emailAdd{rameswar.s@iopb.res.in}

\abstract{We explore the collider phenomenology of the fat-brane realization of the Minimal Universal Extra Dimension (mUED) model, where Standard Model (SM) fields propagate in a small extra dimension while gravity accesses additional large extra dimensions. This configuration allows for gravity-mediated decay (GMD) of Kaluza-Klein (KK) particles, resulting in unique final states with hard photons, jets, massive SM bosons, and large missing transverse energy due to invisible KK gravitons. We derive updated constraints on the model’s parameter space by recasting ATLAS mono-photon, di-photon, and multi-jet search results using 139 inverse femtobern of integrated luminosity data. Recognizing that current LHC search strategies are tailored for supersymmetric scenarios and may not fully capture the distinct signatures, we propose optimized strategies using machine learning algorithms to tag boosted SM bosons and enhance signal discrimination against SM backgrounds. These methods improve sensitivity to fat-brane mUED signatures and offer promising prospects for probing this model in future LHC runs.}

\begin{document} 

\tikzset{
  vector/.style={decorate, decoration={snake,amplitude=.4mm,segment length=2mm,post length=1mm}, draw},
  tes/.style={draw=black,postaction={decorate},
    decoration={snake,markings,mark=at position .55 with {\arrow[draw=black]{>}}}},
  provector/.style={decorate, decoration={snake,amplitude=2.5pt}, draw},
  antivector/.style={decorate, decoration={snake,amplitude=-2.5pt}, draw},
  fermion/.style={draw=black, postaction={decorate},decoration={markings,mark=at position .55 with {\arrow[draw=blue]{>}}}},
  fermionbar/.style={draw=black, postaction={decorate},
    decoration={markings,mark=at position .55 with {\arrow[draw=black]{<}}}},
  fermionnoarrow/.style={draw=black},
  % gluon/.style={decorate, draw=black,decoration={coil,amplitude=4pt, segment length=5pt}},
  scalar/.style={dashed,draw=black, postaction={decorate},decoration={markings,mark=at position .55 with {\arrow[draw=blue]{>}}}},
  scalarbar/.style={dashed,draw=black, postaction={decorate},decoration={marking,mark=at position .55 with {\arrow[draw=black]{<}}}},
  scalarnoarrow/.style={dashed,draw=black},
  electron/.style={draw=black, postaction={decorate},decoration={markings,mark=at position .55 with {\arrow[draw=black]{>}}}},
  bigvector/.style={decorate, decoration={snake,amplitude=4pt}, draw},
  particle/.style={thick,draw=blue, postaction={decorate},
    decoration={markings,mark=at position .5 with {\arrow[blue]{triangle 45}}}},
  % fermion/.style={thick,draw=blue, postaction={decorate},
  %   decoration={markings,mark=at position .56 with {\arrow[blue]{triangle 45}}}},
  gluon/.style={decorate, draw=black,
    decoration={coil,aspect=0.3,segment length=3pt,amplitude=3pt}}
}

\maketitle
\flushbottom

\section{Introduction}
\label{sec:intro}

Notwithstanding the remarkable success of the Standard Model (SM), it continues to suffer from several cavities, particularly its inability to explain observational facts like the existence of Dark Matter (DM), non-zero neutrino masses/mixings, etc., on the one hand, and theoretical issues like the stability of the Higgs boson mass, the related naturalness/hierarchy problems, etc. on the other. These issues have led to a plethora of new dynamics beyond the SM (BSM). Theories with one or more extra space-like dimension(s) accessible to all or a few SM fields are of interest for various reasons. For example, the ADD \footnote{Arkani-Hamed, Dimopoulos, and Dvali} \cite{Arkani-Hamed:1998jmv, Antoniadis:1990ew, Antoniadis:1998ig} (seemingly) and RS \footnote{Randall and Sundrum} \cite{Randall:1999ee, Randall:1999vf} models provide solutions to the long-standing naturalness/hierarchy problem by postulating the existence of compactified extra-dimension(s) accessible only to gravity with the SM fields being confined to a 3-brane embedded in the extra-dimensional bulk. In the case of ADD model \cite{Arkani-Hamed:1998jmv, Antoniadis:1990ew, Antoniadis:1998ig}, for instance, gravity is allowed to propagate into '$N$' number of large, flat, and compactified extra dimensions. The four-dimensional Planck mass is then diluted by the volume of the extra-dimensional space $V_N \sim r^N$, where $N$ and $r$ are the number and size of large extra dimensions, resulting in higher dimensional Planck mass around a few tens of TeV and hence offering a solution to naturalness/hierarchy problem. The introduction of warped metric in RS model \cite{Randall:1999ee, Randall:1999vf} provides an alternative solution to the same problem. On the other hand, there is a class of models wherein some or all SM fields can access the extended space-time manifold \cite{Appelquist:2000nn}, whether fully or partially. Such extra-dimensional scenarios could address a plethora of issues, including some exciting ones like the existence of DM \cite{Servant:2002aq, Kong:2005hn, Dobrescu:2007ec}, an explanation for the number of fermion generations \cite{Dobrescu:2001ae}, the long lifetime of the proton \cite{Appelquist:2001mj}, etc. As a result, over decades, extra-dimensional models remain some of the most extensively studied BSM scenarios. And search for the extra-dimension(s) is one of the prime goals of the collider \cite{Appelquist:2000nn, Belyaev:2012ai, Kakuda:2013kba, Belanger:2012mc, Ghosh:2018mck, Ghosh:2008dp, Ghosh:2008ix, Beuria:2017jez, Dey:2014ana, Edelhauser:2013lia, Flacke:2012ke, Huang:2012kz, Flacke:2011nb, Nishiwaki:2011gm, Murayama:2011hj, Choudhury:2011jk, Ghosh:2010tp, Bhattacherjee:2010vm, Bertone:2010ww, Freitas:2007rh, Dobrescu:2007ec, Macesanu:2002db, Ghosh:2012zc, Choudhury:2009kz, Rizzo:2001sd, Muck:2003kx, Bhattacharyya:2005vm, Battaglia:2005zf, Bhattacherjee:2005qe, Datta:2005zs, Datta:2005vx} (including the Large Hadron Collider (LHC) \cite{CMS:2014jvv}) and non-collider \cite{Datta:2013nua, Flacke:2017xsv} experiments. The ongoing consistency of the LHC data with SM predictions is putting significant pressure on the simplest extra-dimensional scenarios, which are highly predictive. Some of these scenarios have been completely ruled out, including the minimal version of the Universal Extra-Dimension (mUED) model \cite{Avnish:2020atn}. In light of the exclusion of mUED, attention turns to non-minimal \cite{Avnish:2020atn, Flacke:2008ne, Datta:2012tv, Flacke:2013nta, Datta:2013yaa, Flacke:2013pla, Shaw:2017whr} and next-to-minimal \cite{DeRujula:2000he, Donini:1999px, Antoniadis:1999bq, Dicus:2000hm, Macesanu:2002ew, Macesanu:2005wj, Macesanu:2004nb, Gabrielli:2002hn, Macesanu:2003jx} scenarios\footnote{Non-minimal setups maintain identical gauge/Lorentz symmetry and field content as mUED but introduce non-vanishing boundary localized terms at the cut-off scale ($\Lambda$). Next-to-minimal UED scenarios, on the other hand, deviate from mUED by featuring different gauge/Lorentz symmetry or field content.}. Here, we are interested in a specific next-to-minimal version of the UED scenario, namely the 'fat-brane' realization of UED. In addition to the usual TeV$^{-1}$ size extra dimension(s) (universally accessible to all SM fields and the gravity) of UED scenarios, the fat-brane realization of UED includes large ($\sim$ eV$^{-1}$ to keV$^{-1}$ size) extra dimension(s) (accessible only to the gravity).

The simplest version of the UED scenario is characterized by a single flat universal (accessible to all the SM particles) extra dimension ($y$), compactified on a $S_1/Z_2$ orbifold with a radius $R$, known as {\em One Universal Extra-Dimension (OneUED)} model. The particle spectrum of {\em oneUED} contains infinite towers of Kaluza-Klein (KK) modes (identified by an integer $n$, called the KK-number) for each of the SM fields. The zero modes are identified as the corresponding SM particles. From a 4-dimensional perspective, the conservation of the momentum along the fifth direction implies the conservation of the KK-number. However, the additional $Z_2$ symmetry $(y \leftrightarrow -y$), required to obtain the chiral structure of the SM fermions, breaks the translational invariance along the 5$^{\rm th}$ dimension. As a result, KK-number conservation breaks down at the loop level, leaving behind only a conserved KK-parity, defined as $(-1)^n$, which is an automatic outcome of the $S_1/Z_2$ orbifolding and has several consequences, including some exciting ones like the stability of the lightest KK particle (LKP) which can be a good candidate for cold dark matter (CDM) \cite{Dobrescu_2007, Arun:2018yhg, Servant:2002aq, Cheng:2002ej, Kong:2005hn, Hooper:2007qk, ColomiBernadich:2019upo, Kakizaki:2006dz, Burnell:2005hm, Ishigure:2016kxp, Cornell:2014jza}. {{\em OneUED}, being a higher dimensional theory, is non-renormalizable and should be treated as an effective theory valid up to a cut-off scale $\Lambda$. Apart from the usual SM kinetic, Yukawa, and scalar potential terms for the 5D fields, the {\em oneUED} Lagrangian also includes additional SM gauge and Lorentz invariant terms like the vector-like bulk mass terms \cite{Kong:2010qd, Flacke:2011nb, Chen:2009gz, Huang:2012kz, Park:2009cs} for the 5D fermions and kinetic (and Yukawa) terms (boundary localized terms (BLTs)) for all the 5D fields at the orbifold fixed points, {\em i.e.,} the boundaries of the bulk and the brane \cite{delAguila:2003bh, Carena:2002me}. In the {\em minimal} version of {\em oneUED} ({\em mUED} \cite{Cheng:2002iz}), all BLTs are assumed to vanish at the cut-off scale ($\Lambda$) and are radiatively generated at the low scale, which ultimately appears as corrections to the masses of the KK particles. The phenomenology of {\em mUED} \cite{Appelquist:2000nn, Cheng:2002ab} is determined by only two additional parameters, namely, $R$ and $\Lambda$. Hence, its predictions are precise and easily testable in different experiments. A recent study \cite{Avnish:2020atn} on the consistency of {\em mUED} scenario with the DM relic density (RD) as well as collider experiments, concluded that the region of $R^{-1}$-$\Lambda R$ plane, which is consistent with WMAP/PLANCK measured RD \cite{WMAP:2010qai, Planck:2015fie}, has already been ruled out by the ATLAS multijet $+~E_T\!\!\!\!\!/~$~searches \cite{ATLAS:2019vcq}. Present status of {\em mUED} naturally motivates several variants of UED beyond the minimal one, namely the non-minimal UED ({\em nmUED}) model \cite{Avnish:2020atn}, the 'fat-brane' realization of UED \cite{DeRujula:2000he, Donini:1999px, Antoniadis:1999bq, Dicus:2000hm, Macesanu:2002ew, Macesanu:2005wj, Macesanu:2004nb, Gabrielli:2002hn, Macesanu:2003jx}, etc.  While the additional parameters (couplings associated with the non-vanishing BLTs) help reconcile the tension between the dark matter relic density and the LHC results in the case of the former, the decay of the LKP via gravity-matter interactions removes the constraints from the WMAP/PLANCK measurement of DM RD in the case of the latter. 

The 'fat-brane' realization of UED (FB-mUED)\cite{DeRujula:2000he, Donini:1999px, Antoniadis:1999bq, Dicus:2000hm, Macesanu:2002ew, Macesanu:2005wj, Macesanu:2004nb, Gabrielli:2002hn, Macesanu:2003jx} represents an intriguing extension of the ADD scenario. In this framework, SM particles are confined to a $(3 + m)$-brane, which denotes a $(3 + m + 1)$-dimensional manifold embedded in a $(4 + N)$-dimensional bulk \cite{Donini:1999px, Antoniadis:1999bq}. Given that $m$ spatial dimensions are compact, the effective 4-dimensional theory encompasses KK excitations of SM fields. The dimensions of the $m$ spatial dimensions within the bulk are restricted by the constraints set by the experimental lower limit on KK-mode masses. The term 'fat-brane' is derived from the fact that these $m$ small spatial dimensions, accessible to both matter and gravity, can be analogized to the thickness of the SM 3-brane in the $(4 + N)$-dimensional bulk \cite{DeRujula:2000he, Dicus:2000hm, Macesanu:2002ew, Macesanu:2003jx, Macesanu:2005wj}. In this context, gravity spans across $N$ large extra dimensions with a size around eV$^{-1}$, while the propagation of matter is confined to a small length scale (approximately TeV$^{-1}$), corresponding to the thickness of the SM 3-brane along these extra dimensions. In this work, we focus on the collider signatures, particularly those at the LHC, of the fat brane realization of the minimal Universal Extra-Dimension (mUED) model. The collider signatures of the fat brane realization of mUED are very different from those of the standard mUED model. While the pair production and decay of level-1 KK excitations in mUED typically result in soft leptons and jets along with missing transverse energy, the gravity-mediated decays of KK particles in the fat brane realization of mUED lead to the production of hard photons, jets, Z/W bosons, top quarks, etc. at the LHC. Therefore, the search strategies proposed for mUED in the literature are not applicable to the fat brane realization of mUED. However, supersymmetric (SUSY) scenarios with gauge-mediated supersymmetry breaking (GMSB) \cite{Dine:1981za, Dimopoulos:1981au, Dine:1981gu, Nappi:1982hm, Alvarez-Gaume:1981abe, Dimopoulos:1982gm, Dine:1993yw, Dine:1994vc, Dine:1995ag, Giudice:1998bp}, where SUSY particles decay into gravitinos alongside the corresponding SM particles, produce final state signatures similar to those of the fat brane realization of mUED. Consequently, LHC searches for GMSB signatures can indirectly help constrain the parameter space of the fat brane realization of mUED. It is important to note that while the final state signatures of GMSB and the fat brane realization of mUED are similar, the kinematics of gravity-mediated decays differ. In contrast to GMSB, which contains a single gravitino, in the fat brane mUED, there is a tower of graviton excitations. Therefore, LHC bounds on SUSY particles in GMSB are not directly applicable to the level-1 KK excitations of fat brane mUED. However, model-independent bounds on the visible cross-sections of final state signatures, such as mono-photon, di-photon, and multi-jet events associated with large missing transverse energy, can be used to derive constraints on the parameter space of the fat brane realization of mUED. In this work, we obtain the most updated bounds from the LHC searches. 

The persistent consistency of LHC searches with SM background predictions has raised the mass threshold for strongly interacting BSM particles to above a few TeV. Decays of such massive BSM resonances into SM particles result in highly boosted W/Z bosons, Higgs bosons, or top quarks in the final state. Recently, tagging hadronically decaying boosted W/Z or Higgs bosons or top quarks and using them for event selection at the LHC has gained popularity. This approach offers several advantages, such as larger hadronic branching ratios, leading to higher signal rates, suppression of the QCD background due to tagging, and the possibility of kinematic reconstruction because of the absence of missing transverse energy in the final state. After establishing bounds on the level-1 KK particles in FB-mUED from existing LHC searches, we propose new search strategies involving boosted tagged hadronically decaying W/Z/H bosons or top quarks optimized for detecting level-1 KK particles in FB-mUED in future LHC runs.\\

The rest of the paper is organized as follows: Section \ref{sec:model} provides a brief overview of the fat-brane realization of mUED. Section \ref{sec:phenomenology} contains the key findings of our analysis. In Section \ref{sec: kkncsec} and \ref{sec: gmdsec}, we discuss the KK-number conserving and gravity-mediated decays of the level-1 KK particles, respectively. In Section \ref{sec:signatures}, we discuss the collider signatures of the fat-brane mUED scenario. In Section \ref{sec:bounds}, we present the bounds on the model parameter space from three existing ATLAS analyses. In Section \ref{sec: optimization}, we propose a novel analysis strategy to constrain the parameter space even further during future runs of the LHC. Finally, Section \ref{sec: sumandout} concludes the paper with a summary of our findings and a discussion of potential future directions.

\section{The fat brane realization of mUED secnario}
\label{sec:model}
In this work, we focus on the phenomenology of the fat brane realization of the minimal Universal Extra Dimensions (mUED) scenario. In this framework, all the SM fields are allowed to propagate in one small, compactified extra dimension ($y$), which is embedded in a $(4 + N)$-dimensional bulk fully accessible only to gravity. Embedding the mUED scenario into a gravity-accessible $(4 + N)$-dimensional bulk leads to gravity-matter interactions, which allow gravity-mediated decays for the KK particles. The theoretical structure of mUED has already been briefly discussed in the introduction. For a detailed discussion on the KK decomposition of SM fields in five dimensions on the \( S^1/Z_2 \) orbifold and the resulting effective four-dimensional Lagrangian, please refer to Ref. \cite{Cheng:2002iz, Appelquist:2000nn, Cheng:2002ab}. The phenomenology of mUED has been discussed in Ref. \cite{Bhattacherjee:2010vm, Datta:2011vg, Kong:2010mh, Choudhury:2009kz, Bandyopadhyay:2009gd, Datta:2005zs, Battaglia:2005zf}. The fat brane realization of mUED and the consequences of gravity-matter interactions have been studied in Ref. \cite{DeRujula:2000he, Donini:1999px, Antoniadis:1999bq, Dicus:2000hm, Macesanu:2002ew, Macesanu:2005wj, Macesanu:2004nb, Gabrielli:2002hn, Macesanu:2003jx}. However, for the sake of completeness, we provide a brief introduction to the gravity matter interaction in the framework of fat brane realization of mUED in the following.

\subsection{Matter-Gravity Interaction}
\label{sec:gint}
In the fat-brane scenario, gravity is permitted to propagate into $N$ large extra dimensions, which are subsequently compactified on an $N$-dimensional torus $T^N$ with a volume $V_N \sim r^N$, where $r$ represents the size of the $N$ large extra dimensions. The 4D Planck mass $M_{\text{Pl}}$ can be derived from the fundamental $(4 + N)$-dimensional Planck mass $M_D$ as:
\[ M_{\text{Pl}}^2 = M^{N+2}_D \left( \frac{r}{2\pi} \right)^N. \]
Assuming there are $N$ such large extra dimensions denoted by $x^5, \ldots, x^{4+N}$ with a common size of $r \sim \text{eV}^{-1}$, and one small extra dimension denoted by $y = x^4$ with a size of $\text{TeV}^{-1}$, one can express the interaction of SM fields and the graviton in the higher dimension as:
\[ S_{\text{int}} = \int dx^{4+N} \, \delta(x^5) \ldots \delta(x^{4+N}) \sqrt{-\hat{g}} \, \mathcal{L}_m, \]
where $\mathcal{L}_m$ represents the Lagrangian density for SM fermions, gauge bosons, and the Higgs. The higher dimensional flat metric $\hat{g}$ can be expressed as $\hat{g}_{\hat{\mu}\hat{\nu}} = \eta_{\hat{\mu}\hat{\nu}} + \hat{k}\hat{h}_{\hat{\mu}\hat{\nu}}$. The hat denotes quantities that live in $(4+N)$ dimensions, i.e., $\hat{\mu},\hat{\nu}=0,1,2,3,5,...,4+N$. Here $\hat{k}^2=16\pi G^{4+N}$ with $G^{4+N}$ being the gravitational constant in $(4+N)$ dimensions. The KK-decomposition of the $(4+N)$ dimensional graviton field $\hat{h}_{\hat{\mu}\hat{\nu}}$ can be written as,
\begin{equation}\label{eq:2_4}
    \hat{h}_{\hat{\mu}\hat{\nu}}(x,y) = \sum_{\vec{n}} \hat{h}_{\hat{\mu}\hat{\nu}}^{\vec{n}}(x) \exp\left(i\frac{2\pi\vec{n} \cdot \vec{y}}{r}\right),
\end{equation}
where $\hat{h}_{\hat{\mu}\hat{\nu}}^{\vec{n}}(x)$ is the $\vec{n}^{\rm th}$ KK-excitation of the graviton field with $\vec{n}=\{n_5,\ldots,n_{4+N}\}$ being the KK-number representing the $\vec{n}^{\rm th}$ KK-excitation. The $\vec{n}^{\rm th}$ KK-excitation of the graviton field, $\hat{h}_{\hat{\mu}\hat{\nu}}^{\vec{n}}(x)$, can be decomposed into a four-dimensional tensor $h_{\mu\nu}^{\vec{n}}$ (the graviton), $N$ vectors $A_{\mu i}^{\vec{n}}$ (the graviphotons), and $N^2$ scalar fields $\phi_{ij}^{\vec{n}}$ (the graviscalars) as,
\begin{equation}
    \hat{h}_{\hat{\mu}\hat{\nu}}^{\vec{n}} = V_N^{-1/2} \begin{pmatrix}
    h_{\mu\nu}^{\vec{n}} + \eta_{\mu\nu}\phi^{\vec{n}} & A_{\mu i}^{\vec{n}} \\
    A_{\nu j}^{\vec{n}} & 2 \phi_{i,j}^{\vec{n}}
    \end{pmatrix}
\end{equation}
where $V_N$ is the volume of the $N$ dimensional torus, $i, j = 5,6,\ldots,(4+N)$, and $\phi^{\vec{n}} = \phi_{ii}^{\vec{n}}$. All the $\frac{(N+4)(N+5)}{2}$ components of the symmetric $(N+4)$ dimensional rank-2 tensor, $\hat{h}_{\hat{\mu}\hat{\nu}}^{\vec{n}}(x)$, are not independent. In order to eliminate the unphysical degrees of freedom, one has to make particular gauge choices, which have been done in \cite{MACESANU_2006, PhysRevD.59.105006, Giudice_1999, PhysRevD.68.084008} and discussed briefly in the following. The de Donder condition $\partial^{\hat{\mu}}\left(\hat{h}_{\hat{\mu}\hat{\nu}} - \frac{1}{2} \eta_{\hat{\mu}\hat{\nu}} \hat{h}\right)=0$, where $\hat{h}=\hat{h}^{\hat{\mu}}_{\hat{\mu}}$, in addition with the gauge conditions, $n_iA_{\mu i}^{\vec{n}} = 0, n_i \phi_{ij}^{\vec{n}}=0$, eliminate $2(N+4)$ spurious degrees of freedom leaving $\frac{(N+4)(N+1)}{2}$ physical degrees of freedom at level $\vec{n}$, associated with a massive spin-2 graviton $\tilde{h}_{\mu\nu}^{\vec{n}}$ (5 degrees of freedom), $N-1$ massive vector bosons $\tilde{A}^{\vec{n}}_{\mu i}$ (3 $\times$ $(N-1)$ degrees of freedom) known as graviphotons, plus $\frac{N(N-1)}{2}$ massive scalars $\tilde{\phi}^{\vec{n}}_{ij}$ known as graviscalars with 1 degree of freedom each:
\begin{eqnarray}
    \tilde{h}_{\mu\nu}^{\vec{n}} &=& h^{\vec{n}}_{\mu \nu} - \omega \left( \frac{\partial_{\mu}\partial_{\nu}}{m_{\vec{n}}^2}-\frac{1}{2} \eta_{\mu \nu} \right) \tilde{\phi}^{\vec{n}}, \nonumber\\
    \tilde{A}^{\vec{n}}_{\mu i} &=& A^{\vec{n}}_{\mu i}, \nonumber\\
    \frac{1}{\sqrt{2}}\tilde{\phi}^{\vec{n}}_{ij} &=& \phi^{\vec{n}}_{ij} + \frac{3 \omega a}{2}\left(\delta_{ij} - \frac{n_i n_j}{\vec{n}^2}\right)\tilde{\phi}^{\vec{n}}
\end{eqnarray}
where $\tilde{\phi}^{\vec{n}} = 2\phi^{\vec{n}}_{ii}/(3\omega)$, $\omega = \sqrt{2/(3N+6)}$ and $a$ satisfies $3(N-1)a^2 + 6a -1 = 0$. For a given $\vec{n}$ all the gravity excitations (graviton, graviphotons, or graviscalars) share the common mass $m_{\vec{n}} = 2 \pi |\vec{n}|/r$.

The $(4+N)$ dimensional action representing the gravity-matter interaction is given by,
\begin{equation}
    S_{\text{int}} = -\frac{\hat{k}}{2} \int \,d^{N+4}x \, \delta(x^5)\ldots\delta(x^N)\,\hat{h}^{\hat{\mu}\hat{\nu}}T_{\hat{\mu}\hat{\nu}},
\end{equation}
where $\hat{k}= V_N^{1/2}\sqrt{16\pi G_N}$\footnote{Here $G_N$ is the Universal Gravitational Constant} is the $(4+N)$ dimensional gravitational coupling and $T_{\hat{\mu}\hat{\nu}}$ is the energy-momentum tensor, which in terms of the matter Lagrangian $\mathcal{L}_m$ has the following form,
\begin{equation}\label{eq:2_8}
    T_{\hat{\mu}\hat{\nu}} = \left(-\hat{\eta}_{\hat{\mu}\hat{\nu}} + 2 \frac{\mathrm{d}\mathcal{L}_m}{\mathrm{d}\hat{g}^{\hat{\mu}\hat{\nu}}}\right)_{\hat{g}=\hat{\eta}}
\end{equation}
Following equation \ref{eq:2_4} and \ref{eq:2_8}, we can express the gravity-matter action in terms of the KK modes of the physical gravity field,
\begin{equation}\label{eq:2_13}
\begin{split}
    S_{\text{int}} = &-\frac{k}{2} \int \,d^4x \sum_n \Biggl\{ \left[\tilde{h}_{\mu \nu}^{\vec{n}} + \omega \left(\eta_{\mu \nu} + \frac{\partial_{\mu}\partial_{\nu}}{m_{\vec{n}}^2}\right)\tilde{\phi}^{\vec{n}}\right] T_{n_5}^{\mu \nu} \\
    &- 2 \tilde{A}^{\vec{n}}_{\mu 5} T_{n_5 5}^{\mu} + \left(\sqrt{2} \tilde{\phi}^{\vec{n}}_{55} - 3 \omega a \left(1-\frac{n_5^2}{\vec{n}^2}\right) \tilde{\phi}^{\vec{n}}\right) T^{n_5}_{55} \Biggr\}
\end{split}
\end{equation}
Where we have defined,
\begin{equation}
    T_{MN}^{n_5}(x) = \int_0^{\pi R} \,dy \, T_{MN}(x,y) \, e^{\frac{2\pi i n_5 y}{r}}
\end{equation}
Using equation \ref{eq:2_13}, it is straightforward to calculate the Feynman rules for gravity-matter interactions. The resulting expressions are lengthy. We redirect the readers to \cite{Giudice_1999} for a detailed discussion.

\section{Collider Phenomenology}
\label{sec:phenomenology}

After briefly introducing the model and the gravity-matter interactions for the KK particles, we are now equipped to discuss the collider signatures of this model. Due to the conservation of KK-parity, the level-1 KK-particles (which are odd under KK-parity) can only be pair-produced at the LHC. The particle spectrum of level-1 KK fields includes excited fermions (SU(2)$_L$-doublets: $Q_1$ and $L_1$; SU(2)$_L$-singlets: $u_1$, $d_1$, and $e_1$), Higgses, and gauge bosons (excited gluon: $g_1$, W-boson: $W_1^\pm$, Z-boson: $Z_1$, and photon: $\gamma_1$). The pair production of the level-1 KK-quarks/gluons, being driven by QCD processes, has the highest production cross-sections among the level-1 KK particles. KK-parity forbids the decay of the level-1 KK-particles completely into SM particles (which are even under KK-parity). In the framework of the fat-brane realization of mUED, the decays of the level-1 KK-particles can be categorized into two categories: \textbf{Category I:} The level-1 KK-particles can decay into lighter level-1 KK-particles in association with one or more SM particles. These decays are KK-number conserving, and hence, we will denote these decays as KKNC decays for the rest of the article. In the absence of any kinematically allowed KKNC decay modes for the lightest level-1 particle, the KKNC decay for the lightest level-1 particle is forbidden. \textbf{Category II:} In the presence of gravity-matter interaction, the level-1 KK particles can also decay into lighter gravity excitations (KK graviton, graviphoton, or graviscalars) in association with the corresponding SM particles. For the rest of the article, we denote these decays as gravity-mediated decays (GMD). The collider signatures of the level-1 KK-particles in the framework of the fat-brane realization of mUED crucially depend on the decays of the level-1 KK-particles, which will be discussed in the following.

\subsection{KK-Number Conserving (KKNC) Decays }
\label{sec: kkncsec}
The KK parity allows the decay of level-1 KK particles into lighter level-1 KK particles. Therefore, the mass spectrum of level-1 KK particles plays a crucial role in determining the decay of these particles. In the absence of electroweak symmetry breaking, the masses of all level-1 KK particles are given by $R^{-1}$. However, radiative corrections~\cite{Cheng:2002iz} remove this degeneracy. KK-fermions receive positive mass corrections from both gauge interactions (with KK-gauge bosons) and Yukawa interactions. The gauge fields receive mass corrections from self-interactions and gauge interactions (with KK-fermions). Gauge interactions give a negative mass shift, while self-interactions give a positive mass shift. However, the mass of the hypercharge gauge boson ($\gamma_1$) receives only negative corrections from fermionic loops. Numerical computations show that the lightest KK particle is the hypercharge gauge boson $\gamma_1$, and the heaviest level-1 KK particle is the excited gluon ($g_1$), followed by the excited quarks, level-1 electroweak gauge bosons, and leptons. The radiative corrections are proportional to $\ln(\Lambda^2/\mu^2)$, where $\Lambda$ is the cutoff scale. The perturbativity of the $U(1)_Y$ gauge coupling requires $\Lambda \leq 40R^{-1}$. However, much stronger bounds arise from the running of the Higgs-boson self-coupling and the stability of the electroweak vacuum~\cite{Datta:2012db, Datta:2013xwa}. Throughout this analysis, we choose $\Lambda = 5R^{-1}$.

For fixed $R^{-1}$ and $\Lambda$, $g_1$, being the heaviest particle in the spectrum, can decay into doublet $Q_1$ and singlet ($u_1$, $d_1$) quarks with almost equal branching fractions. The singlet quarks can only decay into $\gamma_1$ and an SM quark. On the other hand, the doublet quarks can mostly decay into level-1 KK electroweak gauge bosons, namely $Z_1$ and $W_1$. The hadronic decays of $W_1$ and $Z_1$ are kinematically closed. Therefore, they can only decay into level-1 KK leptons and the corresponding SM lepton. These level-1 leptons then decay into SM leptons and $\gamma_1$. Being the lightest level-1 KK particle, $\gamma_1$ does not have any KKNC decay. Note that the masses and the KKNC decay widths of level-1 particles do not depend on the number of large extra dimensions, $N$, and are determined only by the size of the small extra dimension $R^{-1}$ and the cutoff scale of the model $\Lambda$.

\subsection{Gravity Mediated Decays (GMD)}
\label{sec: gmdsec}
In the fat-brane scenario, the matter fields are confined to a small distance along the fifth dimension. This results in breaking translation invariance along the fifth dimension at the boundaries $y=0$ and $y=\pi R$. Consequently, the interaction between gravity and matter does not respect KK number conservation, and the KK excitations of SM particles can decay directly into the corresponding SM particles with the emission of a gravity excitation. The gravity-mediated decay widths of the level-1 KK particles are discussed in detail in \cite{MACESANU_2006, Giudice_1999}, and for the sake of completeness, we summarize their results in Appendix \ref{sec:appendix-1}. The decay width into individual graviton modes is suppressed by a factor of $1/M_{\text{Pl}}^2$ and hence is negligible. However, each matter field can decay into any kinematically allowed gravity excitation. Since the mass difference between individual gravity excitations is of the order $eV^{-1}$, the number of allowed gravity-mediated decay modes is large. Ergo, the GMD width of the matter fields can be significant and compete with the corresponding KKNC decay width. The total decay width can be expressed as the sum of widths into individual gravity excitations, i.e.,
\begin{equation}
    \Gamma = \sum_{\vec{n}} \Gamma_{\vec{n}} = \sum \Gamma_{h^{\vec{n}}} + \Gamma_{A^{\vec{n}}} + \Gamma_{\phi^{\vec{n}}}
\end{equation}
Since the mass difference between individual gravity excitations is almost negligible, the sum can be replaced by an integral over the graviton density of states (see \cite{Ghosh_2012}), i.e.,
\begin{equation}
    \Gamma = \frac{M_{\text{Pl}}^2}{M_D^{N+2}} \int \,dm \,d\Omega \, \Gamma_{\vec{n}} m_{\vec{n}}^{N-1}
\end{equation}
where $M_{\text{Pl}}$ is the four-dimensional Planck mass, $M_D$ is the fundamental Planck mass, and $m_{\vec{n}}$ is the mass of the level $\vec{n}$ gravity excitation. Figure \ref{fig:grav_mass_dist} shows the normalized gravity-mediated partial widths of a level-1 KK-gluon as a function of the masses of the gravity excitations for three different values of $N$. In Figure \ref{fig:grav_mass_dist}, we have assumed $R^{-1}=500$ GeV. For $N = 2$, the light gravity modes mostly contribute, while for $N = 6$, the massive gravitons dominate the decay width. 

\begin{figure}[htb!]
    \centering
    \includegraphics[width=0.7\columnwidth]{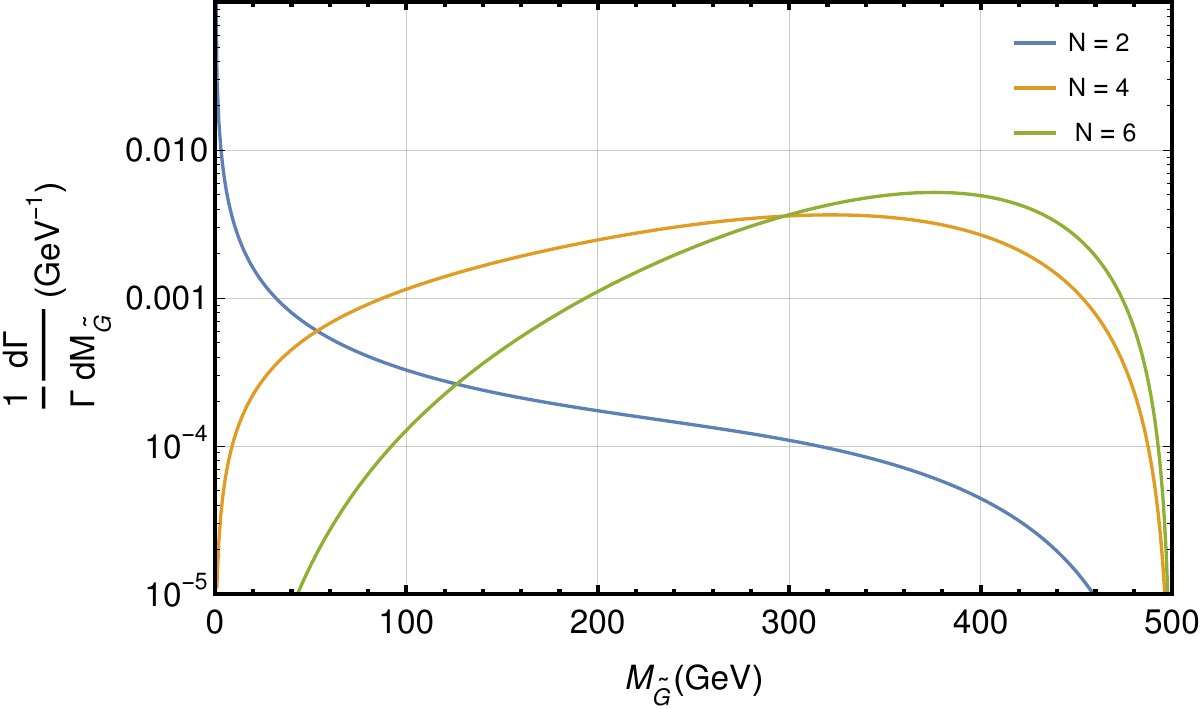}
    \caption{\label{fig:grav_mass_dist} The normalized partial decay width for level-1 KK-gluon into a gluon and gravity excitations as a function of the mass of the gravity excitations, for $R^{-1} = 500$ GeV and $M_D = 5$ TeV.}
\end{figure}

\subsection{Collider Signatures}
\label{sec:signatures}

Level-1 KK particles, after being pair-produced at colliders, can decay via KKNC or gravity-mediated processes. While KKNC decays result in relatively soft \footnote{The KKNC decays allow heavier level-1 KK particles to decay into lighter level-1 KK particles in association with one or more SM quarks or leptons. The hardness of the SM quarks/leptons depends on the mass splitting between the parent and daughter level-1 KK particles. Due to the small mass splitting between level-1 KK particles resulting from radiative corrections, the final state resulting from the KKNC decays typically features soft jets and leptons.} jets and leptons, gravity-mediated decays produce hard \footnote{Gravity-mediated processes allow level-1 KK particles to decay into a gravity excitation accompanied by the corresponding SM particle. The gravity excitations, resulting from the compactification of the large extra dimensions, can have a significant mass difference compared to the level-1 KK particles. Consequently, the SM jets, leptons, photons, W/Z/Higgs bosons, and top quarks, produced from the decays of level-1 light quarks/gluons, leptons, W/Z/Higgs bosons, and top quarks, respectively, are usually energetic.
} jets, leptons, photons, and high-$p_T$ massive SM bosons (W/Z/Higgs bosons) or top quarks. Therefore, the collider signature of the fat brane minimal Universal Extra Dimensions (mUED) scenario crucially depends on the relative strengths of KKNC decays and gravity-mediated decays.

\begin{figure}[htb!]
    \centering
    \includegraphics[width=0.45\columnwidth]{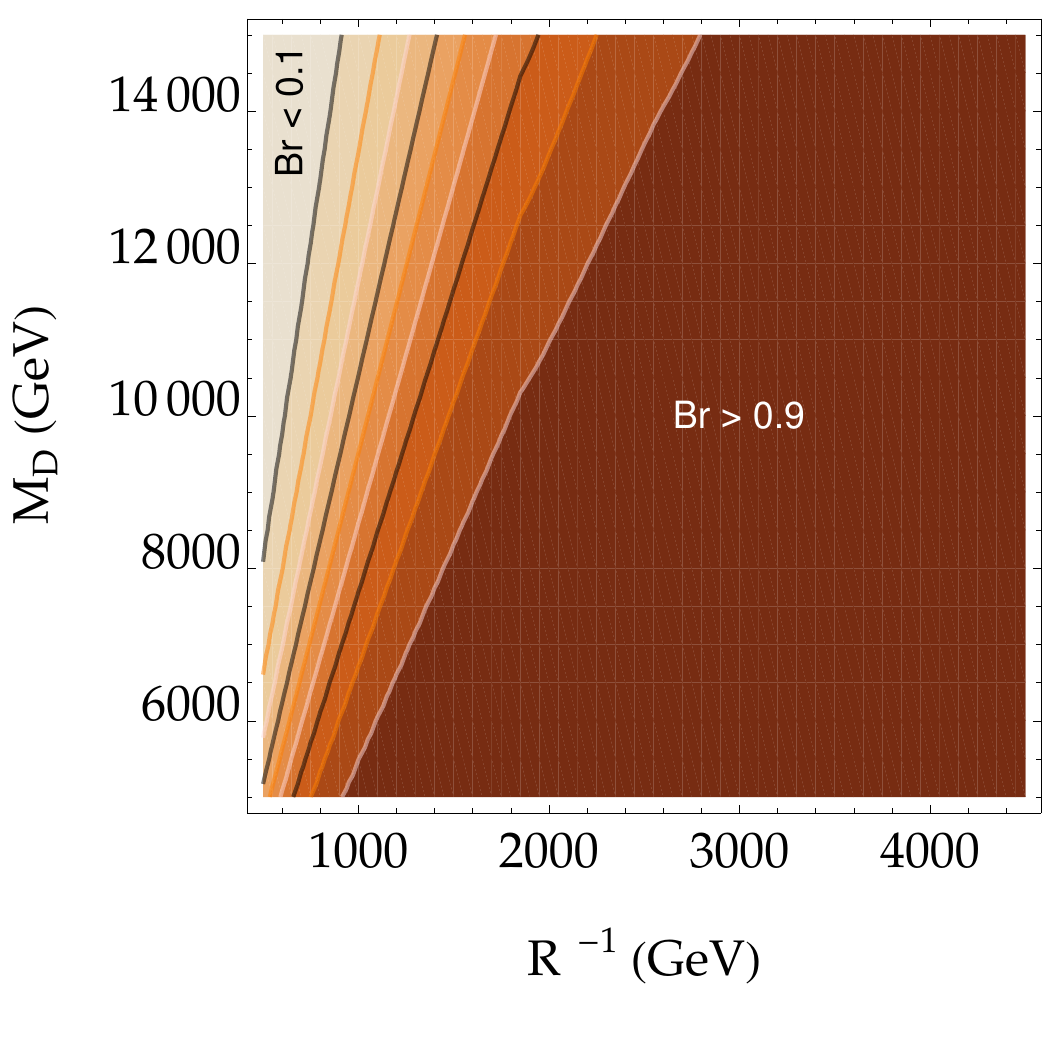} \qquad
    \includegraphics[width=0.45\columnwidth]{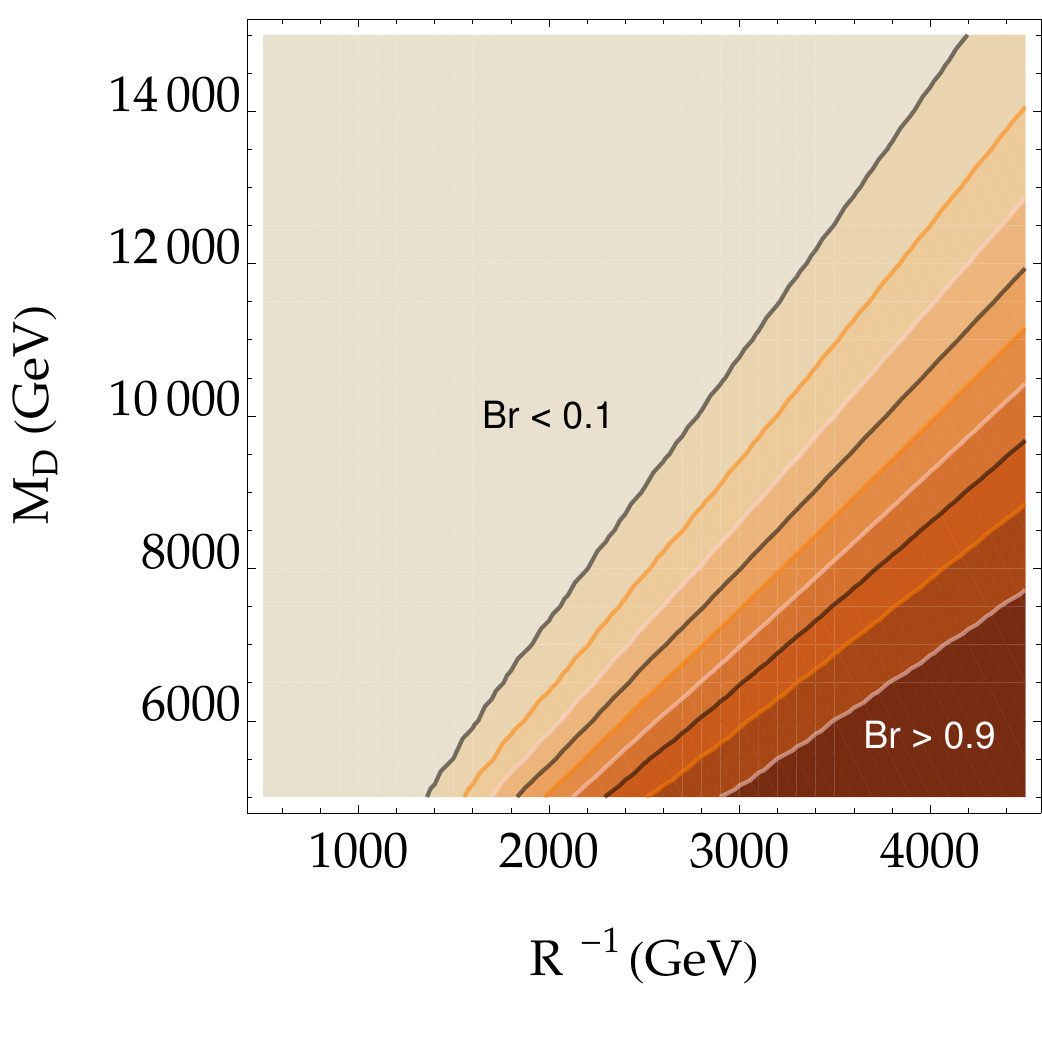} \qquad
    \includegraphics[width=0.45\columnwidth]{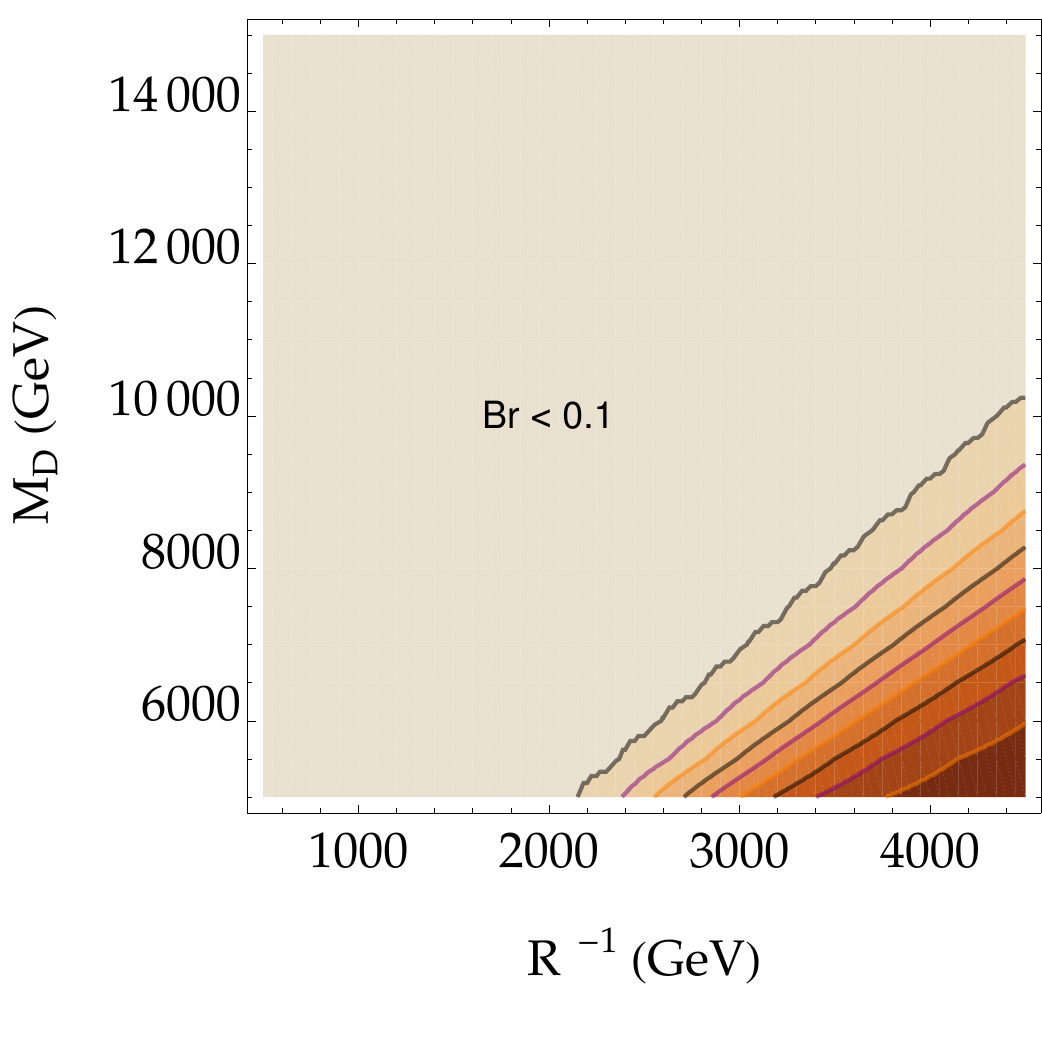}
    \caption{\label{fig:branch_ratios} The branching ratios for the gravity-mediated decays of the level-1 KK-gluons (indicated by the color gradient) on the $R^{-1}$--$M_D$ plane for three different values of the number of large extra dimensions: $N=2$ (top left panel), $N=4$ (top right panel), and $N=6$ (bottom panel).}
\end{figure}

The KKNC decays are determined by SM parameters as well as the masses and mass-splitting between the KK particles, which are controlled by the radius of compactification of the small extra dimension, $R$, and the cut-off scale, $\Lambda$. On the other hand, the gravity-mediated decay widths depend on the masses of the KK particles as well as the fundamental (4+N)-dimensional Planck mass, $M_D$, and the number of large extra dimensions, $N$. For a fixed cut-off scale $\Lambda = 5 R^{-1}$, the branching ratios of gravity-mediated decays and KKNC decays depend only on $R^{-1}$ and $M_D$. In Fig. \ref{fig:branch_ratios}, we present the branching ratios for the gravity-mediated decays of the level-1 KK-gluons (indicated by the color gradient) on the $R^{-1}$--$M_D$ plane for three different values of the number of large extra dimensions: $N=2$ (top left panel), $N=4$ (top right panel), and $N=6$ (bottom panel). Fig. \ref{fig:branch_ratios} shows that for $N=2$, gravity-mediated decays dominate over almost the entire $R^{-1}$--$M_D$ plane, while for $N=6$, KKNC decays become more significant. The different possible final state signatures resulting from the pair production of level-1 quarks/gluons at the LHC are summarized in the following. 
\begin{figure}[htb!]
\begin{tikzpicture}
\begin{feynman}
\vertex[large, dot] (c) {};
\vertex[right=0.4cm of c] (cr);
\vertex[left=0.4cm of c] (cl);
\vertex[right=0.7cm of cr] (cr1);
\vertex[right=1.4cm of cr] (r1);
\vertex[right=0.3cm of r1] (r10);
\vertex[right=1.4cm of r1] (r2);
\vertex[right=0.3cm of r2] (r20);
\vertex[right=1.6cm of r2] (r3);
\vertex[right=0.3cm of r3] (r30);
\vertex[right=1.4cm of r3] (r4);
\vertex[right=0.3cm of r4] (r40);
\vertex[right=1.4cm of r4] (r5);
\vertex[right=0.7cm of r5] (r50);

\vertex[above=1cm of r10] (r100){$g$};
\vertex[above=1cm of r20] (r200){$Q$};
\vertex[above=1cm of r30] (r300){$e^{\mp}/\nu$};
\vertex[above=1cm of r40] (r400){$e^{\pm}$};
\vertex[above=1cm of r50] (r500){$\gamma /Z$};
\vertex[below=1cm of r50] (r501){$G^{n}$};

\vertex[left=0.7cm of cl] (cl1);
\vertex[left=1.4cm of cl] (l1);
\vertex[left=0.3cm of l1] (l10);
\vertex[left=1.4cm of l1] (l2);
\vertex[left=0.3cm of l2] (l20);
\vertex[left=1.4cm of l2] (l3);
\vertex[left=0.7cm of l3] (l30);

\vertex[above=1cm of l10] (l100){$g$};
\vertex[above=1cm of l20] (l200){$q$};
\vertex[above=1cm of l30] (l300){$\gamma /Z$};
\vertex[below=1cm of l30] (l301){$G^{n}$};

\vertex[right=0.2cm of c] (u10);
\vertex[above=0.2cm of u10] (u1);
\vertex[above=1.0cm of cr1] (u2) {$p$};
\vertex[left=0.2cm of c] (d10);
\vertex[below=0.2cm of d10] (d1);
\vertex[below=1.0cm of cl1] (d2) {$p$};

\diagram*{
(cr) -- [gluon, edge label=\(g^{1}\), near end] (r1) -- [fermion, edge label=\(Q^1\), near end] (r2),
(r2) -- [boson, edge label=\(W^{1}/Z^1\)] (r3) -- [fermion, edge label=\(E^{1\pm}\), near end](r4),
(r4) -- [photon, edge label=\(\gamma^1\), near end] (r5) -- [photon] (r500), (r5) -- [scalar] (r501),
r1 --[gluon](r100), r2 --[fermion](r200), r3 --[fermion](r300), r4 --[fermion](r400),

(cl) -- [gluon, edge label=\(g^{1}\), near end] (l1) -- [fermion, edge label=\(q^1\), near end] (l2),
(l2) -- [photon, edge label=\(\gamma^{1}\), near end] (l3) -- [photon] (l300), (l3) -- [scalar] (l301),
l1 --[gluon](l100), l2 --[fermion](l200),

(u2) -- [fermion] (u1),
(d2) -- [fermion] (d1)
};
%
%\draw [decoration={brace}, decorate] (r41.north west) -- (r71.south west) node [pos=0.5, right] {\(H^{++}\operatorname{-jet}\)};
\end{feynman}
\end{tikzpicture}
\caption{\label{fig: KKNC_Feyn} Schematic Feynman diagram for $p p \to g^1 g^1$ and its subsequent decays in the scenario when KKNC decays dominate.}
\end{figure}
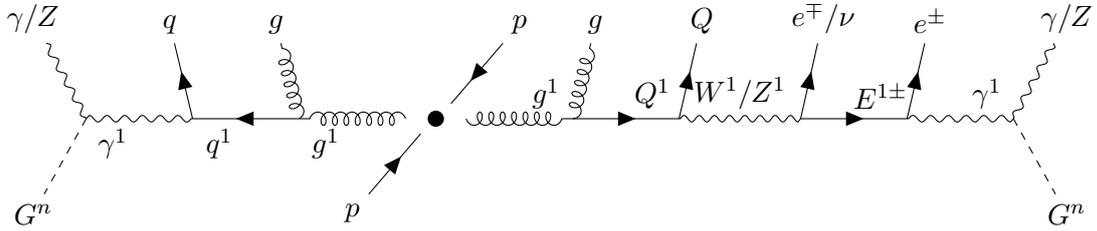

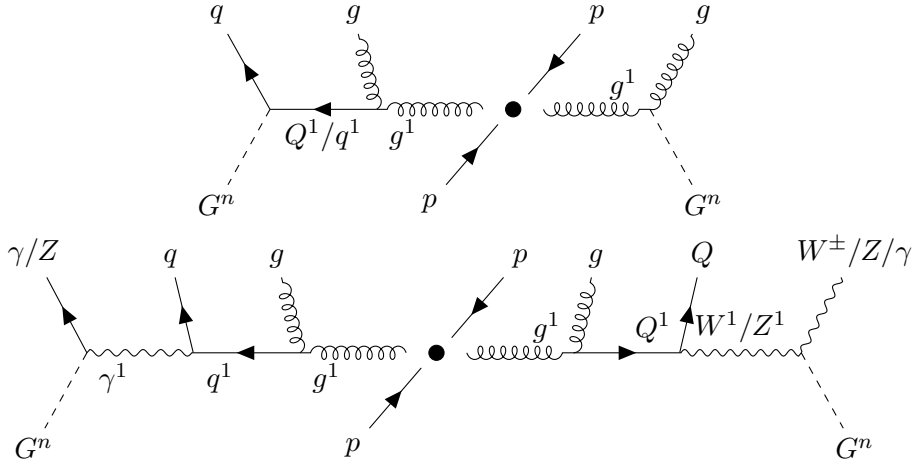
\begin{figure}[htb!]
\begin{center}
\begin{tikzpicture}
\begin{feynman}
\vertex[large, dot] (c) {};
\vertex[right=0.4cm of c] (cr);
\vertex[left=0.4cm of c] (cl);
\vertex[right=0.7cm of cr] (cr1);
\vertex[right=1.4cm of cr] (r1);
\vertex[right=0.7cm of r1] (r10);

\vertex[above=1cm of r10] (r100){$g$};
\vertex[below=1cm of r10] (r101){$G^{n}$};

\vertex[left=0.7cm of cl] (cl1);
\vertex[left=1.4cm of cl] (l1);
\vertex[left=0.3cm of l1] (l10);
\vertex[left=1.4cm of l1] (l2);
\vertex[left=0.7cm of l2] (l20);

\vertex[above=1cm of l10] (l100){$g$};
\vertex[above=1cm of l20] (l200){$q$};
\vertex[below=1cm of l20] (l201){$G^{n}$};

\vertex[right=0.2cm of c] (u10);
\vertex[above=0.2cm of u10] (u1);
\vertex[above=1.0cm of cr1] (u2) {$p$};
\vertex[left=0.2cm of c] (d10);
\vertex[below=0.2cm of d10] (d1);
\vertex[below=1.0cm of cl1] (d2) {$p$};

\diagram*{
(cr) -- [gluon, edge label=\(g^{1}\), near end] (r1) -- [scalar] (r101),r1 --[gluon](r100),

(cl) -- [gluon, edge label=\(g^{1}\), near end] (l1) -- [fermion, edge label=\(Q^1/q^1\)] (l2) --[fermion] (l200),(l2) --[scalar](l201),
l1 --[gluon](l100),

(u2) -- [fermion] (u1),
(d2) -- [fermion] (d1)
};
%
%\draw [decoration={brace}, decorate] (r41.north west) -- (r71.south west) node [pos=0.5, right] {\(H^{++}\operatorname{-jet}\)};
\end{feynman}
\end{tikzpicture}

\begin{tikzpicture}
\begin{feynman}
\vertex[large, dot] (c) {};
\vertex[right=0.4cm of c] (cr);
\vertex[left=0.4cm of c] (cl);
\vertex[right=0.7cm of cr] (cr1);
\vertex[right=1.4cm of cr] (r1);
\vertex[right=0.3cm of r1] (r10);
\vertex[right=1.4cm of r1] (r2);
\vertex[right=0.3cm of r2] (r20);
\vertex[right=1.6cm of r2] (r3);
\vertex[right=0.7cm of r3] (r30);

\vertex[above=1cm of r10] (r100){$g$};
\vertex[above=1cm of r20] (r200){$Q$};
\vertex[above=1cm of r30] (r300){$W^{\pm}/Z/\gamma$};
\vertex[below=1cm of r30] (r301){$G^{n}$};

\vertex[left=0.7cm of cl] (cl1);
\vertex[left=1.4cm of cl] (l1);
\vertex[left=0.3cm of l1] (l10);
\vertex[left=1.4cm of l1] (l2);
\vertex[left=0.3cm of l2] (l20);
\vertex[left=1.4cm of l2] (l3);
\vertex[left=0.7cm of l3] (l30);

\vertex[above=1cm of l10] (l100){$g$};
\vertex[above=1cm of l20] (l200){$q$};
\vertex[above=1cm of l30] (l300){$\gamma/Z$};
\vertex[below=1cm of l30] (l301){$G^{n}$};

\vertex[right=0.2cm of c] (u10);
\vertex[above=0.2cm of u10] (u1);
\vertex[above=1.0cm of cr1] (u2) {$p$};
\vertex[left=0.2cm of c] (d10);
\vertex[below=0.2cm of d10] (d1);
\vertex[below=1.0cm of cl1] (d2) {$p$};

\diagram*{
(cr) -- [gluon, edge label=\(g^{1}\), near end] (r1) --[fermion, edge label=\(Q^{1}\), near end]  (r2),
(r2) --[boson,edge label=\( W^1/Z^1\)](r3) -- [boson](r300), (r3) --[scalar](r301),
r1 --[gluon](r100),r2 --[fermion](r200),

(cl) -- [gluon, edge label=\(g^{1}\), near end] (l1) -- [fermion, edge label=\(q^1\), near end] (l2) --[boson, edge label=\(\gamma^{1}\), near end] (l3),
l3 --[fermion](l300),l3 --[scalar](l301),
l1 --[gluon](l100),l2 --[fermion](l200),
(u2) -- [fermion] (u1),
(d2) -- [fermion] (d1)
};
%
%\draw [decoration={brace}, decorate] (r41.north west) -- (r71.south west) node [pos=0.5, right] {\(H^{++}\operatorname{-jet}\)};
\end{feynman}
\end{tikzpicture}

\caption{\label{fig: GMD_Feyn} Schematic Feynman diagram for $p p \to g^1 g^1$ and subsequent decays in the scenario when GMD decays dominate.}
\end{center}
\end{figure}

\begin{enumerate}
    \item {\bf Scenarios with KKNC decays dominating over the  GMDs:} 
In the scenario where KKNC decays dominate over gravity-mediated decays, the pair-produced level-1 KK quarks/gluons decay into the lightest level-1 KK particle, namely the level-1 KK excitation of the $U(1)_Y$ gauge boson denoted by $\gamma_1$, through cascades involving other level-1 KK particles. Such decay cascades give rise to multiple quarks and leptons in the final state. In the absence of any kinematically allowed KKNC decay for the lightest level-1 KK particle, $\gamma_1$ undergoes gravity-mediated decay into a photon or SM Z-boson in association with a gravity excitation. Therefore, in this scenario, the pair production of level-1 quarks/gluons at the LHC leads to hard photons/Z-bosons plus soft jets/leptons in association with large missing transverse energy resulting from the invisible gravity excitations. The Feynman diagram depicting the pair production and subsequent decay of level-1 gluons is presented in Fig. \ref{fig: KKNC_Feyn}.
\item {\bf Scenarios with GMDs dominating over the KKNC decays:} In the scenario where gravity-mediated decays dominate over KKNC decays, the pair-produced level-1 KK quarks/gluons undergo gravity-mediated decays into a gravity excitation in association with a SM quark or gluon. Therefore, the pair and associated production of level-1 KK excitations of the light SM quarks and gluons at the LHC gives rise to hard jets and a large missing transverse energy (resulting from the invisible gravity excitations) signature. On the other hand, the pair production of level-1 top quarks results in boosted top quarks plus missing transverse energy in the final state. The Feynman diagram in Fig. \ref{fig: GMD_Feyn} (top panel) schematically shows the production and possible decays of level-1 gluons.

\item {\bf Scenarios with comparable gravity mediated and KKNC decay widths:} The Feynman diagram in Fig. \ref{fig: GMD_Feyn} (bottom panel) schematically shows the production and possible decays of level-1 gluons in a scenario with comparable decay widths for gravity-mediated decays (GMDs) and KKNC decays. Note that the KKNC decay widths are enhanced by color factors for the strongly interacting level-1 particles. The Feynman diagram in Fig. \ref{fig: GMD_Feyn} (bottom panel) corresponds to a region of parameter space with comparable gravity-mediated and KKNC decay widths for the color singlet level-1 particles. In this case, level-1 quarks/gluons follow the KKNC decay modes, while level-1 weak gauge bosons or leptons undergo gravity-mediated decays, leading to interesting high $p_T$ photons/massive SM gauge bosons or Higgs bosons in the final state.
\end{enumerate}

To summarize the signatures of fat brane realization of mUED at the LHC, the pair and associated production of level-1 quarks/gluons at the LHC within the framework of the fat brane realization of mUED results in conventional mono-photon, di-photon, and multi-jet final states associated with large missing transverse energy. These signatures have been extensively studied by the CMS and ATLAS collaborations of the LHC in the context of different supersymmetric scenarios. Depending on the region of the $R^{-1}$--$M_D$ plane and the number of large extra dimensions $N$, the production and decay of level-1 quarks/gluons might also give rise to interesting final states with boosted (high-$p_T$) W/Z/Higgs bosons or top quarks. In the following, we first obtain bounds on the $R^{-1}$--$M_D$ plane for different values of $N$ by recasting the LHC results from mono-photon, di-photon, and multi-jet plus large missing transverse energy searches in the context of our model. After establishing the bounds on the $R^{-1}$--$M_D$ plane, we focus on novel search strategies that use machine learning-based algorithms to tag boosted W/Z/Higgs bosons or top quarks. These strategies aim to suppress the SM background and thus enhance the sensitivity of future LHC runs.

\subsection{Bounds from the existing LHC searches}
\label{sec:bounds}
Final state topologies with multiple hard photons (mono-photon or di-photon signatures) or jets (multi-jet signatures) plus large missing transverse momentum are common in various BSM scenarios, particularly in different versions of supersymmetric models. These signatures have been extensively studied by the CMS and ATLAS collaborations at the LHC. Although LHC studies are designed to search for specific BSM scenarios, the model-independent bounds on visible signal cross-sections in different signal regions can be used to constrain the parameter space of other BSM scenarios with similar signatures. In this work, we study three ATLAS searches at the 13 TeV LHC: the di-photon plus missing transverse energy search with 36.1 fb$^{-1}$ luminosity data, the multi-jets plus missing transverse energy search, and the mono-photon plus missing transverse energy search with 139 fb$^{-1}$ integrated luminosity data. These analyses are designed to search for SUSY particles in various SUSY scenarios. We use the model-independent bounds on visible signal cross-sections in different signal regions to constrain the $R^{-1}$--$M_D$ plane for different values of the number of large extra dimensions $N$. Before presenting the constraints on the $R^{-1}$--$M_D$ plane, we briefly discuss the ATLAS searches, the technical details of our implementation and validation of these searches, and the simulation of events in the framework of the fat brane mUED model.

\subsubsection{ATLAS  mono-photon plus missing transverse energy search \cite{ATLAS:2022ckd}}
\label{sec: monophoton}
 Several experimental searches at the LHC \cite{ATLAS:2016poa, CMS:2017qca, CMS:2017brl, ATLAS:2015dcr, ATLAS:2018nud} have investigated the diphoton/mono-photon final states. In the absence of any excess over the SM background predictions, stringent constraints have been set on the model-independent visible cross-section. In this work, we use the ATLAS mono-photon search result presented in Ref. \cite{ATLAS:2022ckd} to constrain the parameter space of our model. In Ref. \cite{ATLAS:2022ckd}, the ATLAS collaboration searched for events with at least one hard ($p_T > 145$ GeV) photon in association with large missing transverse energy ($E_T^{\text{miss}}$) in the context of SUSY scenarios with gauge-mediated SUSY breaking (GMSB). In GMSB-type SUSY scenarios with gravitino as the lightest SUSY particle (LSP), the decay of the lightest neutralino ($\tilde\chi_1^0$), the next-to-lightest SUSY particle (NLSP), into the gravitino LSP in association with a photon or Z/Higgs boson gives rise to final states with hard photons and large $E_T^{\text{miss}}$ resulting from the invisible gravitinos in the LHC detector.

\begin{table}[h!]
    \centering
    \begin{tabular}{|l|l|}
        \hline\hline
        \multicolumn{2}{c}{{\bf Object Reconstruction Criteria} \cite{ATLAS:2022ckd}}\\
        \hline\hline
        \textbf{Object} & \textbf{Criteria} \\
        \hline
        Photon & 
        \begin{tabular}[c]{@{}l@{}}
            $p_T > 50$ GeV, $|\eta| \in [0, 1.37] \cup [1.52, 2.37]$;
            Isolation criteria \cite{ATLAS:2022ckd}
        \end{tabular} \\
        \hline
        Electron & 
        \begin{tabular}[c]{@{}l@{}}
            $p_T > 25$ GeV, $|\eta| \in [0, 1.37] \cup [1.52, 2.47]$;
            Loose isolation criteria
        \end{tabular} \\
        \hline
        Muon & 
        \begin{tabular}[c]{@{}l@{}}
            $p_T > 25$ GeV,
            $|\eta| < 2.7$; 
            Loose isolation criteria
        \end{tabular} \\
        \hline
        Jet & 
        \begin{tabular}[c]{@{}l@{}}
            Anti-$k_T$ algorithm \cite{Cacciari:2008gp, Cacciari:2011ma} with $R=0.4$,
            $p_T > 30$ GeV, 
            $|\eta| < 2.5$
        \end{tabular} \\
        \hline
        $E_T^{\text{miss}}$ & 
        \begin{tabular}[c]{@{}l@{}}
            The negative vector sum of transverse momenta of all visible objects, including\\
            the calorimeter energy deposits not matching any reconstructed objects
        \end{tabular} \\
        \hline\hline
        \multicolumn{2}{c}{{\bf Object Isolation Criteria} \cite{ATLAS:2022ckd}}\\
        \hline\hline
        $\gamma$-Lepton & Photon candidates within $\Delta R_{\gamma l} < 0.4$ of a lepton are removed \\
        \hline
        e-Jet & 
        \begin{tabular}[c]{@{}l@{}}
            Jets removed if an electron is within $\Delta R_{ej} < 0.2$ \\
            Electrons removed for $0.2 < \Delta R_{ej} < 0.4$
        \end{tabular} \\
        \hline
        $\mu$-Jet & 
        \begin{tabular}[c]{@{}l@{}}
            Muons within $\Delta R_{\mu j} < 0.4$ of a jet are removed if the jet has \\
            at least three good-quality tracks; otherwise, the jet is removed
        \end{tabular} \\
        \hline
        $\gamma$-Jet & Jets within $\Delta R_{\gamma j} < 0.4$ of a photon are eliminated \\
        \hline\hline
        \multicolumn{2}{c}{{\bf Electron/Jets faking photon}}\\
        \hline\hline
        \multicolumn{2}{|c|}{The possibilities of electrons and jets faking as photons are considered according} \\ \multicolumn{2}{|c|}{to Ref. \cite{CMS:2012ad, Haase:2011eja}}\\
        \hline\hline
    \end{tabular}
    \caption{Summary of object reconstruction criteria used in Ref. \cite{ATLAS:2022ckd} as well as in our implementation.}
    \label{tab:criteria}
\end{table}

As discussed in the previous section, the pair-production and decay of the level-1 KK particles in the framework of fat brane realization of mUED also give rise to similar final state topologies with photons plus $E_T^{\text{miss}}$. However, the kinematics for the decay of SUSY particles in GMSB and KK-particles in fat brane realization of mUED are very different. For example, while the NLSP $\tilde\chi_1^0$ in the GMSB scenario decays into a single light gravitino by emitting a hard photon/Z/Higgs boson, the next-to-lightest KK particle (NLKP) $\gamma_1$ in fat brane realization of mUED can decay into any kinematically allowed member of the tower of gravity excitations. Therefore, the $p_T$ spectrum of the photon or Z-boson resulting from the NLKP decay is not necessarily always very hard. As a result, the bounds obtained in Ref. \cite{ATLAS:2022ckd} on the gluino-neutralino mass plane are not applicable to constrain the $m_{g_1}$--$m_{\gamma_1}$ plane of fat brane mUED. However, the model-independent bounds on visible signal cross-sections in different mono-photon signal regions in Ref. \cite{ATLAS:2022ckd} are applicable after proper simulation of the fat brane mUED events following the object reconstruction and event selection criteria of Ref. \cite{ATLAS:2022ckd}. A detailed discussion about object reconstruction, event selection, and mono-photon signal region definitions, which we have implemented in our event simulation, can be found in Ref. \cite{ATLAS:2022ckd} and is also tabulated in Table \ref{tab:criteria} and \ref{tab: Mono_Photon}. Tables \ref{tab:criteria} and \ref{tab: Mono_Photon} provide a summary of the criteria used in our analysis, as implemented by the ATLAS collaboration. Table \ref{tab:criteria} details the object reconstruction and isolation criteria for different physics objects, while Table \ref{tab: Mono_Photon} outlines the criteria for the three signal regions, SRL, SRM, and SRH, used to probe different areas of the gluino-$\tilde{\chi}_1^0$ mass plane in GMSB in Ref. \cite{ATLAS:2022ckd}. In our analysis, we have used the model-independent bounds presented in Table \ref{tab: Mono_Photon} on the visible signal cross-section in the signal regions SRL, SRM, and SRH to constrain the parameter space of the fat-brane realization of mUED. The implementation of object reconstruction criteria in Table \ref{tab:criteria}, and signal region definition in Table \ref{tab: Mono_Photon}, as well as the validation of our implementation, are discussed in the following.

\begin{table}[htb!]
	\begin{center}
		\begin{tabular}{l|c|c|c}
			\toprule 
			 & SRL & SRM & SRH \\ 
			\toprule
			$N_{photons}$ & $\ge 1$& $\ge 1$& $\ge 1$\\
			
			$p_T^{leading-\gamma}$ &$> 145$ GeV &$ > 300$ GeV& $>400$ GeV \\
			
			$N_{leptons}$ &0 & 0& 0\\
			
			$N_{jets}$ &$\ge 5$ &$\ge 5$& $\ge 3$\\
			
			$\Delta \phi(jet,E_T^{miss})$ &$> 0.4$ &$> 0.4$&$> 0.4$\\
			
			$\Delta \phi(\gamma,E_T^{miss})$ &$> 0.4$ &$> 0.4$ &$> 0.4$\\
			
			$E_T^{miss}$ &$> 250$ GeV &$> 300$ GeV& $> 600$ GeV\\
			
			$H_T$ &$> 2000$ GeV &$> 1600$ GeV& $> 1600$ GeV\\
			
			$R_T^4$ &$< 0.90$ &$< 0.90$& - \\
            \bottomrule
            \bottomrule
            $\left\langle \epsilon \sigma\right\rangle^{95}_{obs}$(fb) &$0.034$ &$0.022$& $0.054$ \\

			\bottomrule
		\end{tabular} 
		\caption{\label{tab: Mono_Photon} Cuts used by the ATLAS collaboration to define the three signal regions along with the model-independent 95 \% Confidence Level upper bound on the visible cross-section. Here, $H_T$ represents the scalar sum of the transverse momentum of all signal region jets and the leading photon, while $R_T^4$ represents the ratio of the scalar sum of the transverse momentum of the four leading signal region jets to the scalar sum of the transverse momentum of all signal region jets in the event. The other symbols carry their usual meaning.}
	\end{center}
\end{table}

\paragraph{Implementation \& Validation:} The results of the mono-photon plus $E_T^{\text{miss}}$ analysis in Ref.~\cite{ATLAS:2022ckd} are interpreted as bounds on the gluino NLSP-neutralino mass plane in the context of a simplified GMSB scenario. To validate our implementation of the object reconstruction and event selection procedure described in Ref.~\cite{ATLAS:2022ckd}, we reproduced the cut-flow tables from Ref.~\cite{ATLAS:2022ckd} for a given set of parameters of the simplified GMSB scenario\footnote{While reproducing the limits and cut-flow tables in Ref.~\cite{ATLAS:2022ckd}, we identified several simplified assumptions adopted in Ref.~\cite{ATLAS:2022ckd} that are not theoretically consistent. These assumptions lead to bounds on the gluino NLSP-neutralino mass plane that should be interpreted cautiously. For example, Ref.~\cite{ATLAS:2022ckd} considers the pair production and SUSY cascade decays of gluinos to the neutralino NLSP, followed by gravity-mediated decays of the NLSP to a gravitino in association with a photon or a Z/Higgs boson. However, we found that, for the given set of parameters, the gravity-mediated decays of different SUSY particles in the cascade, which were neglected in Ref.~\cite{ATLAS:2022ckd}, can be significant in different regions of the gluino NLSP-neutralino mass plane. This allows particles to decay into gravitinos instead of lighter SUSY particles in the cascade, leading to a reduced cross-section for the signal regions defined in Ref.~\cite{ATLAS:2022ckd}. Obtaining a realistic bound on the gluino NLSP-neutralino mass plane, considering all possible complexities associated with the decays of SUSY particles in the cascade is beyond the scope of this article. We plan to address these issues in a separate publication.
}. We used the default minimal supersymmetric standard model (MSSM) file provided with SARAH \cite{Staub:2013tta} and generated the particle spectrum and decay tables in SPheno \cite{Porod:2011nf}. Gluino pairs with up to two additional partons were generated in MG5\_AMC@NLO \cite{Alwall:2014hca}, and their subsequent decays, showering, and hadronization were simulated in \textsc{pythia8} \cite{Bierlich:2022pfr}. The gravitino decay modes of supersymmetric particles were implemented in the \textsc{Pythia} code using the analytical partial decay width expressions from Refs.~\cite{Ambrosanio:1996jn, Dutta:2017jpe}. Reconstruction of different physics objects was performed in the fast detector simulator \textsc{Delphes} \cite{deFavereau:2013fsa} using the prescription discussed in Section \ref{sec: monophoton}. To validate our implementation, we replicated the cut-flow table found in HEPData \cite{ATLAS:2022ckd}. The outcome for the signal region SRL is provided in Table \ref{tab: SRL}. Similar results were obtained for the SRM and SRH signal regions, and we present them in Appendix \ref{sec:appendix-2}. In Table \ref{tab: SRL}, we present our simulation results alongside those of the ATLAS collaboration in Ref.~\cite{ATLAS:2022ckd} to facilitate comparison, and there is a notable concurrence between the two sets of numbers.

\begin{table}[htb!]
	\begin{center}
		\begin{tabular}{l|c|c|c|c}
			\toprule 
            &\multicolumn{4}{r}{$m_{\Tilde{g}}=2000 GeV, m_{\Tilde{\chi_1^0}}=250 GeV$} \\ 
            \midrule
			  Cuts& \multicolumn{2}{c|}{$\gamma/Z$} & \multicolumn{2}{c}{$\gamma/h$} \\ 
			\midrule
             & Hepdata& Our Result & Hepdata  &  Our Result\\
            \midrule
			Trigger (one photon $p_T >$ 140 Gev) & $47.26$& $ 49.07 $ & $49.12$&  $ 46.21 $\\

             At least one photon                 & $47.19$& $ 49.07 $ & $49.02$&  $ 46.21 $\\

             Lepton Veto                         & $29.80$& $ 31.97 $ & $33.36$&  $ 33.07 $\\

             Leading photon $p_T >$ 145 Gev      & $26.42$& $ 31.49 $ & $30.54$&  $ 32.56 $\\

             $E_T^{miss} >$ 250 Gev             & $18.96$& $ 21.28 $ & $21.18$&  $ 21.89 $\\

             Number of Jets $\ge$ 5             & $18.65$& $ 21.21 $ & $20.82$&  $ 21.78 $\\

             $\Delta \phi(jet, E_T^{miss}) >$ 0.4 & $15.86$& $ 17.68 $ & $17.65$&  $ 17.59 $\\

             $\Delta \phi(\gamma, E_T^{miss}) >$ 0.4& $12.04$& $ 13.94 $ & $13.51$&  $ 14.10 $\\

            $H_T >$ 2000 GeV                        & $10.27$& $ 10.87 $ & $11.77$&  $ 11.19 $\\

            $R_T^4 <$ 0.9                         & $8.08$& $ 9.52 $ & $9.40$&  $ 9.61 $\\

			\bottomrule
		\end{tabular} 
		\caption{\label{tab: SRL} Cut-Flow table for the SRL signal region. The entries in the second and fourth columns are the results provided by the ATLAS collaboration \cite{ATLAS:2022ckd} in the form of Hepdata. The entries in the third and fifth columns represent our simulated results.}
	\end{center}
\end{table}

\subsubsection{ATLAS  multi-jet plus missing transverse energy search \cite{ATLAS:2019vcq}}
\label{sec:multijet}
The fat-brane realization of the mUED model can lead to final states with multiple hard jets and large missing transverse momentum when gravity-mediated decays dominate over the KKNC decay for level-1 quarks/gluons. In such scenarios, pair-produced level-1 quarks/gluons directly decay to a graviton excitation in association with an SM quark/gluon, giving rise to multiple hard jets at the LHC. Multi-jets plus $E_T^{\text{miss}}$ final states have been extensively searched at the LHC \cite{ATLAS:2019vcq, CMS:2016eep, CMS:2016eju, CMS:2016igb, CMS:2017abv, CMS:2017okm, CMS:2018ncl} as a signature of different BSM scenarios. We have considered the recent multi-jet search by the ATLAS collaboration in Ref. \cite{ATLAS:2019vcq} with the full (139 fb$^{-1}$ integrated luminosity) run-II data of the LHC at 13 TeV center-of-mass energy. Although the analysis in Ref. \cite{ATLAS:2019vcq} is dedicated to the search for squarks and gluinos in the context of supersymmetry, the model-independent 95\% CL upper limits on the visible $n_j + E_T^{\text{miss}}$ cross-sections ($\langle \sigma \rangle_{95}^{\text{obs}}$) for different signal regions (SRs) can be used to constrain the parameter space of other BSM scenarios like our scenario, which also gives rise to similar final states. In Ref. \cite{ATLAS:2019vcq}, the ATLAS collaboration defined ten signal regions for their model-independent study of multijet plus missing transverse energy final states. The signal regions are defined by varying numbers of jet multiplicities (between 2–6) along with the minimum value of the effective mass $m_{\text{eff}}$. In view of the high level of agreement between the predicted background and observed yield in all signal regions, a model-independent 95\% CL upper limit is set on the visible BSM contribution to the multijet cross-section ($\langle \sigma \rangle_{95}^{\text{obs}}$) for each signal region. In our analysis, we have used the ATLAS-derived bounds on $\langle \sigma \rangle_{95}^{\text{obs}}$ in each signal region to constrain the parameter space of the fat-brane realization of mUED.  We have closely followed the object reconstruction, event selection, and signal region definitions presented in Ref. \cite{ATLAS:2019vcq} to obtain bounds on the $R^{-1}$--$M_D$ plane for different values of $N$. The ATLAS multi-jet search in Ref. \cite{ATLAS:2019vcq} has already been implemented and validated in a previous work in Ref. \cite{Ghosh:2018mck} in the context of obtaining bounds on minimal and non-minimal UED models. For brevity, we omit the technical details of the search in Ref. \cite{ATLAS:2019vcq}. Interested readers are referred to Ref. \cite{Ghosh:2018mck} for more information.

\subsubsection{ATLAS di-Photon plus missing transverse energy search \cite{ATLAS:2018nud}}
\label{sec:diphoton}
Final state topology with two photons, many soft jets/leptons, and a large missing transverse momentum are common in models with general gauge mediation when the pair-produced squarks and gluinos decay via the cascade involving other lighter SUSY particles to the neutralino NLSP, which in turn, decay into a gravitino LSP in association with a photon. Our fat-brane mUED scenario mimics the above final state topology when the KKNC decays of level-1 KK particles dominate the GMD decay modes. Several analyses at the LHC \cite{ATLAS:2016poa, CMS:2017qca, CMS:2017brl, ATLAS:2015dcr, ATLAS:2018nud} have looked into the diphoton final state in the case of GGM-type models. The most recent ATLAS search for the diphoton plus missing transverse energy final states in Ref. \cite{ATLAS:2018nud} has already been recast in the context of the fat-brane realization of mUED in Ref. \cite{Ghosh:2018mck}, leading to a lower bound on $R^{-1}$ of 2900 (2700) GeV for $N=6(4)$ and $M_D=$ 15 TeV. Note that the ATLAS diphoton search in Ref. \cite{ATLAS:2018nud} corresponds to only 36.1 fb$^{-1}$ integrated luminosity data of the 13 TeV LHC. In this work, we have used the diphoton search results in Ref.~\cite{ATLAS:2018nud} with a twofold motivation:
\begin{enumerate}
    \item To reproduce the bounds on $R^{-1}$ in Ref.~\cite{Ghosh:2018mck} and validate our implementation of gravity-mediated decays of level-1 KK particles in \textsc{PYTHIA8}.
    \item To extrapolate the results in Ref.~\cite{ATLAS:2018nud} to obtain the reach at 139 fb$^{-1}$ integrated luminosity of the LHC. Note that the mono-photon and multi-jet searches discussed in the previous paragraphs correspond to 139 fb$^{-1}$ of data. Extrapolation of the diphoton search will make the diphoton search comparable to the mono-photon and multi-jet search bounds.
\end{enumerate}
The details of the object selection and signal region definition can be found in \cite{ATLAS:2018nud, Ghosh:2018mck}; we omit them here for brevity. However, for the sake of completeness of this article, we have presented the definitions of the signal regions used for the ATLAS diphoton analysis of Ref. \cite{ATLAS:2018nud} in Table \ref{tb:di-photon_cuts}. Table \ref{tb:di-photon_cuts} also shows the 95 \% CL bound on the visible cross-section in different signal regions that has been obtained by the ATLAS collaboration with 36.1 fb$^{-1}$ integrated luminosity data at the 13 TeV LHC. In the last row of Table \ref{tb:di-photon_cuts}, we have presented the 95 \% CL expected upper limits on the visible cross-section in different signal regions with 139 fb$^{-1}$ integrated luminosity of the LHC. To obtain the expected limits in the last row in Table \ref{tb:di-photon_cuts}, we extracted the background cross-section from the expected number of background events provided in Ref. \cite{ATLAS:2018nud} at 36.1 fb$^{-1}$ data for the different signal regions. Then, following Refs.~\cite{Cowan:2010js, Li:1983fv, Cousins:2007yta}, we use the following approximated expressions for the median expected exclusion significance to estimate the expected upper bound on the signal cross-section at 139 fb$^{-1}$ integrated luminosity:
\begin{equation*}
Z_{\rm exc} = \left[ 2 \left\{ s-b \ln \left( \frac{b+s+x}{2b} \right) - \frac{b^2}{\delta_b^2} \ln \left( \frac{b-s+x}{2b} \right) \right\} - (b+s-x)(1+b/\delta_b^2) \right]^{1/2}
\label{eq:zdis2},
\end{equation*}
where $x = \sqrt{(s+b)^2 - 4sb\delta_b^2/(b+\delta_b^2)}$, $s$ and $b$ are the number of signal and background events, respectively, and $\delta_b$ is the uncertainty in the measurement of the background. For our analysis, we adopt a conservative approach and assume an overall 40\% (statistical $+$ systematic) uncertainty in the measurement of the backgrounds. In the absence of any diphoton analysis with the full LHC run II data, we derived the expected reach of the di-photon channel on the $R^{-1}$-M$_D$ mass plane.
 \begin{table}
\begin{center}
\begin{tabular}{|c||c|c|}
\hline
\multirow{1}{*}{\textbf{Cuts}} & \textbf{$SR_{S-L}^{\gamma\gamma} $ } &\textbf{$SR_{S-H}^{\gamma\gamma} $ }\\
\hline
\hline
Number of photons & $\geq 2$& $\geq 2$ \\
\hline
$p_T(\gamma_1)>$ [GeV]& 75& 75\\
\hline
$p_T(\gamma_2)>$ [GeV]& 75& 75\\
\hline
$\slashed E_T>$ [GeV]& 150& 250\\
\hline
$H_T>$ [TeV] & 2.75 & 2.00\\
\hline
$\Delta \phi (\rm{jet},\slashed E_T)>$& 0.5 & 0.5\\
\hline
$\Delta \phi (\gamma,\slashed E_T)>$&- & 0.5\\
\hline
\hline
$<\epsilon\sigma>_{\rm{obs}}^{95}$ [fb] (${\cal L}=36.1$ fb$^{-1}$) \cite{ATLAS:2018nud}& 0.083 & 0.083\\
\hline
\hline
$<\epsilon\sigma>_{\rm{exp}}^{95}$ [fb] (${\cal L}=139$ fb$^{-1}$) & 0.0305 & 0.0305\\
\hline
\end{tabular}
\end{center}
\caption{Signal regions and cuts used by the ATLAS Collaboration \cite{ATLAS:2018nud} in di-photon search along with the observed 95\% C.L. upper limit on model-independent visible beyond the SM cross-section. Here, $H_T$ is the scalar sum of the transverse energy of photons, any additional jets, and leptons in an event. $\Delta \phi(\rm{jet},\slashed E_T)$ is the azimuthal separation between two leading jets with $p_T>75$ GeV and $\vec{\slashed E_T}$ vector. The other symbols carry their usual meaning.}
\label{tb:di-photon_cuts}
\end{table}

\subsubsection{Event Simulation}
\label{sec:Eventsimulation}
 To generate events for the pair/associated production of level-1 KK quarks/gluons in the framework of the fat-brane realization of the mUED scenario, we used the mUED model file \cite{Cheng:2002ab, Cheng:2002iz, Datta:2010us} provided in the Feynrules model database. The production of colored level-1 KK particles is simulated in MG5\_AMC@NLO \cite{Alwall:2014hca} with the  NNPDF21LO\cite{NNPDF:2014otw} parton distribution function (PDF). The subsequent decay, showering, and hadronization are simulated in Pythia8 \cite{Bierlich:2022pfr}. PYTHIA8 does not consider gravity-mediated decay for the KK particles. Moreover, these decays cannot be trivially incorporated by modifying the decay table of the KK particles in the SLHA file since gravity-mediated decays need to be simulated not to a single gravity excitation but to a member of the tower of gravity excitations. We have incorporated the gravity-mediated decay for the KK particles in PYTHIA8 by modifying the \texttt{PYTHIA PYWIDTH} subroutine according to our requirements. We used the fast detector simulator Delphes to simulate different physics objects like jets, leptons, and photons. For the mono-photon analysis, we have followed the object reconstruction criteria described in \ref{sec: monophoton}. Similarly, for the di-photon and multi-jet analysis, we have followed the prescription for object reconstruction as described in \cite{Ghosh:2018mck, Avnish:2020atn} respectively. To generate the bounds on the model parameter space for the mono-photon and multi-jet analysis, we have compared the signal cross-section in the different signal regions with the respective 95 \% confidence level upper limits provided by the ATLAS collaboration in Ref. \cite{ATLAS:2022ckd} and \cite{ATLAS:2019vcq}, respectively. On the other hand, for the di-photon analysis, we have reinterpreted the expected number of background events for 139 $fb^{-1}$ LHC and used it to calculate the 95 \% C.L upper limit on the model parameter space. In the next section, we will discuss the bounds on the model parameter space from the different collider analyses.
\subsubsection{Results}
\label{sec:boundresults}
For convenience, we have presented the collider bounds on the parameter space of fat brane realization of mUED for three distinct values of the number of large extra dimensions, denoted as $N=2$, $N=4$, and $N=6$. Once the value of N is fixed, the phenomenology of the model is governed by three independent parameters: the cut-off scale $\Lambda$, the radius of compactification for the small extra dimensions R, and the fundamental Planck mass $M_D$. Throughout our analysis, we have fixed $\Lambda R$ at 5. We have varied $R^{-1}$ within the 2 to 3.5 TeV range and the fundamental Planck mass $M_D$ within the 5 to 15 TeV range. We present our results in Figure \ref{fig: valplot} as bounds on the $R^{-1} Vs. M_D$ plane. The bounds on the $R^{-1}-M_D$ plane presented in Figure \ref{fig: valplot} for $N=2$ (top left panel), $N=4$ (top right panel), and $N=6$ (bottom panel) are discussed in the following.
\begin{itemize}
    \item {\bf Bounds for $N=2$ (top left panel):} For the $N=2$ scenario, the spacing between the individual gravity excitations is minimum, resulting in a higher density of graviton states compared to cases where N equals 4 and 6. Consequently, the level-1 KK particles have the freedom to decay into numerous graviton states, leading to a substantial GMD width (see top left panel of Figure \ref{fig:branch_ratios}). At the LHC, the pair produced level-1 quarks/gluons dominantly decay via gravity-mediated processes and result in a final state characterized by high $p_T$ jets and a significant missing transverse energy. This makes the multi-jet + $E_T^{miss}$ search the most sensitive for $N=2$ and effectively excludes a large part of the parameter space. For instance, it sets a lower limit on $R^{-1}$ of around 2975 GeV independent of the value of $M_D$. The mono-photon analysis, though not as effective as the multi-jet one, excludes $R^{-1} <$ 2534(2596) GeV for $M_D = $ 5(15) TeV. The Di-photon analysis, on the other hand, excludes a tiny portion of the $R^{-1}-M_D$ plane in the upper left corner where KKNC decays start becoming significant (see Figure \ref{fig:branch_ratios} top left panel).
    \item {\bf Bounds for $N=4$ (top right panel):} In this case, the GMD and KKNC decay widths have almost comparable strength (see upper right panel of Figure \ref{fig:branch_ratios}). The former decay mode prevails in the small $M_D$ region, leading to a final state characterized by multiple jets and a large $E_T^{miss}$ at the collider. Conversely, the KKNC decays dominate for larger values of $M_D$; hence, mono-photon and diphoton searches can effectively constrain this part of the parameter space. As depicted in \ref{fig: valplot} (top right panel), the multi-jet search excludes $R^{-1} <$ 2898(2800) GeV for $M_D = $ 5(15) TeV. For higher values of $M_D$, though the KKNC decays dominate, the final state $Z$ bosons, followed by its hadronic decay, can contribute to the multi-jet final state. Ergo, we see a comparable performance of the multi-jet SR throughout the range of $M_D$. The Mono-Photon analysis is most effective in the upper $M_D$ region and excludes $R^{-1} <$ 2958 GeV corresponding to $M_D = $ 15 TeV. The di-photon analysis also has comparable performance, and it sets a lower limit of 2874 GeV on $R^{-1}$ for $M_D =$ 15 TeV.

    \item {\bf Bounds for $N=6$ (bottom panel):} For a fixed value of $M_D$,  as we increase N, the density of graviton states decreases, resulting in a reduced GMD decay width. Therefore, for the N=6 case, the KKNC decays dominate over the GMD decays for most parts of the parameter space. In this scenario, the specific decay mode of the level-1 photon determines the resulting final state topology. If both $\gamma^1$ decay via the photon channel, it results in a di-photon final state. If one $\gamma^1$ follows the photon channel, we get the monophoton final state. Conversely, we get a multijet final state if both $\gamma^1$ decay to $Z$-boson followed by the hadronic decay of both the $Z$-boson. For smaller values of $M_D$, the strength of GMD decays increases, resulting in a multijet final state topology at the LHC. In the bottom panel of Figure \ref{fig: valplot}, we summarise our results for the N=6 scenario. The multi-jet signal region excludes $R^{-1} <$ 2871(2761) GeV for $M_D=$ 5(15) TeV. The monophoton signal region sets a lower limit of $R^{-1}=$ 2599(2886) GeV for $M_D = $ 5(15) TeV. The diphoton search is sensitive in the large $M_D$ region and leads to an expected reach of $R^{-1}$ up to 3000 GeV for $M_D = 15$ TeV at the LHC with 139 fb$^{-1}$ integrated luminosity.

\end{itemize}

\begin{figure}[htb!]
	\centering
	\includegraphics[width=0.45\columnwidth]{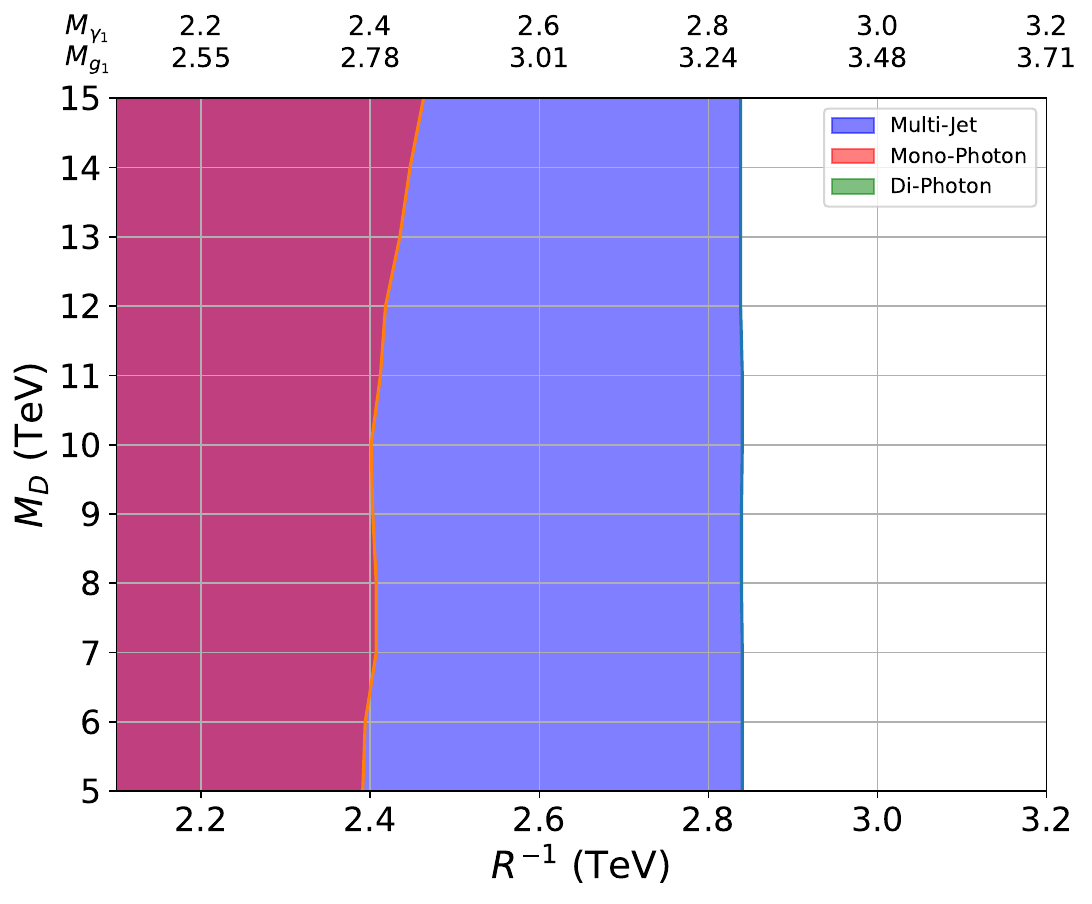} \qquad
	\includegraphics[width=0.45\columnwidth]{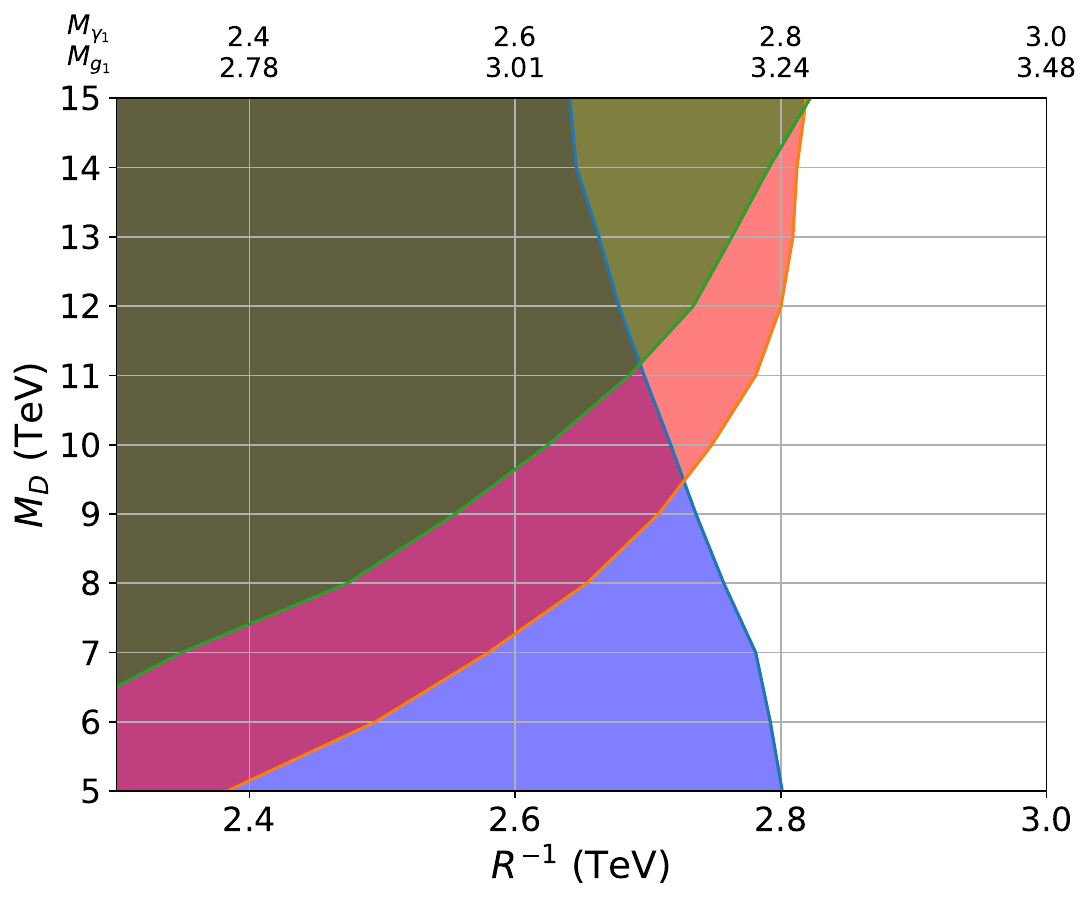} \qquad
    \includegraphics[width=0.45\columnwidth]{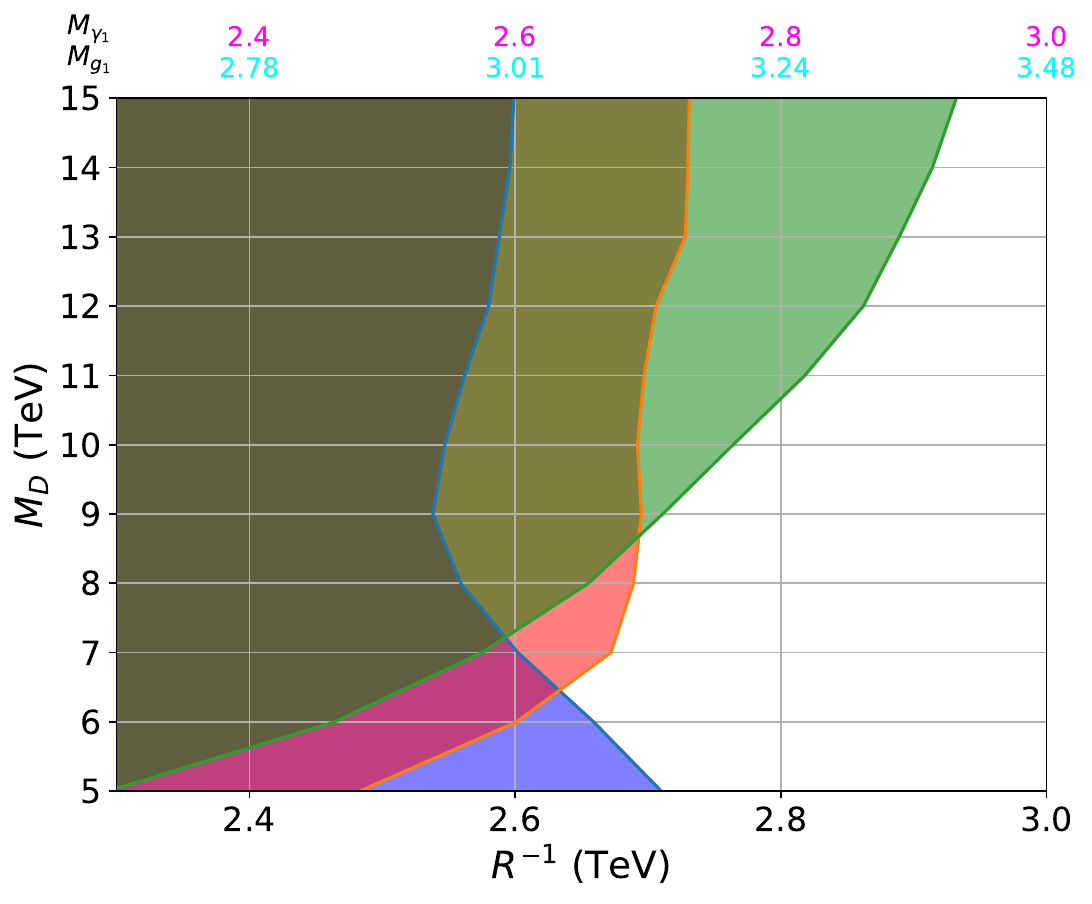}
	\caption{\label{fig: valplot} The exclusion limits for fat-brane MUED parameter space in the $R^{-1}$ Vs $M_D$ plane from the three ATLAS analysis \cite{ATLAS:2022ckd, ATLAS:2018nud, ATLAS:2019vcq}. Throughout this paper, we have fixed $\Lambda R = 5$. The three plots correspond to N=2(top left), N=4(top right), and N=6(bottom). The limits in the mono-photon and multijet cases are derived from the 95 \% C.L. upper limit on the visible cross-section $\left\langle \epsilon \sigma\right\rangle^{95}_{obs}$. For the Di-Photon signal region, we present the expected 2$\sigma$ C.L. lower limit on the model parameters for $139 fb^{-1}$ luminosity of data collected at the 13 TeV LHC. }
\end{figure}
\subsection{Optimized search strategies for the Future LHC Runs}
\label{sec: optimization}
The updated LHC bounds presented in Figure~\ref{fig: valplot} indicate that masses of the level-1 KK-gluon (the heaviest level-1 particle) and level-1 KK photon below approximately 3 TeV have already been excluded, based on the consistency of LHC data with the Standard Model background predictions. These LHC bounds are derived using the ATLAS mono-photon, diphoton, and multijets search results, which are optimized for various supersymmetric scenarios. For instance, the mono-photon and diphoton searches are designed to target GGM models, while the multijet analysis focuses on detecting squarks and gluons in the R-parity conserving MSSM model. Although the final state topologies from these scenarios resemble those arising from the production and decay of level-1 KK quarks and gluons in the fat-brane realization of the mUED model, the kinematics of the final state particles differ significantly. For example, in the GGM scenario, the photon resulting from the decay of a massive lightest neutralino (ranging from a few hundred GeV to a few TeV) into a nearly massless gravitino is always very energetic. In contrast, in the fat brane mUED scenario, the level-1 KK photon can decay into any KK member of the graviton tower with a mass below that of the level-1 KK photon. Consequently, the energy of the photon from this decay depends on the mass of the KK graviton. A very energetic photon results when the level-1 KK photon decays into a light KK graviton, whereas a decay into a heavier KK graviton results in a softer photon. Therefore, search strategies optimized for SUSY scenarios might not be efficient for the fat brane mUED scenario. The remainder of this article focuses on developing optimized mono-photon, diphoton, and multijet search strategies specifically for the fat brane mUED scenario.

\subsubsection{Event Generation and Object Reconstruction}
\label{sec: finaleventgen}
For generating the signal events, we have followed the strategy outlined in Section \ref{sec:Eventsimulation}. The model parameter space studied follows the discussion in Section \ref{sec:boundresults}. We have fixed the value of $\Lambda R$ at five and considered three possible values of the number of large extra dimensions: $N$ = 2, 4, and 6. For generating the signal events, we have varied $R^{-1}$ in the range of 2.7 TeV to 3.5 TeV, and for each value of $R^{-1}$, we have varied $M_D$ in the range of 5 to 15 TeV. To present our numerical results (cross-section, kinematic distributions, etc.), we defined three benchmark points in Table \ref{tab:sig_cross}.\\

\begin{table}[htb!]
	\begin{center}
		\begin{tabular}{l|c|c|c}
			\toprule 
             Process & Cross Section (fb)& k-factor & $M_D$ (TeV) \\
            \midrule
            $N = 2$ &  &  & 5 \\
            $N = 4$ & 0.15 & 1 & 15 \\
            $N = 6$ &  &  & 15\\
            \bottomrule
		\end{tabular} 
		\caption{\label{tab:sig_cross} List of benchmark signal scenarios corresponding to $R^{-1}$ = 3.1 TeV.}
	\end{center}
\end{table}

As for the SM backgrounds, we have considered all the SM processes contributing to the mono-photon, di-photon, and multi-jet final states \cite{ATLAS:2022ckd, ATLAS:2018nud, ATLAS:2019vcq}. The list of SM processes includes single and multi-top production (with associated vector bosons), mono-boson (with additional partons), di-boson, tri boson, and tetra boson production. All background events are generated with up to two additional partons in  MG5\_AMC@NLO \cite{Alwall:2014hca} with the  NNPDF21LO \cite{NNPDF:2014otw} parton distribution function. All background events are generated at leading order (LO), followed by MLM matching using Pythia8 \cite{Bierlich:2022pfr}, and we subsequently scale the LO crosssection with the appropriate next-to-leading order (NLO) k-factors \cite{Campbell:2011bn, Shen:2016ape, Muselli:2015kba, Broggio:2019ewu, LHCHiggsCrossSectionWorkingGroup:2016ypw, Kidonakis:2015nna, Catani:2009sm, Balossini:2009sa, Blok:2020ckm, Balazs:1999yf, Campbell:2018wfu, ATLAS:2015jos}. For completeness, we have listed all the SM background processes considered in our analysis in Table \ref{tab:back_cross1}. \\

\begin{table}[htb!]
	\begin{center}
		\begin{tabular}{l|c||c|c}
			\toprule 
            \multicolumn{2}{c||}{Backgrounds for the photonic SRs}&\multicolumn{2}{c}{Backgrounds for the multi-jet SRs} \\
            \midrule
             Process &  k-factor  & Process &  k-factor \\
            \midrule
			$\gamma + jets$ &   1.66  (Fig.4 of \cite{Chen:2022gpk}) & $t \Bar{t} + jets$   & 1.5 \cite{Muselli:2015kba} \\
            $\gamma$ $\gamma + jets$  & 1.66 (Fig.1 of \cite{Chawdhry:2021hkp})  &$V + jets$ &  1.09 \cite{Catani:2009sm, Balossini:2009sa} \\
            $t \Bar{t}$ $\gamma$  & 1 & $t \Bar{t} V$  & 1.66 \cite{Broggio:2019ewu}\\
            $V V \gamma$  & 1.8 \cite{Bozzi:2009ig} & $t W + jets$ &  1.27 \cite{Kidonakis:2015nna} \\
            $V V \gamma \gamma$ & 1  &$V V + jets$ &  1.66 \cite{Campbell:2011bn}  \\
            $W (\rightarrow l \nu) \gamma + jets$  & 1.31 \cite{Campbell:2011bn}  & $V V V$  & 2.6 \cite{Shen:2015cwj} \\
            $W (\rightarrow l \nu) \gamma \gamma$  & 1  & &  \\
            $Z (\rightarrow \nu \nu) \gamma + jets $   & 1.18 \cite{Campbell:2011bn}  & &\\
            $Z (\rightarrow \nu \nu) \gamma \gamma$  & 1  & & \\
		\bottomrule
		\end{tabular} 
		\caption{\label{tab:back_cross1} List of the dominant SM backgrounds considered in our final analysis.}
	\end{center}
\end{table}

With lower bounds on the KK-quark and KK-gluon masses approaching around 3 TeV (see Figure~\ref{fig: valplot}), decays of such massive KK-quarks and KK-gluons are expected to produce highly boosted SM heavy bosons or top quarks in the final state. In particular, the decay of level-1 top quarks or electroweak (EW) gauge bosons into graviton excitations could lead to highly boosted SM \( W \) or \( Z \) bosons in the final state. In a collider like the LHC, the hadronic decays of these boosted bosons/top quarks appear as a collimated beam of hadrons, making it convenient to reconstruct them as a single large-radius (fat) jet. Along with their reconstruction, proper identification of these fat jets is also crucial, as similar fat jets can also arise from QCD radiation. Efficiently identifying hadronically decaying boosted SM heavy bosons or top quarks over QCD jets is anticipated to enhance the discrimination of the signal from the SM background. To distinguish the fat jets originating from these bosons from jets originating from QCD-initiated quarks and gluon jets, our analysis implements a simple boosted decision tree (BDT) classifier. The details of the classifier are discussed in Appendix \ref{sec: appendix-3}.

% \\

To reconstruct final state jets, leptons, etc., we follow the steps outlined in Section \ref{sec: monophoton} with a slight modification. In addition to the $R=0.4$ radius jets, we also reconstruct fat jets with a radius of 1.0. These fat jets pass through a jet pruning algorithm \cite{Ellis:2009su, Ellis:2009me} to eliminate the broad QCD emissions. For our analysis, we have used the default pruning setup of Delphes with parameters: a $z_{cut}$ value of 0.1 and an $R_{cut}$ value of 0.5. Reconstructed fat jets are required to have a transverse momentum greater than 300 GeV and pseudorapidity $|\eta| < 2.5$. These fat jets are then passed through the BDT classifier and are tagged as either a \( V \) (i.e., \( W/Z \)) jet or a QCD jet. For further analysis, we retain only the \( V \)-tagged \( R=1.0 \) radius jets (denoted by \( j^{V\text{-}tag} \) throughout the rest of the manuscript). Since our analysis considers both \( R=0.4 \) jets and \( V \)-tagged \( R=1.0 \) radius jets, we need a strategy to avoid the possibility of double-counting jets. To address this, we exclude any \( R=0.4 \) jets from our analysis if they lie within a cone of radius 1.0 from the center of a \( V \)-tagged jet. The resolved \( R=0.4 \) jets are denoted by \( j^{0.4} \) in the remainder of the manuscript.
\\

\subsubsection{Event Selection}
\label{sec: event_selection}
The signal and background samples are selected by designing appropriate signal regions based on the expected multiplicity of the final state photons, leptons, and jets in the signal processes. The idea is to design mutually independent signal regions that can be statistically combined and, at the same time, capture most of the signal events, thereby maintaining high signal selection efficiency. We achieve this by designing five signal regions (SRs): SR\_1p0l, SR\_1pnl, SR\_2p0l, SR\_2pnl, and SR\_0p0l. We present the preselection requirements defining these signal regions in Table~\ref{tab: signal_regions} and discuss them in the following sections.\\

\begin{itemize}
    \item \textbf{SR\_1p0l and SR\_1pnl Signal Regions}: The \texttt{SR\_1p0l} signal region is designed to capture final states with a single photon, at least two \( R = 0.4 \) jets, and no leptons. As shown in Table~\ref{tab: signal_regions}, additional requirements are imposed on the transverse momentum of the photons and jets, as well as the missing transverse momentum. This signal region is suitable for cases where one NLKP \( \gamma_1 \) decays into a photon. This can occur in the \( N = 4 \) scenario if one of the intermediate particles in the decay cascade follows the GMD mode, breaking the decay chain, or in both the \( N = 4 \) and \( N = 6 \) scenarios when one of the two NLKPs decays into a \( Z \) boson. The \texttt{SR\_1pnl} signal region has similar requirements on photons and jets but also requires additional leptons. While it captures similar signal scenarios as \texttt{SR\_1p0l}, the presence of leptons in the final state reduces contributions from the SM background.

    \item \textbf{SR\_2p0l and SR\_2pnl Signal Regions}: The \texttt{SR\_2p0l} and \texttt{SR\_2pnl} signal regions are designed to capture final states with two photons and at least two jets, along with additional requirements on the missing transverse momentum and the \( p_T \) of the photons and jets. Like the mono-photon signal regions, these two di-photon signal regions differ in terms of lepton multiplicity. In the \texttt{SR\_2p0l} signal region, a lepton veto is imposed, while the \texttt{SR\_2pnl} signal region requires events with at least one lepton. These signal regions are effective for scenarios where the decay of level-1 quarks/gluons follows the KKNC decay cascade and both NLKP \( \gamma_1 \) particles decay to photons. In the \texttt{SR\_2pnl} signal region, additional leptons are expected from the decay cascade and are anticipated to be soft, so no further \( p_T \) cuts on leptons are imposed.

    \item \textbf{SR\_0p0l Signal Region}: The \texttt{SR\_0p0l} signal region is specifically designed for the \( N = 2 \) scenario, where gravity-mediated decays of level-1 KK quarks/gluons dominate over the KKNC decay mode across most of the parameter space. Consequently, photons from the GMD decay of \( \gamma_1 \) or leptons from the KKNC decay cascade are not expected in the final state, so both a photon veto and a lepton veto are imposed. This signal region requires at least three jets, a large missing transverse momentum, and a minimum transverse momentum for the three leading jets (see Table~\ref{tab: signal_regions}).
\end{itemize}

\begin{table}[htb!]
	\begin{center}
		\begin{tabular}{l|c|c|c|c|c}
			\toprule 
             Cuts& SR\_1p0l& SR\_1pnl & SR\_2p0l  &  SR\_2pnl & SR\_0pnl\\
            \midrule
             $N_{\gamma}$  & 1    & 1 & 2&  2& 0\\

             $N_l$         & 0      & $ \ge 1 $ & 0    & $\ge 1$   & 0 \\

             $N_j$         & $\ge 2$  & $\ge 2$   & $\ge 2$&  $\ge 2$  & $\ge 3$ \\

             $p_T (\gamma_1)$ (GeV) & $>150$ & $>150$ & $>150$&  $>150$ & -- \\
             
             $p_T (j1)$ (GeV)    & $>100$        & $>100$ & $>100$&  $>100$& $>200$ \\

              $p_T (j2)$ (GeV)   & $>50$    & $ >50 $ & $>50$&  $ >50 $&$>100$\\

             $p_T (j3)$ (GeV)    & --    & -- & --& -- & $>50$ \\

             $E_T^{miss}$ (GeV)   & $>200$     & $>200$ & $>200$ &  $>200$ & $>300$ \\
	
			\bottomrule
		\end{tabular} 
		\caption{\label{tab: signal_regions} Pre-selection cuts for the five signal regions implemented in our analysis.}
	\end{center}
\end{table}

To analyze the signal and background events in each signal region, we construct several kinematic variables that capture the kinematic properties of the signal and background events. These variables are designed to encode the differences in behavior between signal and background events. By carefully designing constraints on these variables, the background contribution in these signal regions can be suppressed while retaining a significant fraction of the signal events. Determining the optimal set of kinematic cuts requires examining the distribution of these variables for both signal and background samples. The goal is to choose kinematic constraints that sufficiently reduce the background contribution while maintaining high signal efficiency. This involves inspecting these variables individually and in various combinations. As the number of variables increases, this process becomes time-consuming and cumbersome. A more systematic and automated approach is achievable with the help of a machine learning-based classifier. For our final analysis, we have developed five BDT-based classifiers, one for each signal region. The kinematic variables constructed for the signal and background events serve as input to these classifiers, and the output, i.e., the BDT score, can be used as the final discriminating variable between signal and background. All these classifiers use the same hyperparameters as the classifier discussed in Appendix~\ref{sec: appendix-3}.
 \\

We list the observables used for training the BDT classifiers in the five signal regions in Appendix~\ref{sec: appendix-4}. The list of observables includes missing transverse energy (\( E_T^{\text{miss}} \)), \( H_T \), \( M_{\text{eff}} \), \( E_T^{\text{miss}}/\sqrt{H_T} \), \( p_T \) and \( \eta \) of \( R=0.4 \) jets, \( V \)-tagged jets, leptons, and photons; the azimuthal angle (\( \Delta \phi \)) between the missing transverse energy vector and reconstructed jets, leptons, and photons; and more (see Appendix~\ref{sec: appendix-4} for the full list of observables). In Figures~\ref{fig: var_SR1}–\ref{fig: var_SR5}, we present the distributions of three of the most important kinematic variables for each of the signal regions. All these distributions assume an integrated luminosity of \( 500 \, \text{fb}^{-1} \). To enhance visibility, the number of signal events in each bin is multiplied by a factor of 100. For each signal region, we only show the distribution of the dominant backgrounds, along with the distributions for a few benchmark signal scenarios. Since the \texttt{SR\_1p0l}, \texttt{SR\_1pnl}, \texttt{SR\_2p0l}, and \texttt{SR\_2pnl} regions are most effective for the \( N = 4 \) and \( N = 6 \) scenarios, we present results for these two scenarios with fixed values of \( R^{-1} = 3.1 \, \text{TeV} \) and \( M_D = 15 \, \text{TeV} \). Similarly, for the \texttt{SR\_0pnj} signal region, we present results for the \( N = 2 \) scenario with \( R^{-1} = 3.1 \, \text{TeV} \) and \( M_D = 5 \, \text{TeV} \). As expected for each signal region, the missing transverse momentum is the most important observable for discriminating signal from the background, as the signal events typically exhibit large missing transverse energy due to the invisible KK gravity excitation in the final state. The effective mass (\( M_{\text{eff}} \)), defined as the scalar sum of the transverse momentum of all visible final-state particles and the missing transverse momentum, also plays a crucial role in distinguishing signal from background.

\begin{figure}[htb!]
	\centering
	\includegraphics[width=0.33\columnwidth]{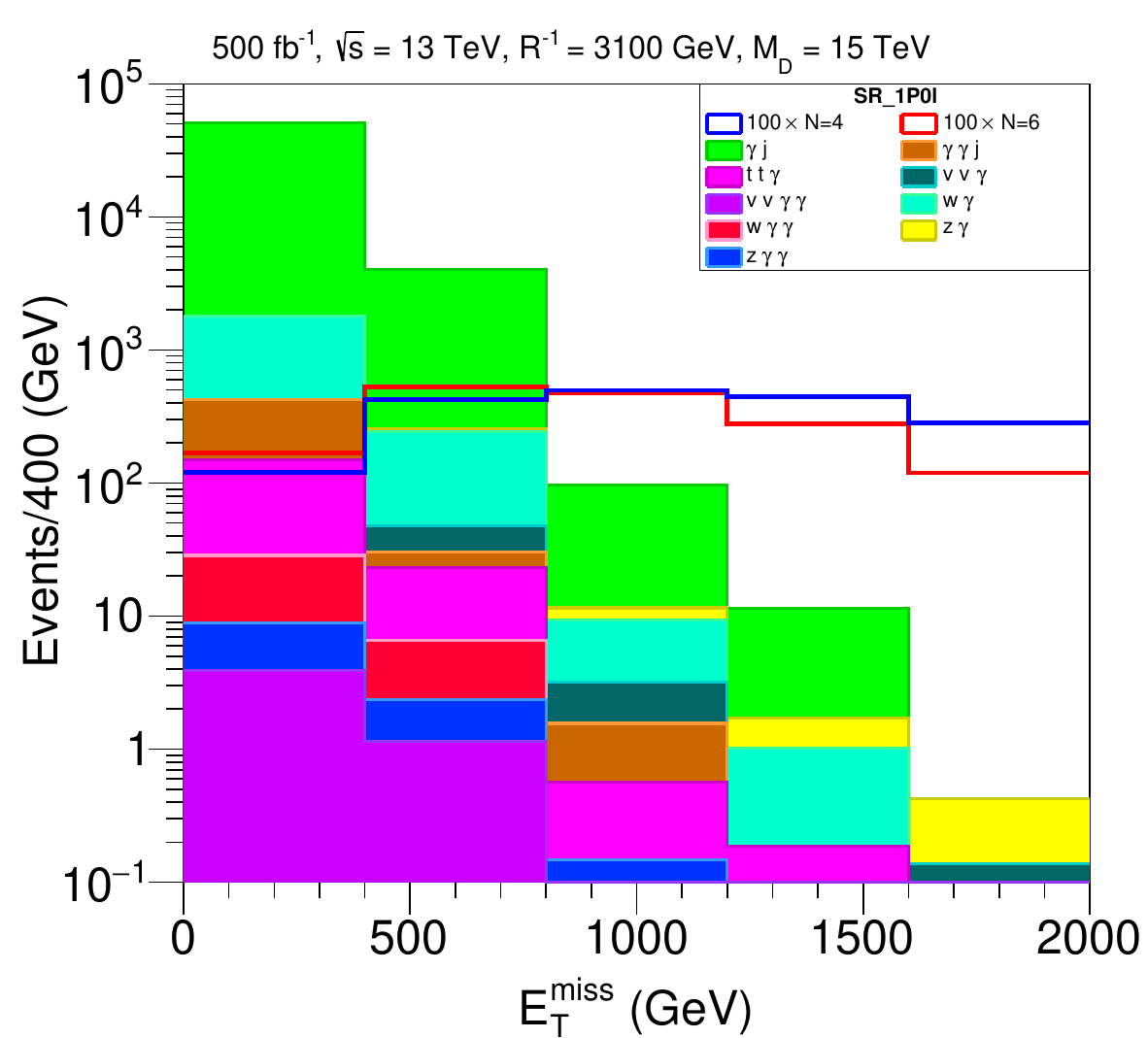} 
	\includegraphics[width=0.33\columnwidth]{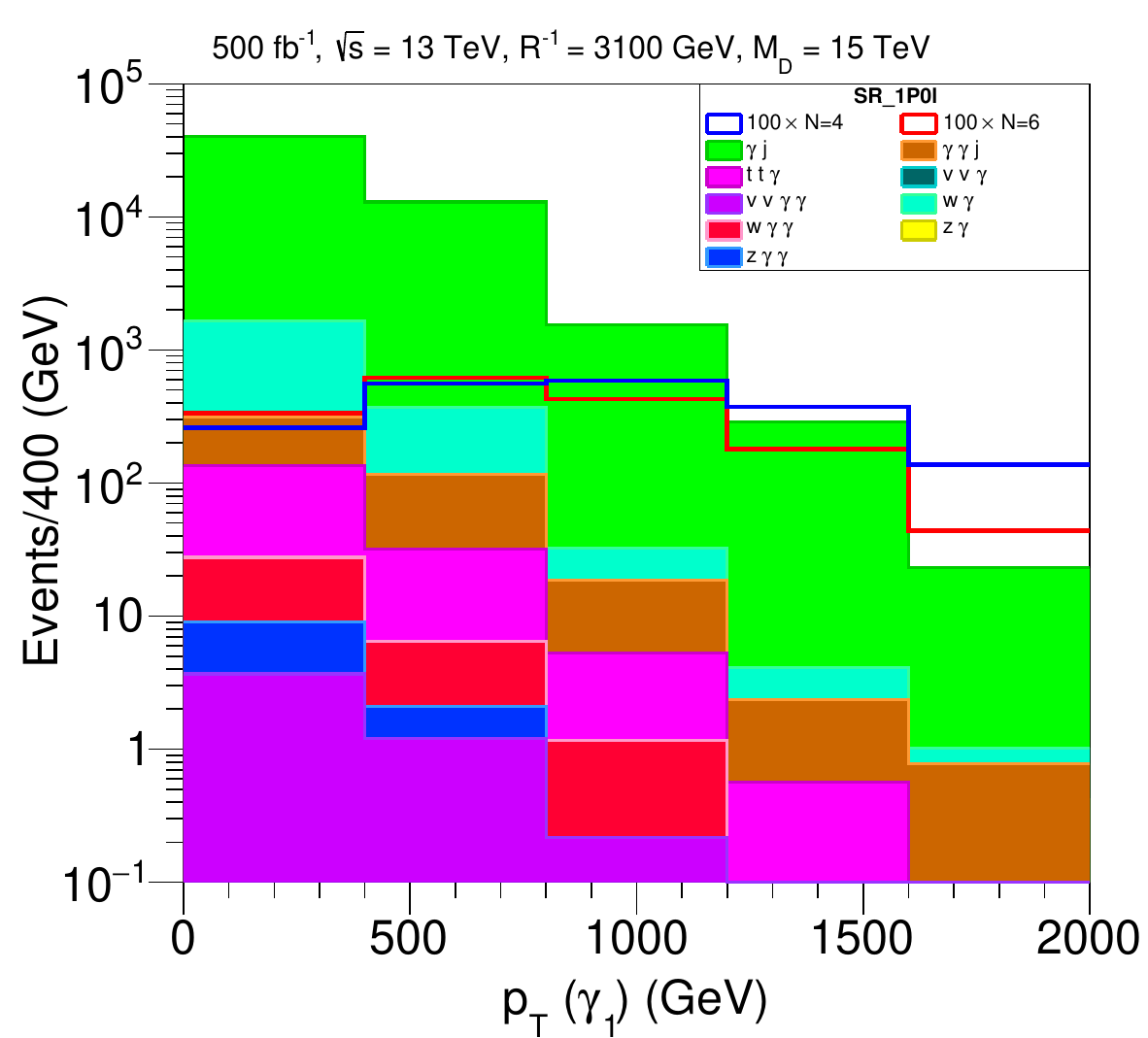}
    \includegraphics[width=0.32\columnwidth]{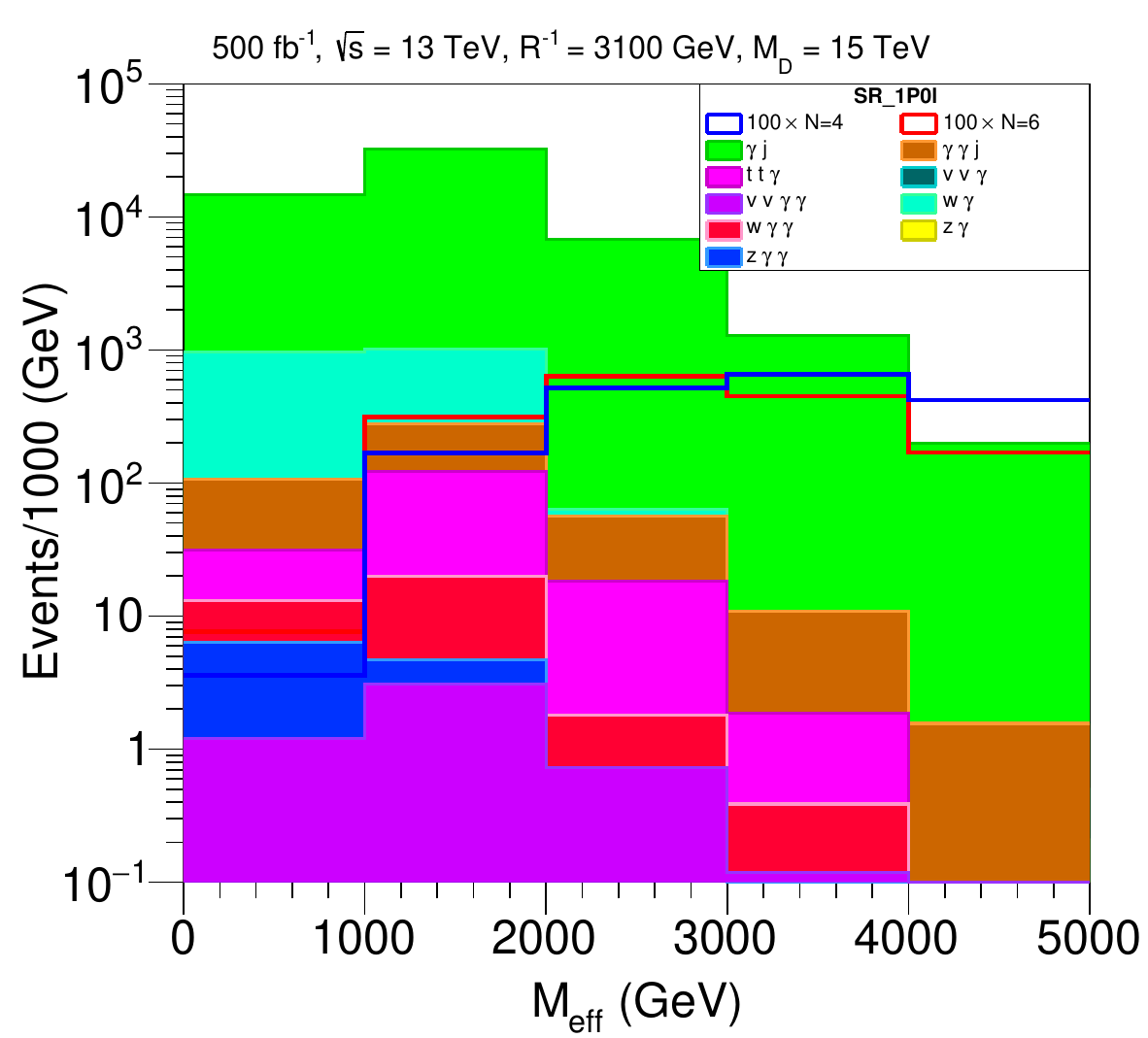}
	\caption{\label{fig: var_SR1} Distribution of the top three important kinematic variables for the \texttt{SR\_1p0l} signal region.  }
\end{figure}

\begin{figure}[htb!]
	\centering
	\includegraphics[width=0.33\columnwidth]{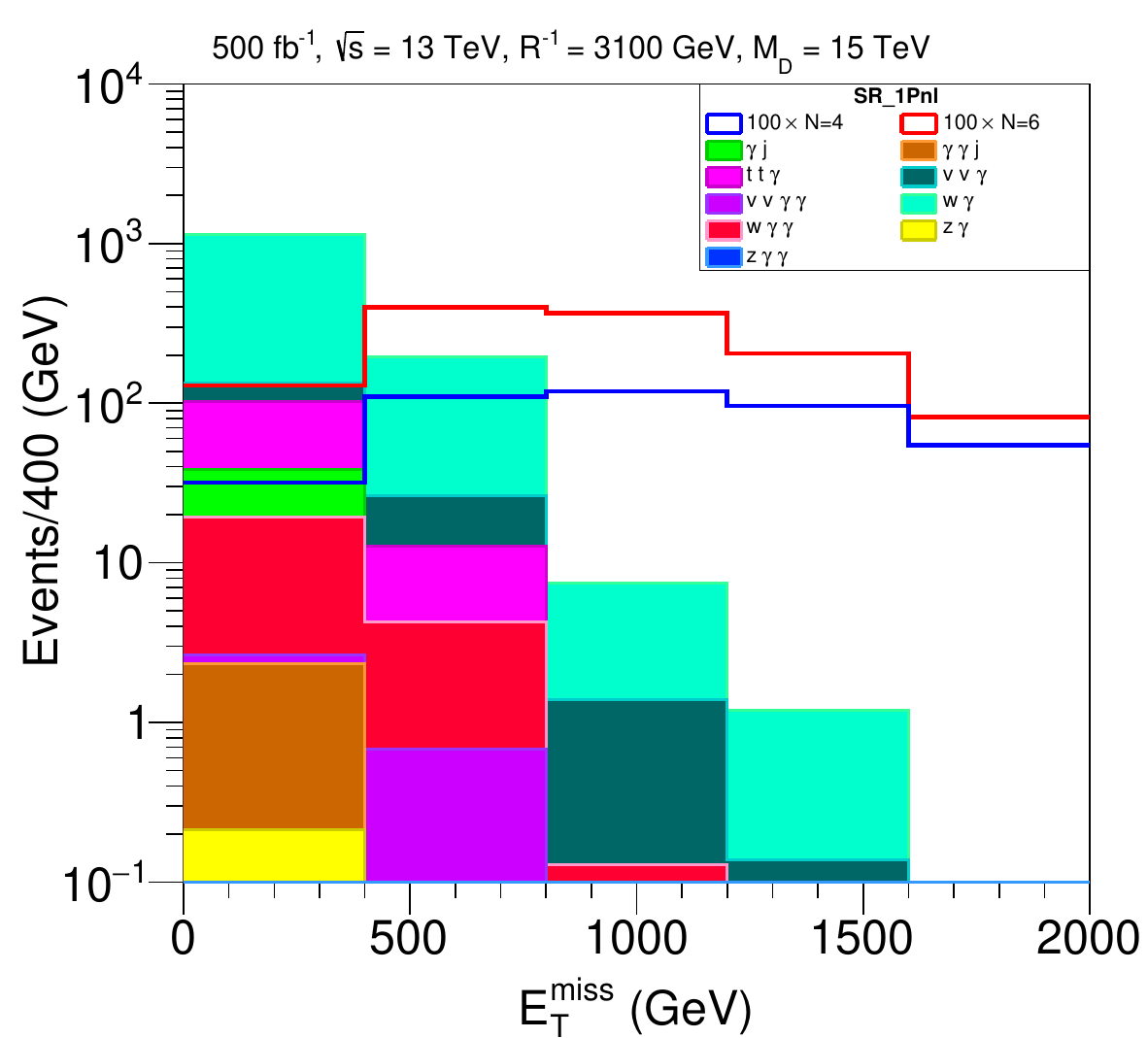} 
	\includegraphics[width=0.33\columnwidth]{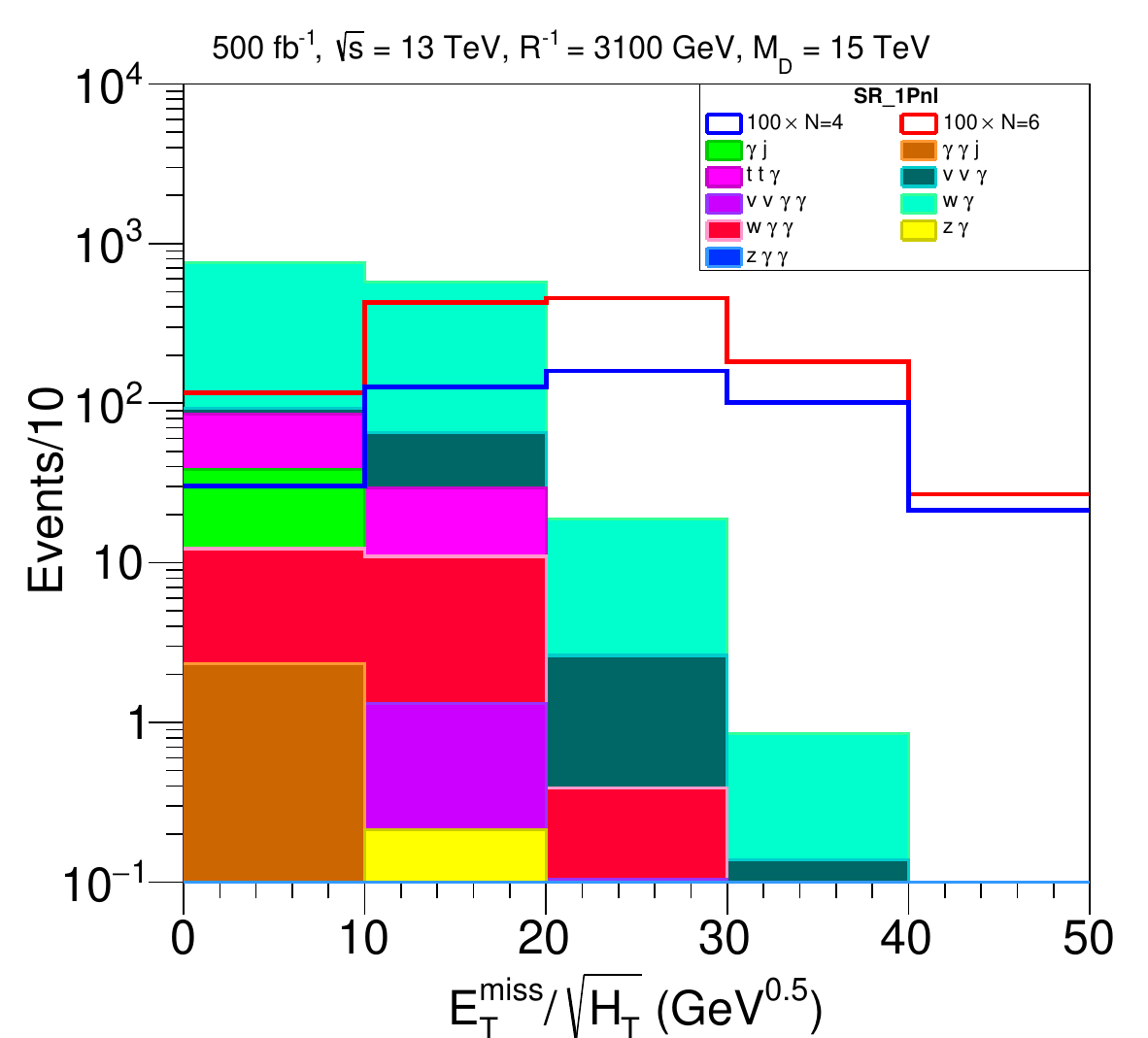}
    \includegraphics[width=0.32\columnwidth]{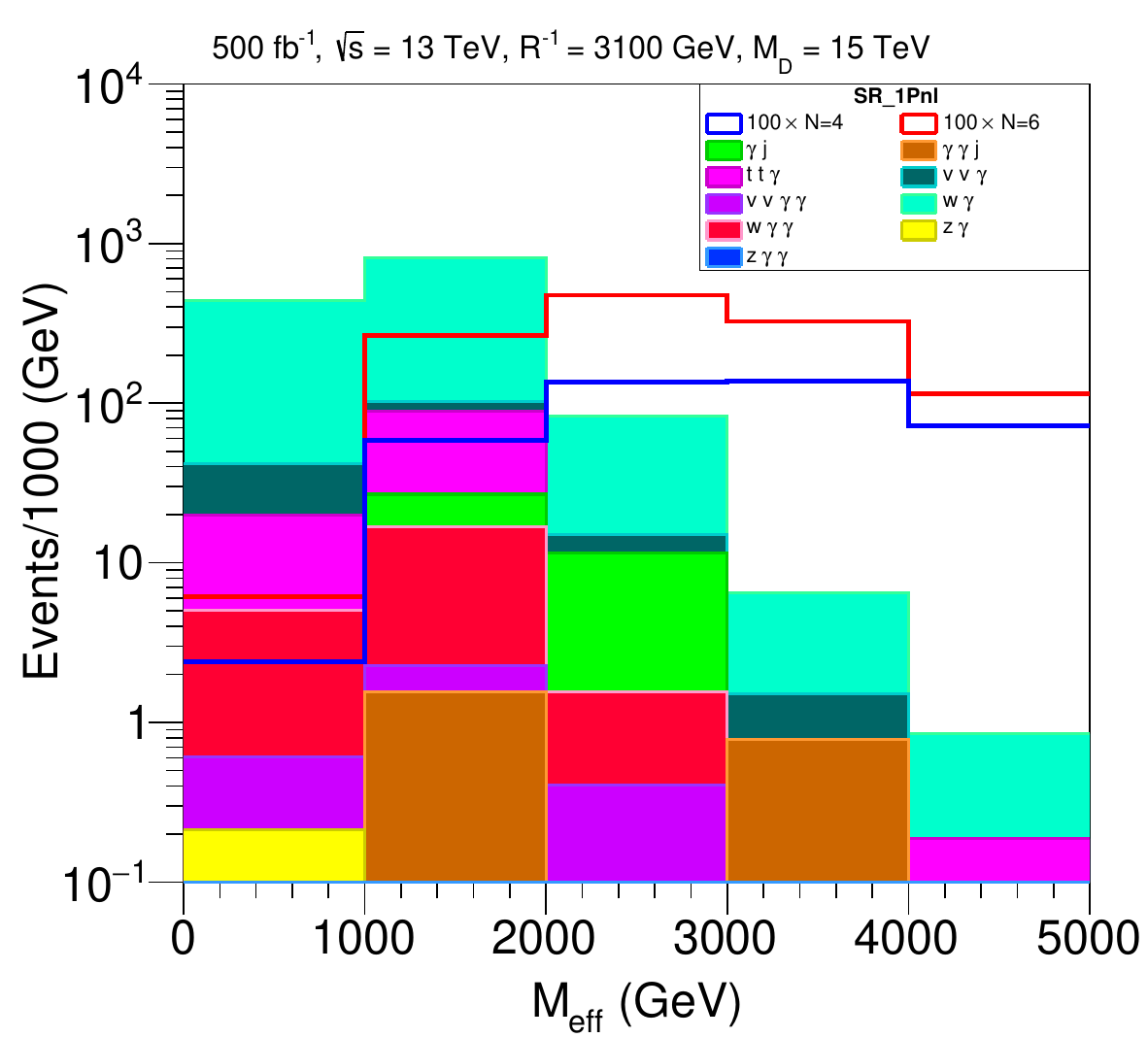}
	\caption{\label{fig: var_SR2} Distribution of the top three important kinematic variables for the \texttt{SR\_1pnl} signal region.  }
\end{figure}

\begin{figure}[htb!]
	\centering
	\includegraphics[width=0.33\columnwidth]{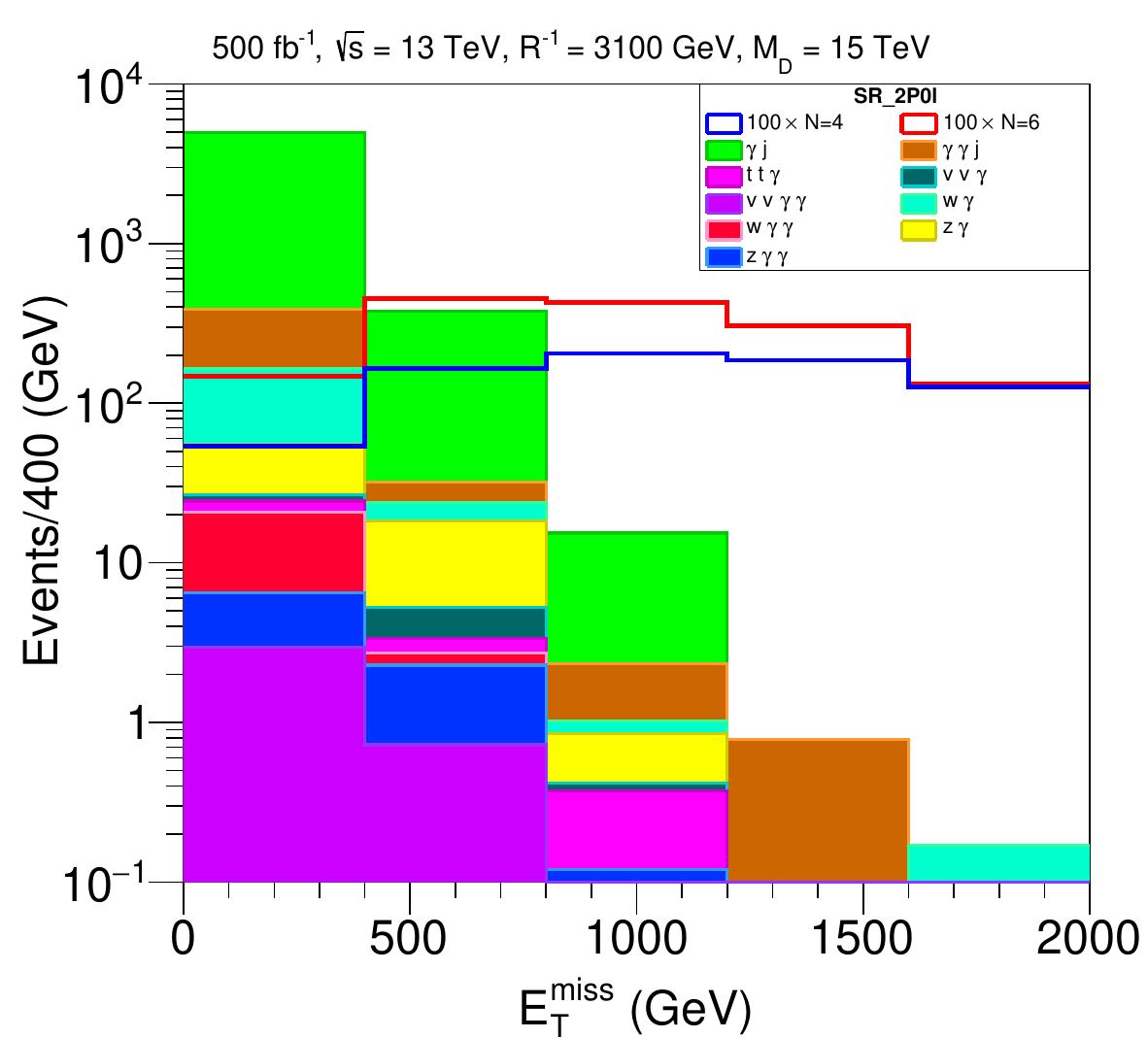} 
	\includegraphics[width=0.33\columnwidth]{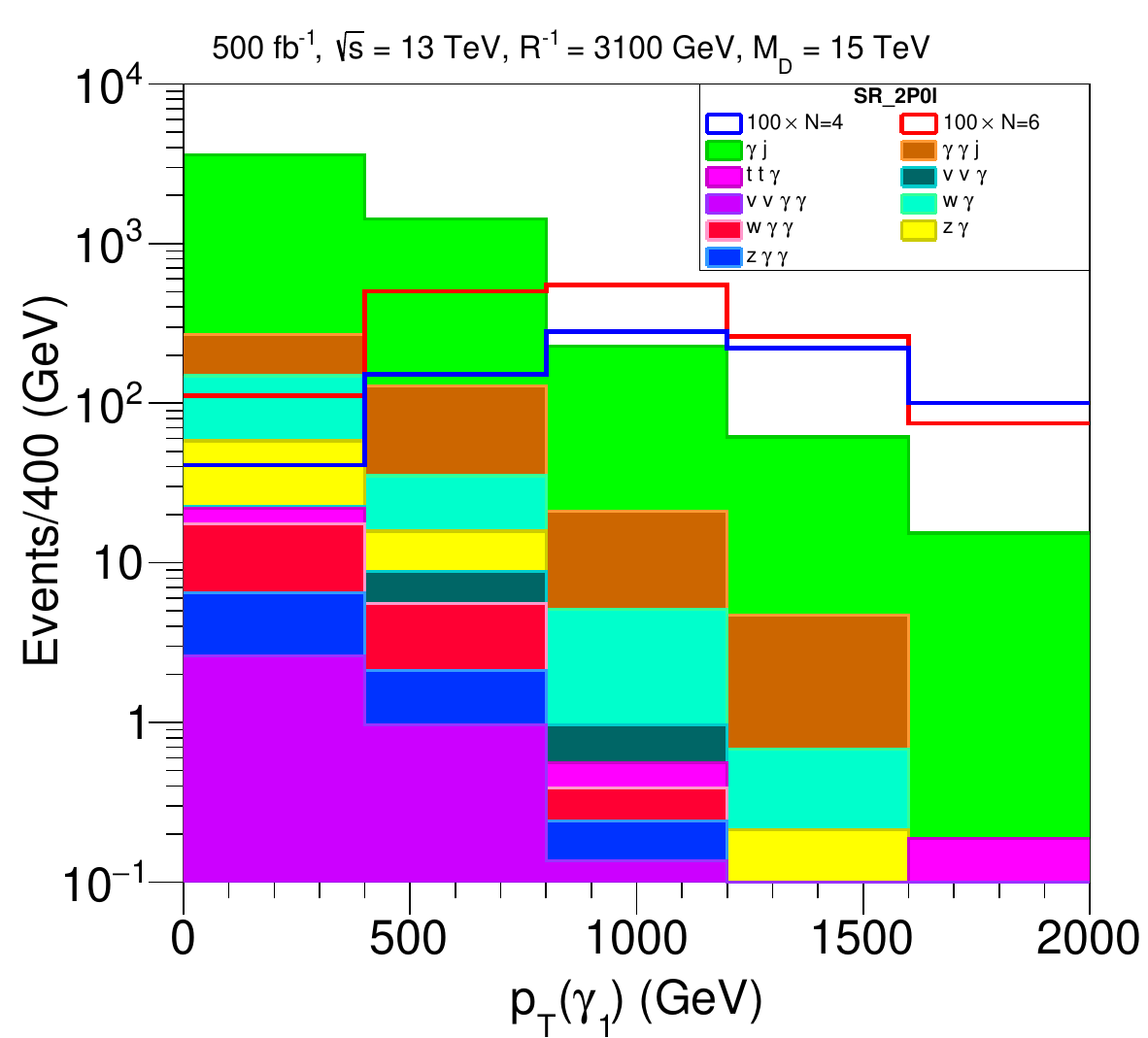}
    \includegraphics[width=0.32\columnwidth]{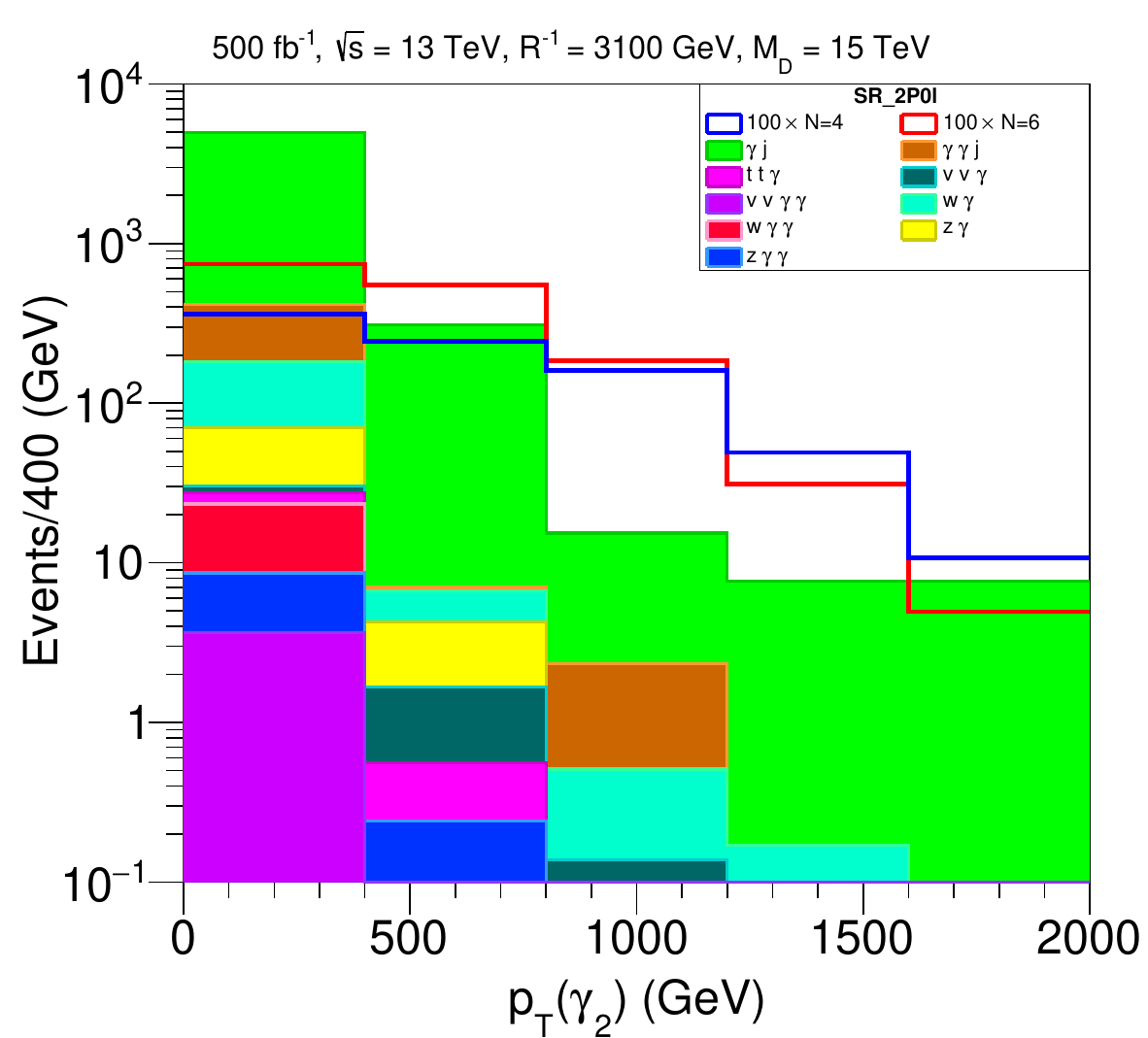}
	\caption{\label{fig: var_SR3} Distribution of the top three important kinematic variables for the \texttt{SR\_2p0l} signal region.  }
\end{figure}

\begin{figure}[htb!]
	\centering
	\includegraphics[width=0.33\columnwidth]{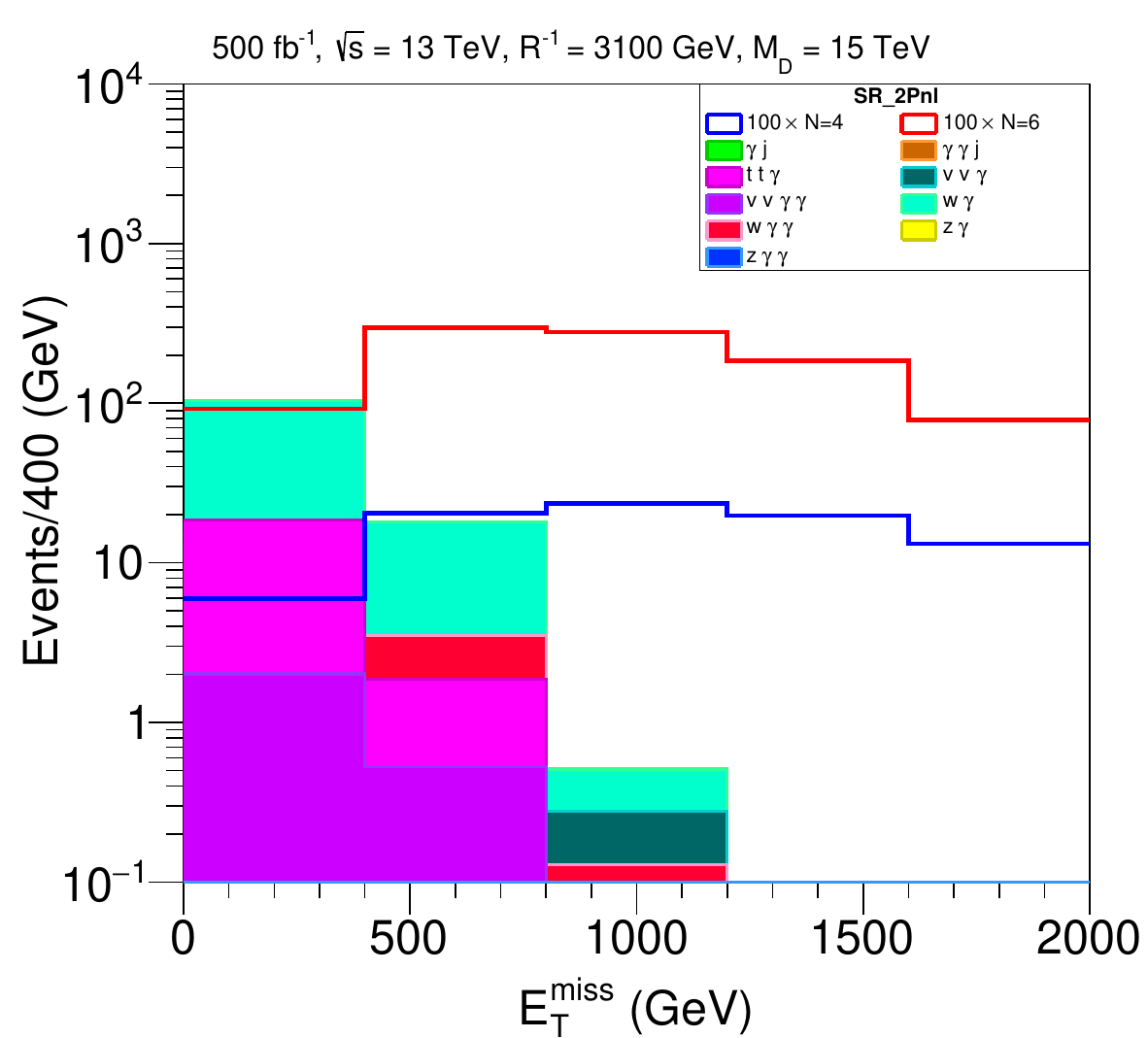} 
	\includegraphics[width=0.33\columnwidth]{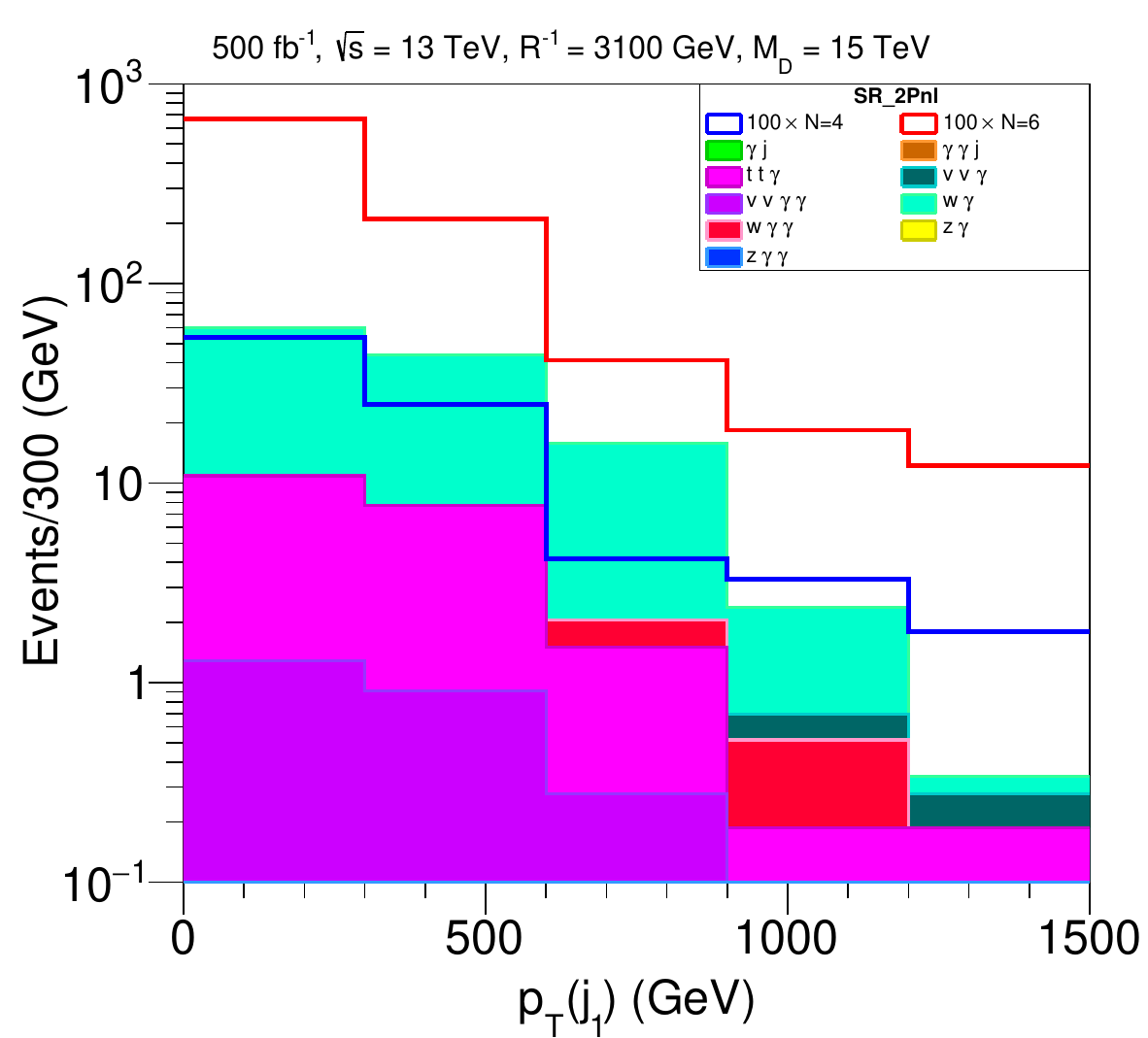}
    \includegraphics[width=0.32\columnwidth]{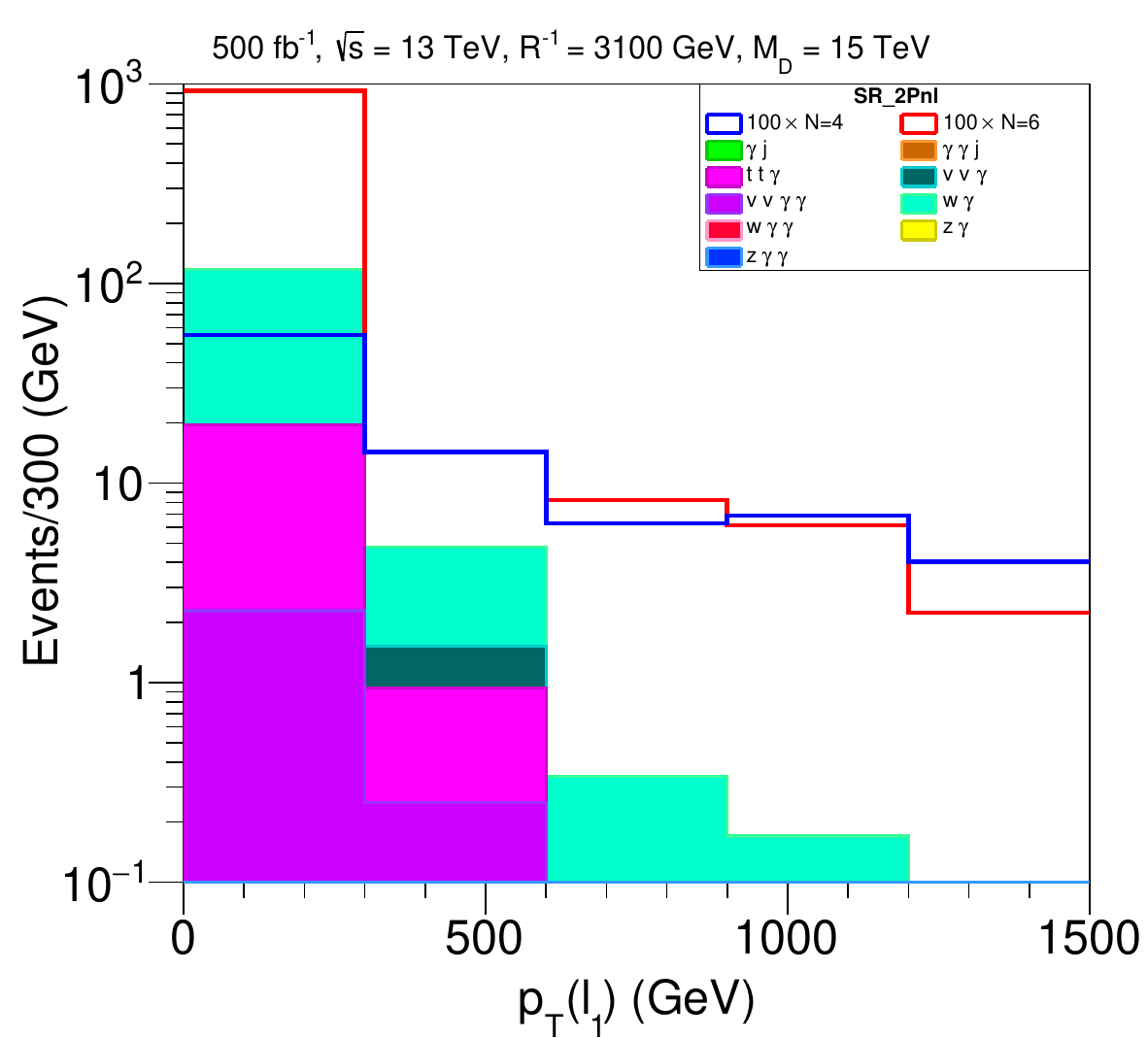}
	\caption{\label{fig: var_SR4} Distribution of the top three important kinematic variables for the \texttt{SR\_2pnl} signal region.  }
\end{figure}

\begin{figure}[htb!]
	\centering
	\includegraphics[width=0.33\columnwidth]{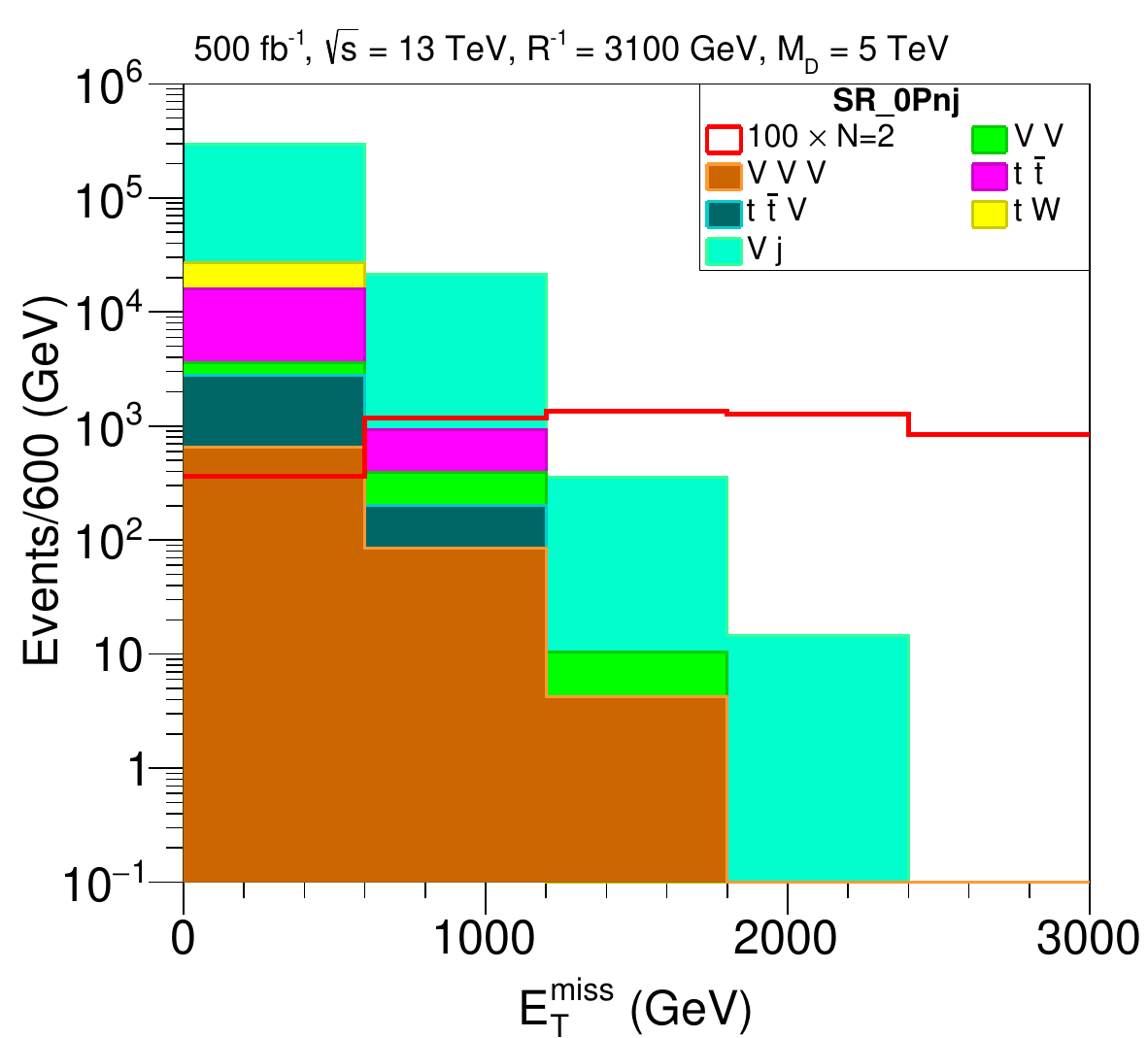} 
	\includegraphics[width=0.33\columnwidth]{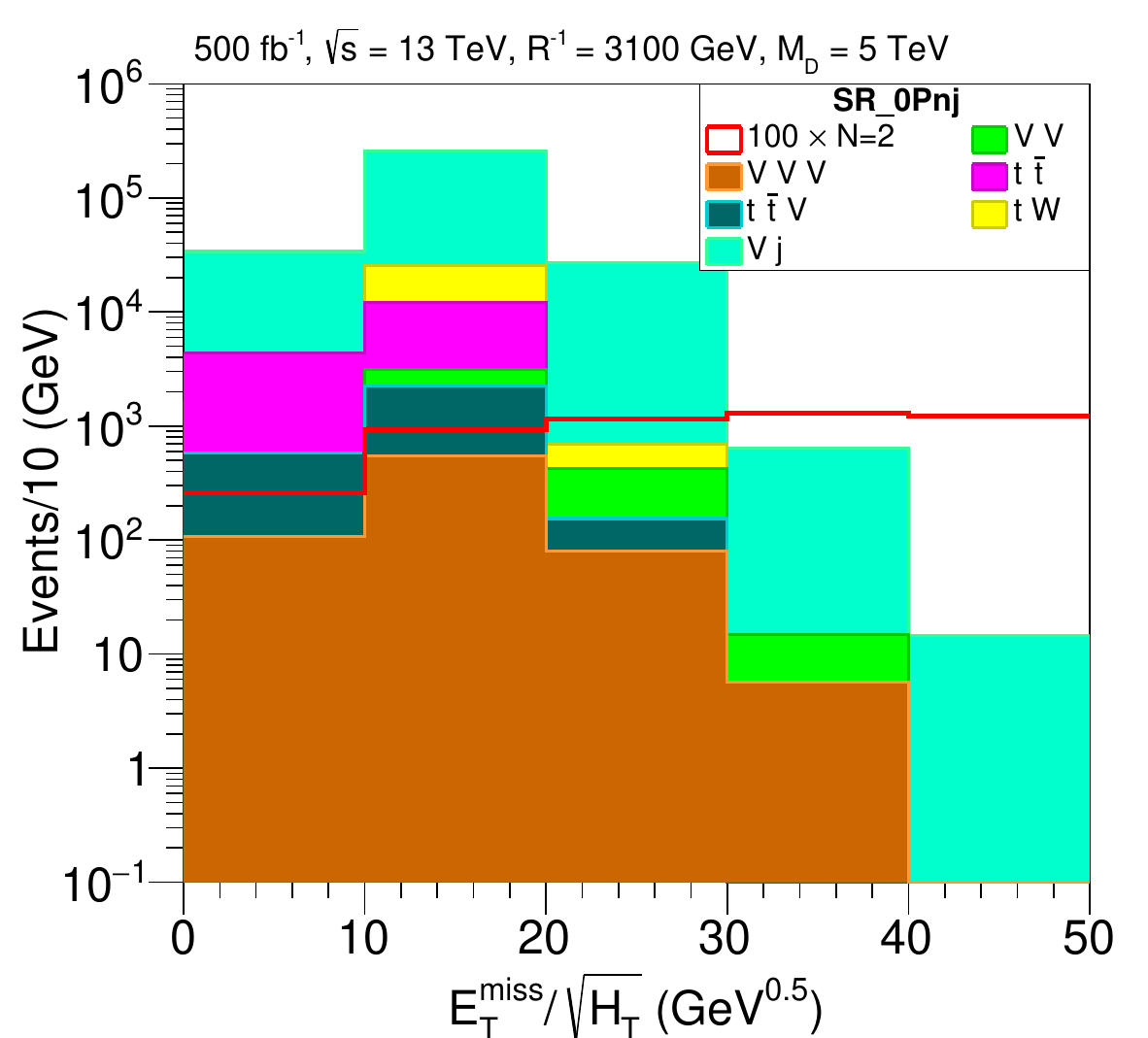}
    \includegraphics[width=0.32\columnwidth]{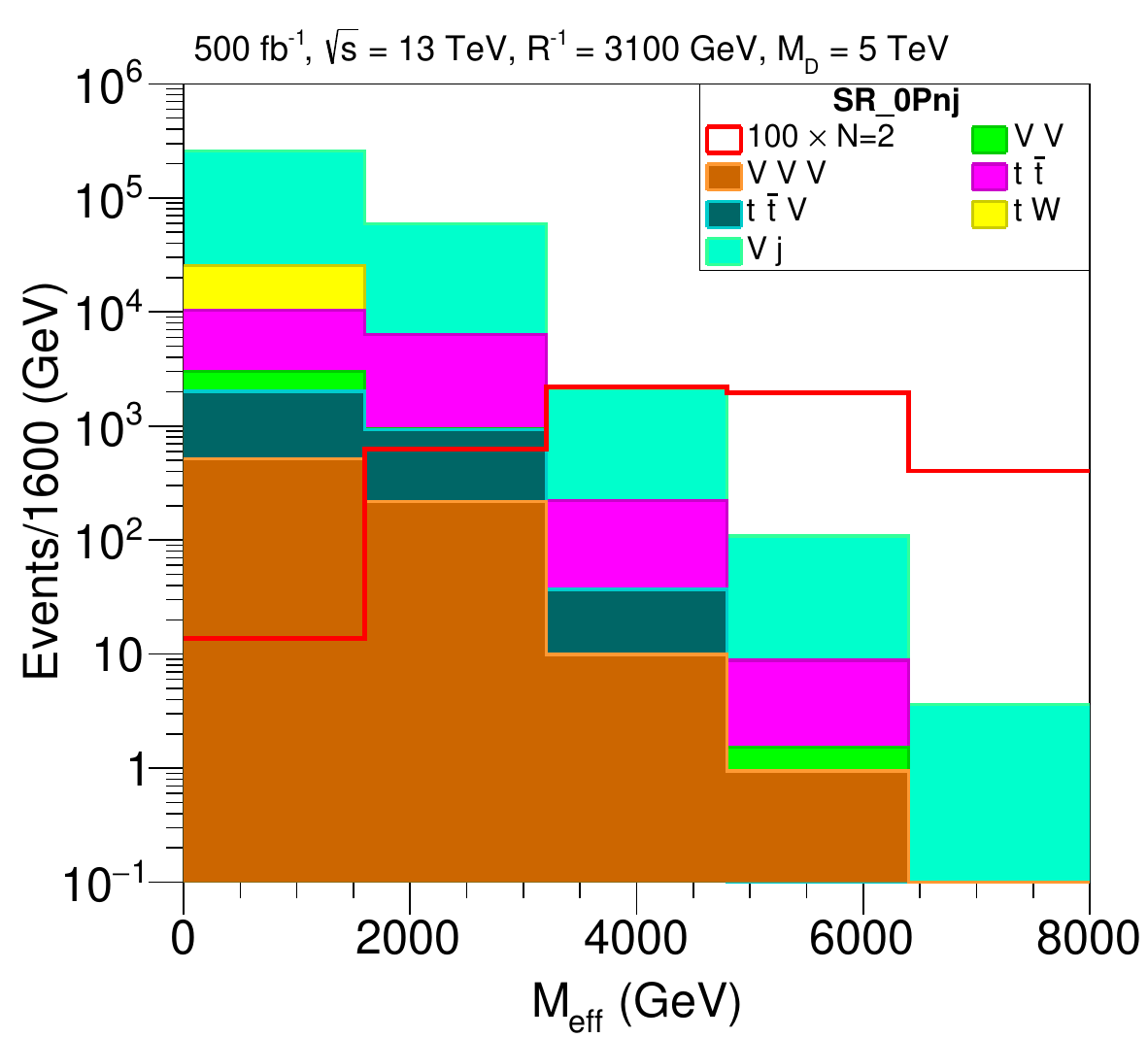}
	\caption{\label{fig: var_SR5} Distribution of the top three important kinematic variables for the \texttt{SR\_0p0l} signal region.  }
\end{figure}

\subsubsection{Results}
In this section, we discuss the future projection of the LHC reach on the parameter space of the fat brane MUED scenario. Figure~\ref{fig: score_srs} shows the distribution of the signal and background BDT scores from the five classifiers trained on variables constructed from the signal and background events across the five signal regions. Similar to Figures~\ref{fig: var_SR1}–\ref{fig: var_SR5}, for the \texttt{SR\_1p0l}, \texttt{SR\_1pnl}, \texttt{SR\_2p0l}, and \texttt{SR\_2pnl} signal regions, we present the BDT score distribution for the \( N = 4 \) and \( N = 6 \) signal scenarios, with fixed values of \( R^{-1} = 3.1 \, \text{TeV} \) and \( M_D = 14 \, \text{TeV} \). For the \texttt{SR\_0p0l} signal region, we show the distribution for the \( N = 2 \) signal scenario with \( R^{-1} = 3.1 \, \text{TeV} \) and \( M_D = 6 \, \text{TeV} \). The results are presented for a luminosity of \( 500 \, \text{fb}^{-1} \). A detailed analysis of the scores suggests that an asymmetric binning of the scores with edges at -1, 0, 0.2, 0.4, 0.6, and 1 provides the most effective separation between signal and background in bins with higher BDT scores. Since the number of background events generated corresponds to the chosen luminosity, the complete absence of background events in the final bin does not present a problem for our analysis.
\\ 

\begin{figure}[htb!]
	\centering
	\includegraphics[width=0.45\columnwidth]{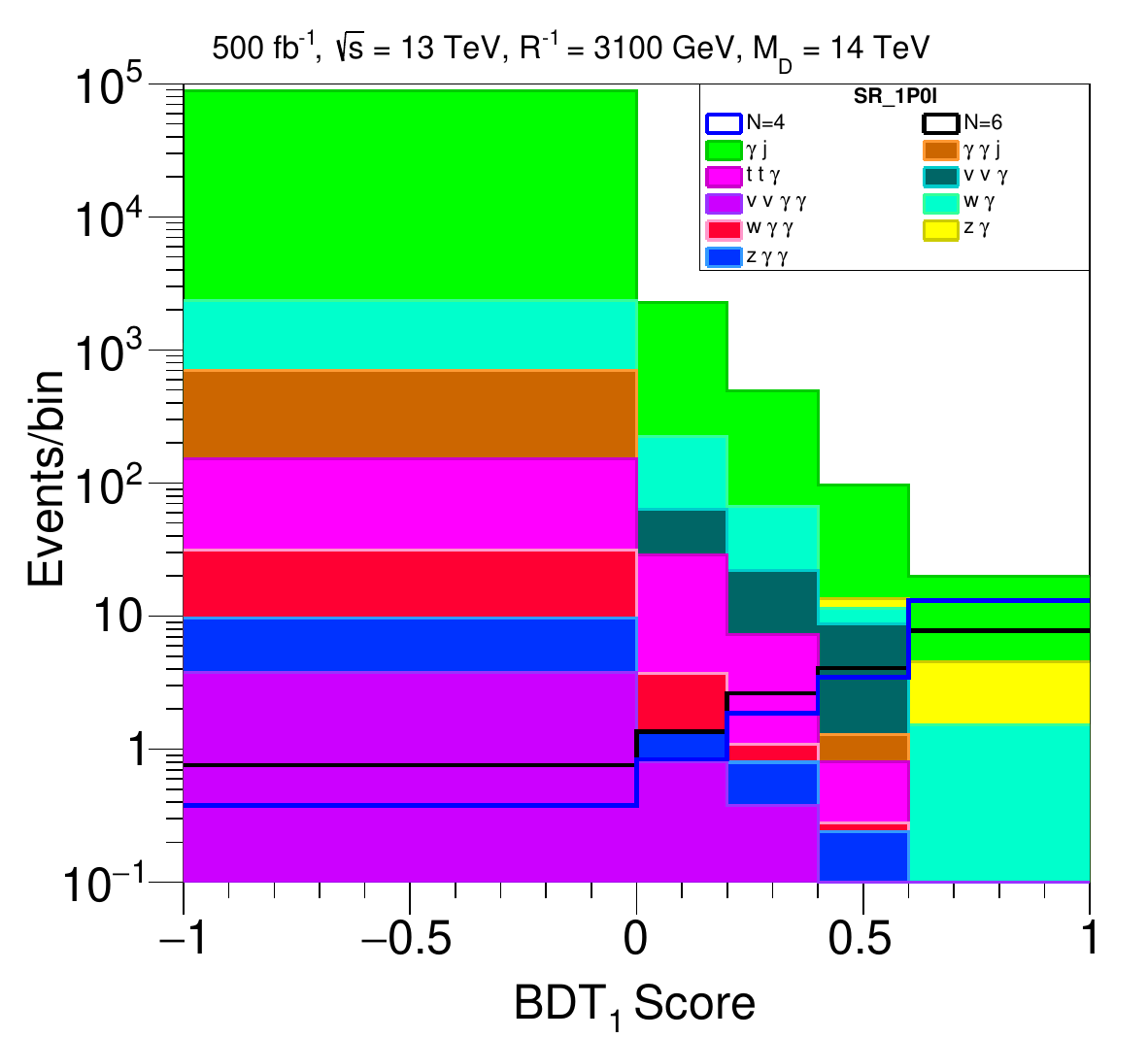} 
	\includegraphics[width=0.45\columnwidth]{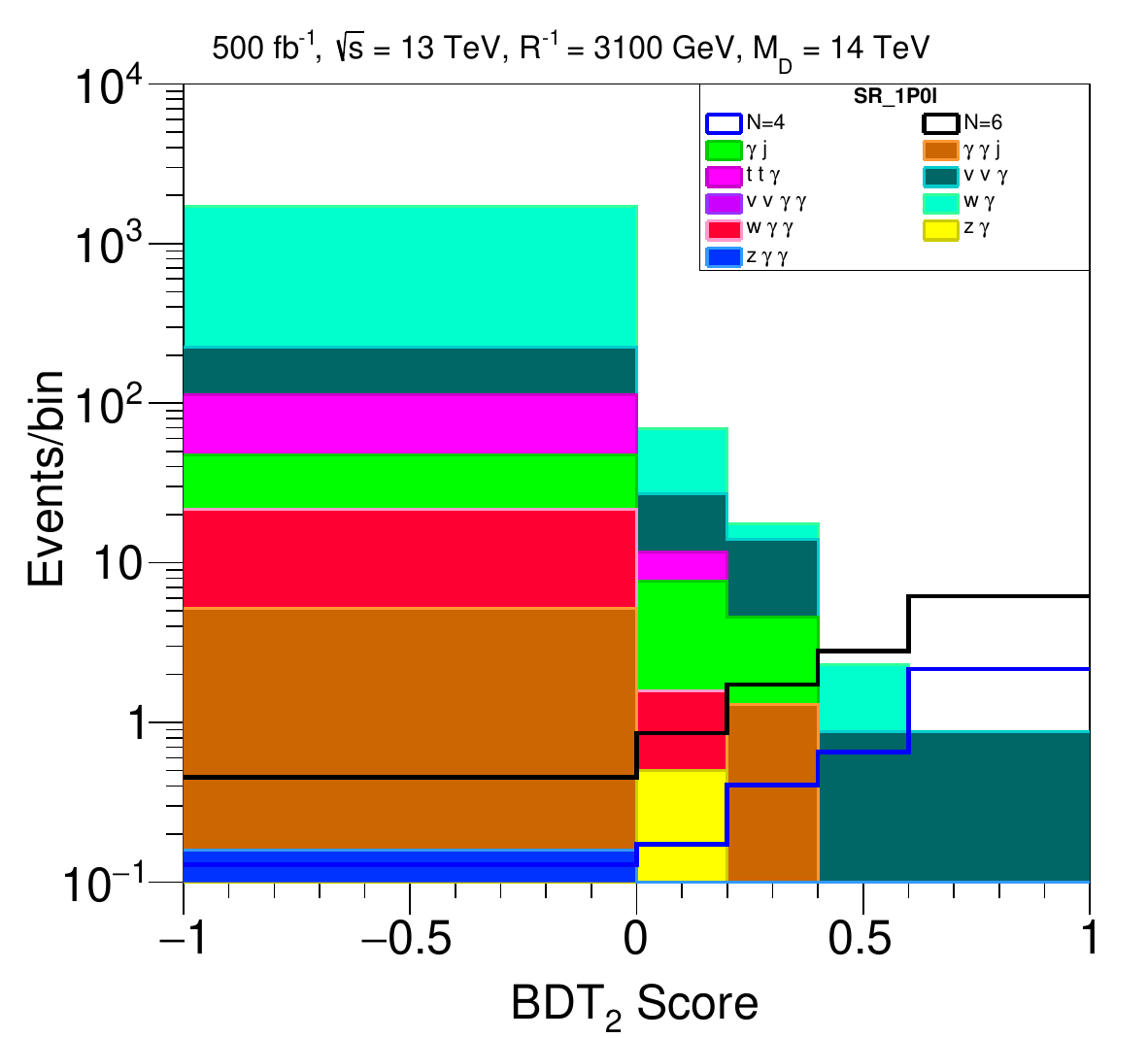}
    \includegraphics[width=0.45\columnwidth]{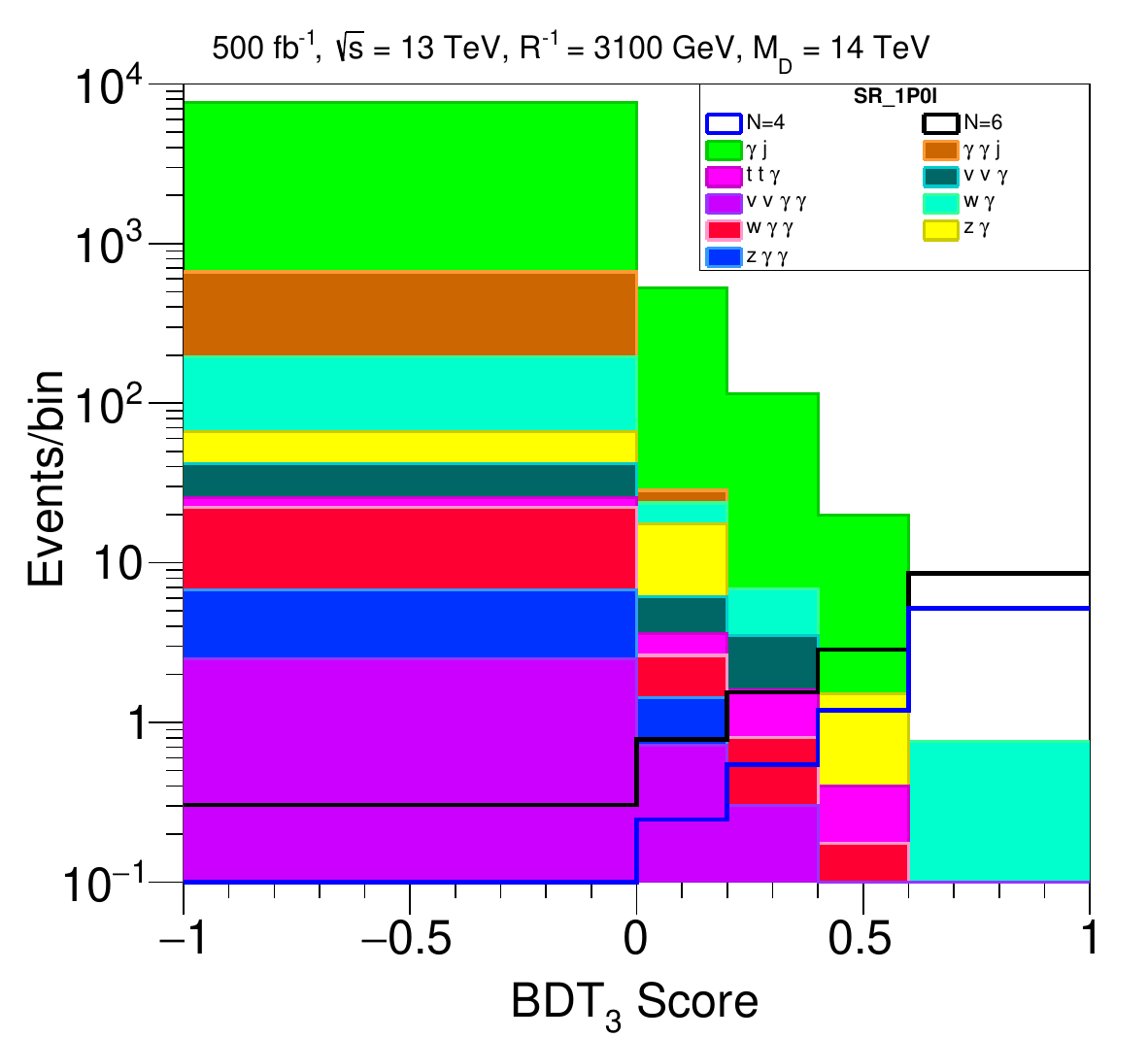}
    \includegraphics[width=0.45\columnwidth]{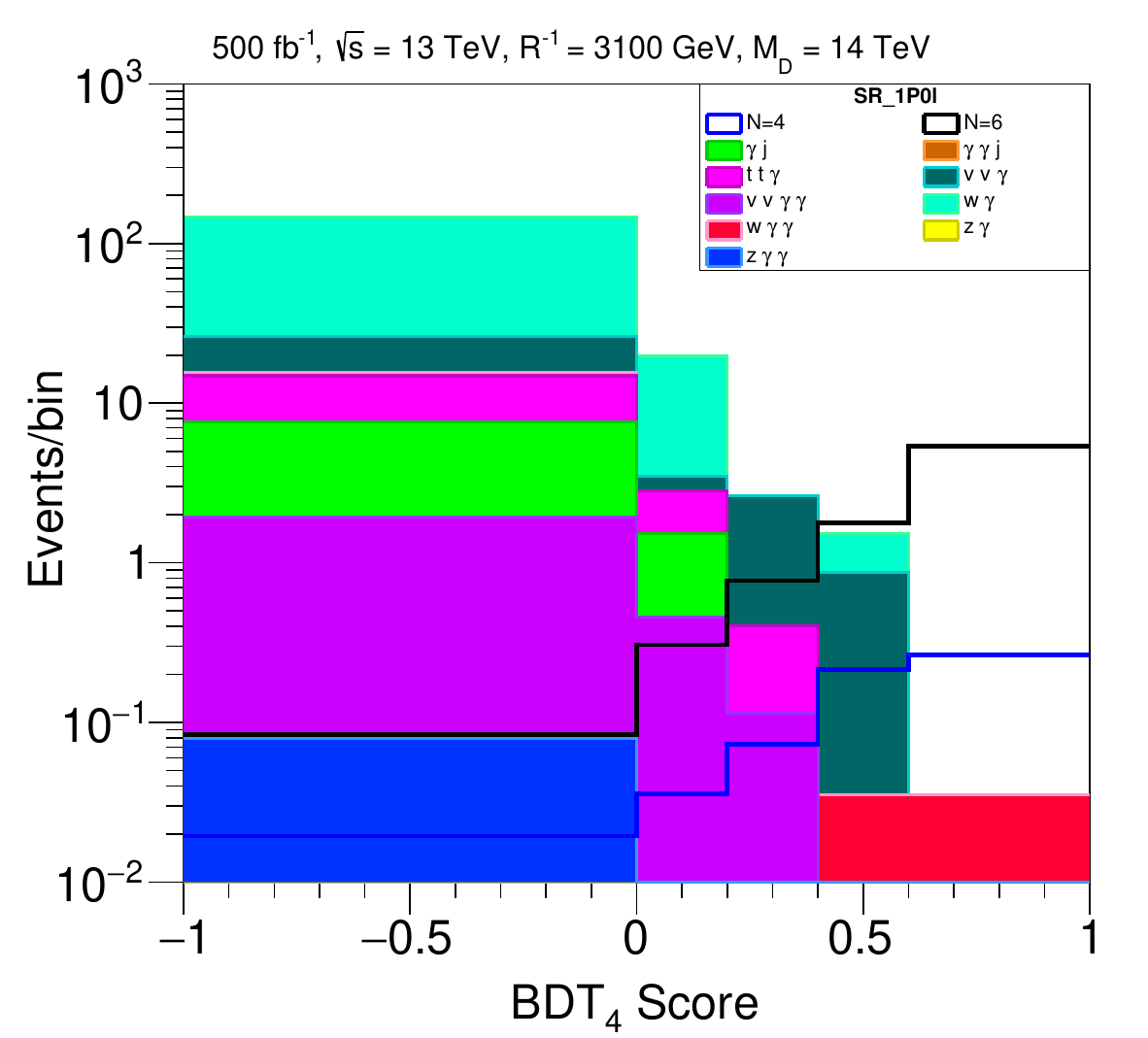}
    \includegraphics[width=0.45\columnwidth]{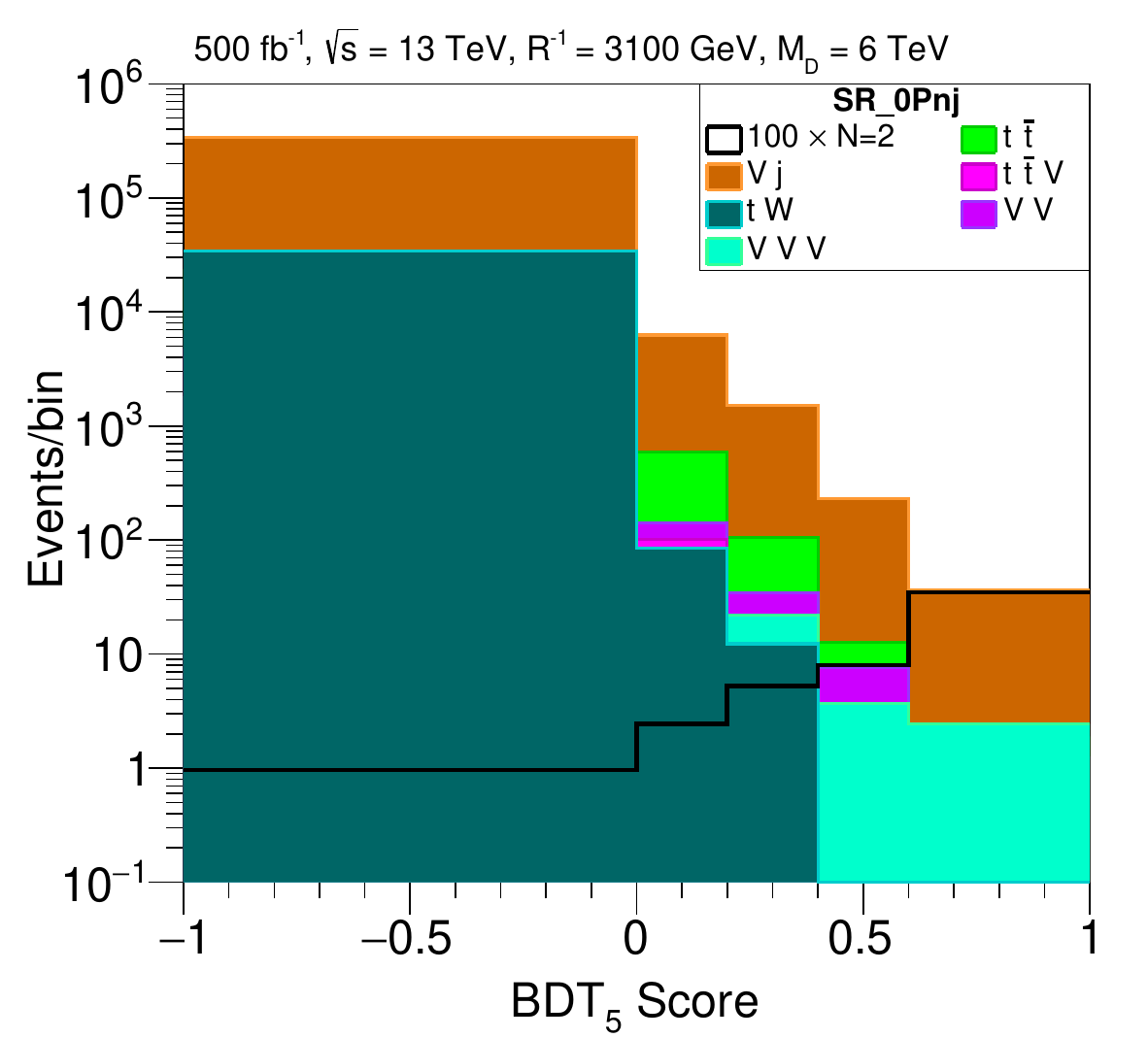}
	\caption{\label{fig: score_srs} Distribution of BDT score for the signal and background events in the five signal regions.  }
\end{figure}

To determine the signal significance in each signal region, we have statistically combined the number of signal and background events in each bin of the score histogram (see Figure~\ref{fig: score_srs}) using a profile likelihood approach with the publicly available Python-based package Spey \cite{Araz:2023bwx}. Estimating the background uncertainty in detail is beyond the scope of this work. Thus, we take a conservative approach and assume an overall uncertainty (statistical + systematic) of 20\% in the background estimation for our analysis. In Figure~\ref{fig: res1}, we present the required luminosity as a function of \( R^{-1} \) to achieve a median significance of 1.645 (equivalently, 95\% C.L.). For convenience, we have fixed \( M_D = 14 \, \text{TeV} \) for these plots. Figure~\ref{fig: res1} provides a quantitative insight into the effectiveness of the signal regions across different signal scenarios. For instance,
\begin{itemize}
    \item   In the \( N = 2 \) scenario, the gravity-mediated decay of level-1 KK particles dominates over the KKNC decay modes. As a result, in most events, the produced level-1 quarks and gluons tend to decay directly into gravity excitations, effectively breaking the decay cascade. Consequently, the likelihood of obtaining events with final-state photons—arising from the decay of the level-1 KK photon at the end of the decay chain—is very low. For this reason, the \texttt{SR\_0pnj} signal region is most effective in constraining the \( N = 2 \) scenario.
    \item   In the \( N = 4 \) signal scenario, the strength of KKNC decay modes becomes comparable to that of the GMD modes for level-1 EW gauge bosons and leptons. This allows pair-produced KK quarks and gluons to undergo KKNC decays into level-1 EW gauge bosons and leptons. The non-negligible GMD branching ratios for level-1 EW gauge bosons and leptons lead to both mono-photon and di-photon final states. However, due to the reduced SM background contribution, the di-photon final state is expected to provide better sensitivity.  This is the behavior we observe in the upper right panel of Figure \ref{fig: res1}.
    \item In the \( N = 6 \) case, the KKNC decay modes dominate over the GMD modes, making it highly probable that KK particles will follow the complete decay cascade. As a result, most events are expected to produce a final state with two photons. This explains the improved sensitivity observed for the \texttt{SR\_2p0l} and \texttt{SR\_2pnl} signal regions in the plot shown in the bottom panel of Figure~\ref{fig: res1}.

\end{itemize}

\begin{figure}[htb!]
	\centering
	\includegraphics[width=0.45\columnwidth]{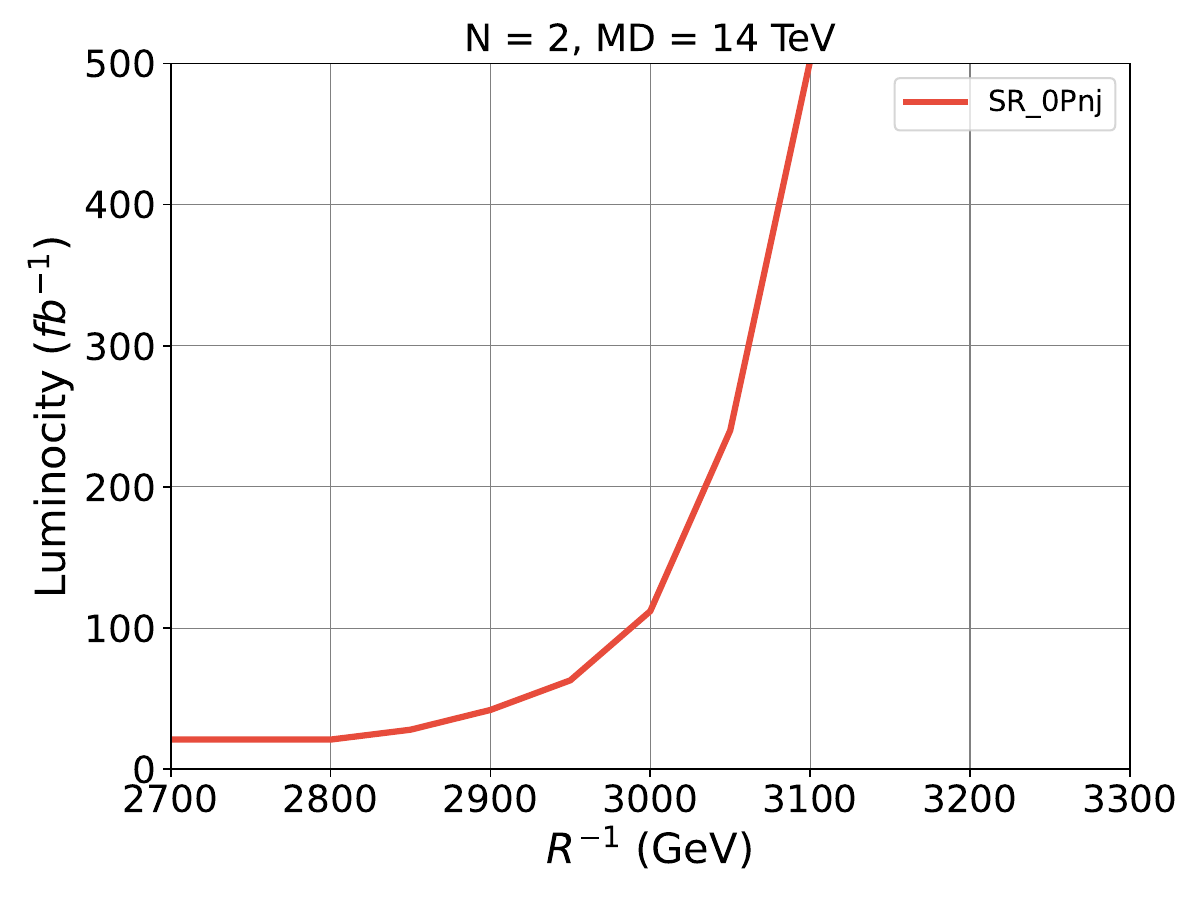} 
	\includegraphics[width=0.45\columnwidth]{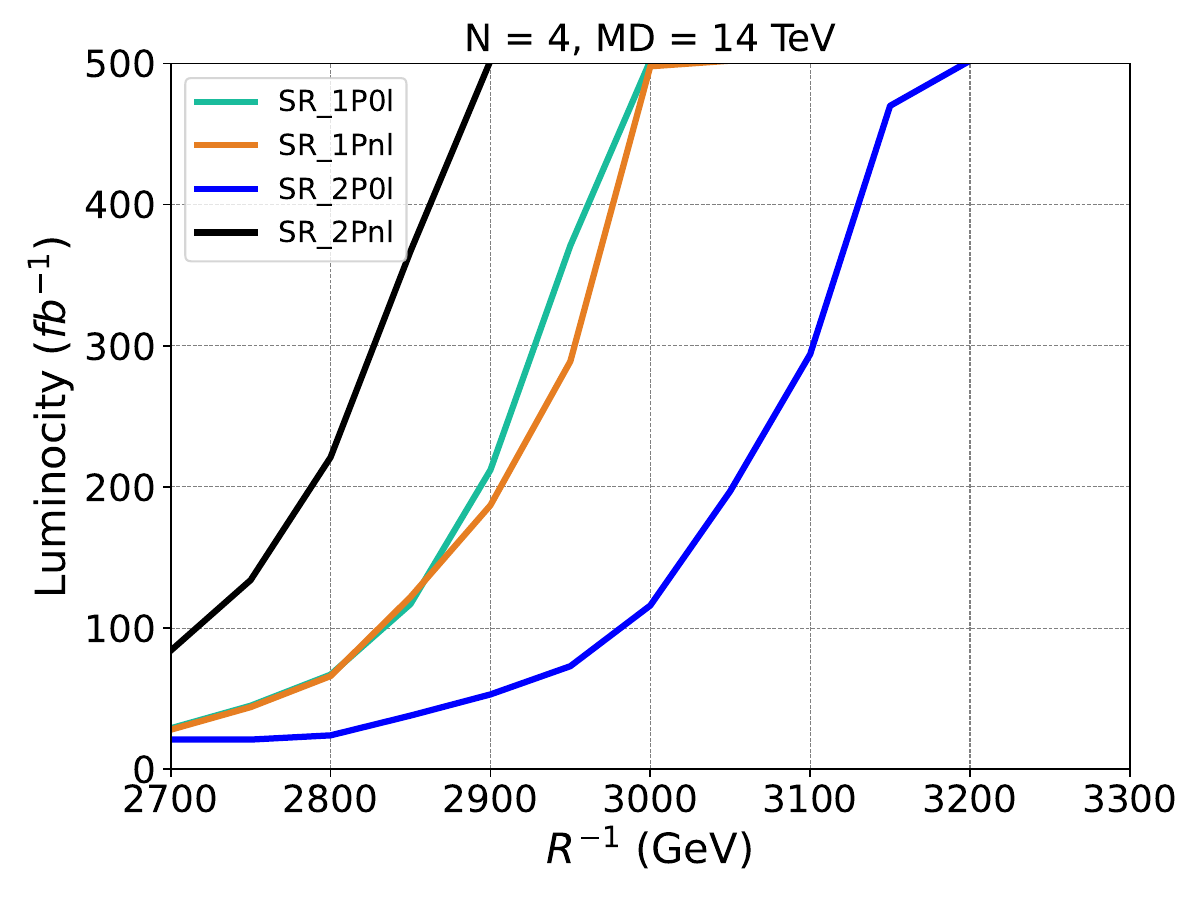}
    \includegraphics[width=0.5\columnwidth]{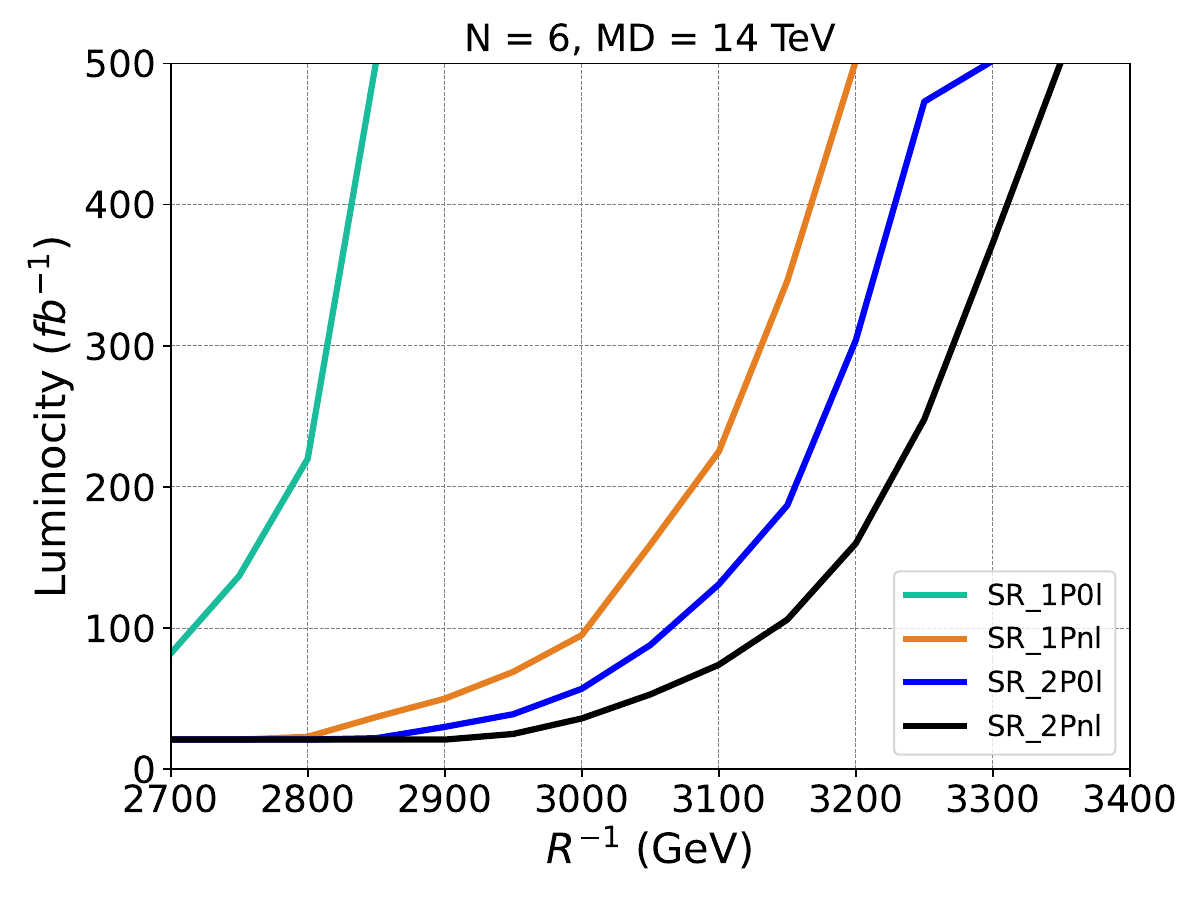}
	\caption{\label{fig: res1} Required luminosity as a function of $R^{-1}$
to achieve a median significance of 1.645 (equivalently, 95\% C.L.).  }
\end{figure}
 
In Figures \ref{fig: res21} and \ref{fig: res22}, we present the expected significance for different values of \( R^{-1} \) and \( M_D \) for the three scenarios: \( N = 2 \), 4, and 6. Based on the previous discussion, we only include results for the key signal regions for each benchmark scenario. All results are computed assuming a luminosity of \( 500~\text{fb}^{-1} \).

\begin{itemize}
    \item For the \( N = 2 \) signal scenario, only the \texttt{SR\_0pnj} signal region is relevant. The corresponding result is shown in the upper left panel of Figure~\ref{fig: res21}. Note that the significance does not depend on \( M_D \). The parameter \( M_D \) determines the density of gravity states, where a higher \( M_D \) corresponds to a lower density, reducing the available gravity states for KK particles to decay into and thus reducing the GMD decay width. However, for the \( N = 2 \) scenario, the GMD decay consistently dominates over the KKNC decays, as shown in the upper left plot of Figure~\ref{fig: res21}.

    \item For the \( N = 4 \) scenario, we show the reach of the \texttt{SR\_1pnl} (top right panel of Figure \ref{fig: res21}), \texttt{SR\_2p0l} (bottom left panel of Figure \ref{fig: res21}), and \texttt{SR\_0pnj} (bottom right panel of Figure \ref{fig: res21}) signal regions. In this case, the GMD and KKNC decay widths are comparable, making the effect of \( M_D \) more pronounced. For higher \( M_D \) values, the KKNC decay modes dominate, enhancing the sensitivity of the mono-photon and di-photon regions, as observed in the top right and bottom left panels of Figure \ref{fig: res21}. For lower \( M_D \) values, the GMD decay width dominates, making the \texttt{SR\_0pnj} region the most efficient. However, the reach of \texttt{SR\_0pnj} is weaker than in the \( N = 2 \) case, as the gravity excitation mass spectrum tends to peak at higher values (See Figure \ref{fig:grav_mass_dist}), producing softer jets that may not meet the signal selection criteria.

    \item Figure~\ref{fig: res22} presents our results for the \( N = 6 \) scenario in the \texttt{SR\_1p0l}, \texttt{SR\_1pnl}, \texttt{SR\_2p0l}, and \texttt{SR\_2pnl} signal regions. As discussed, the two di-photon regions show greater sensitivity due to reduced background and the dominance of the KKNC decay mode.
\end{itemize}

\begin{figure}[htb!]
	\centering
	\includegraphics[width=0.48\textwidth]{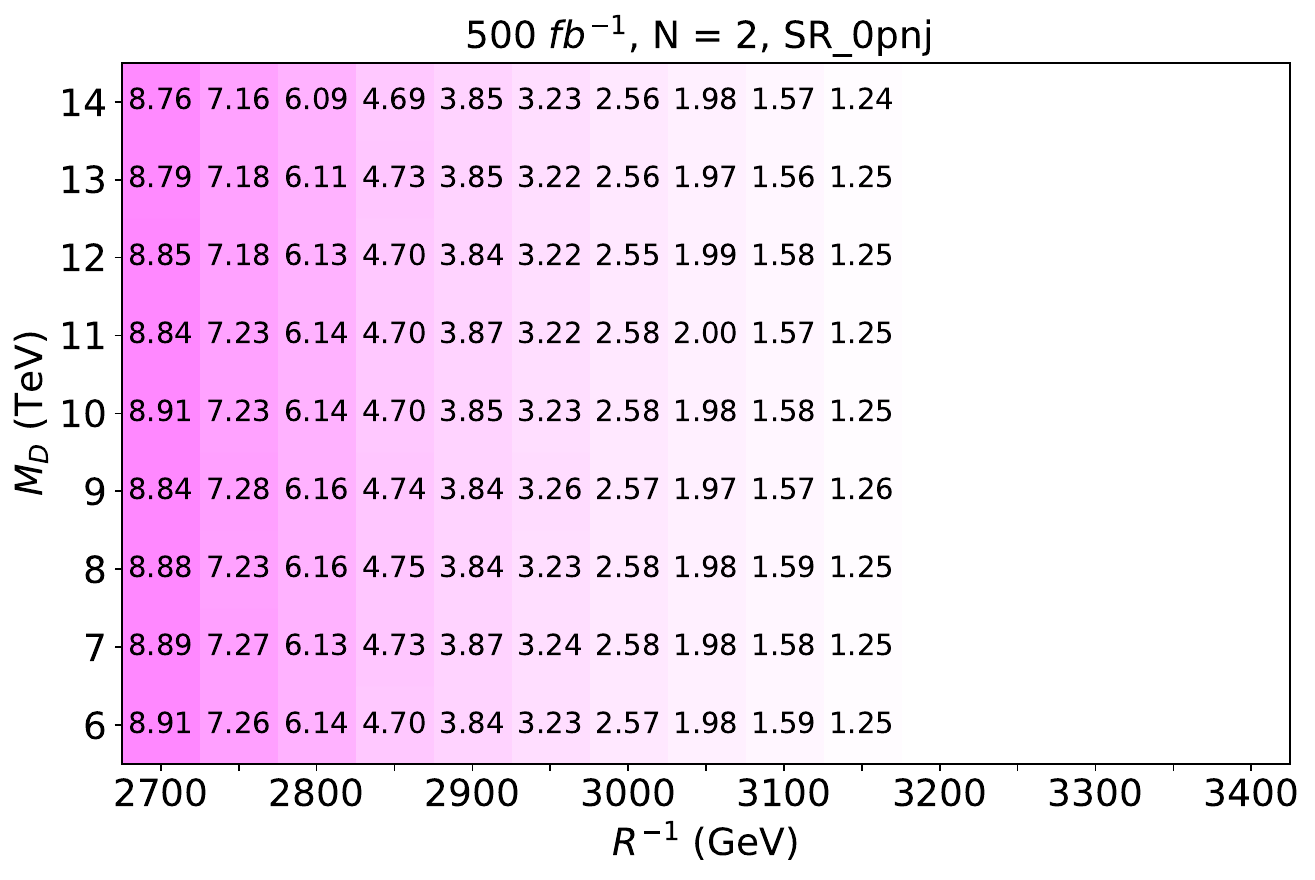} \quad
	\includegraphics[width=0.48\textwidth]{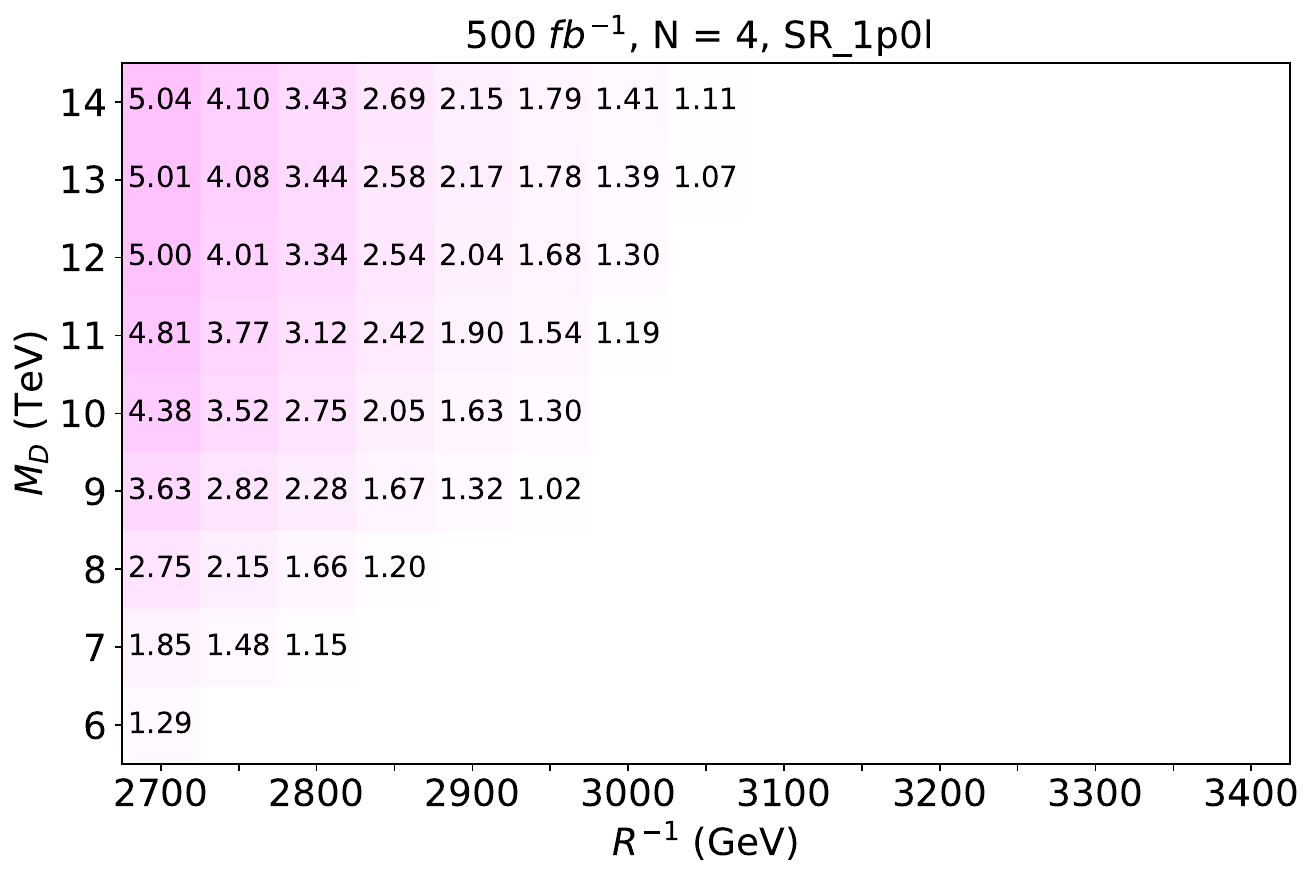} \\
    \includegraphics[width=0.48\textwidth]{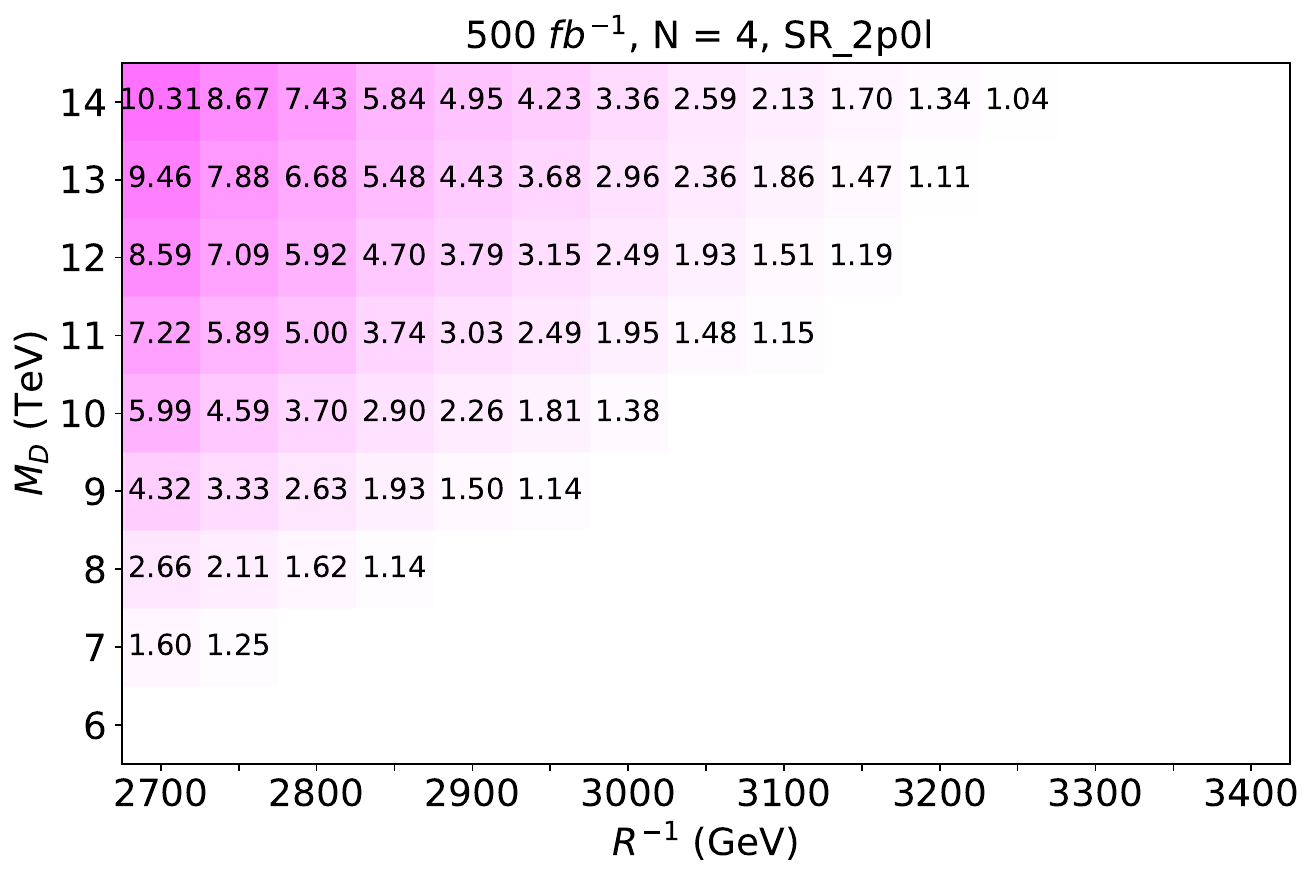} \quad
    \includegraphics[width=0.48\textwidth]{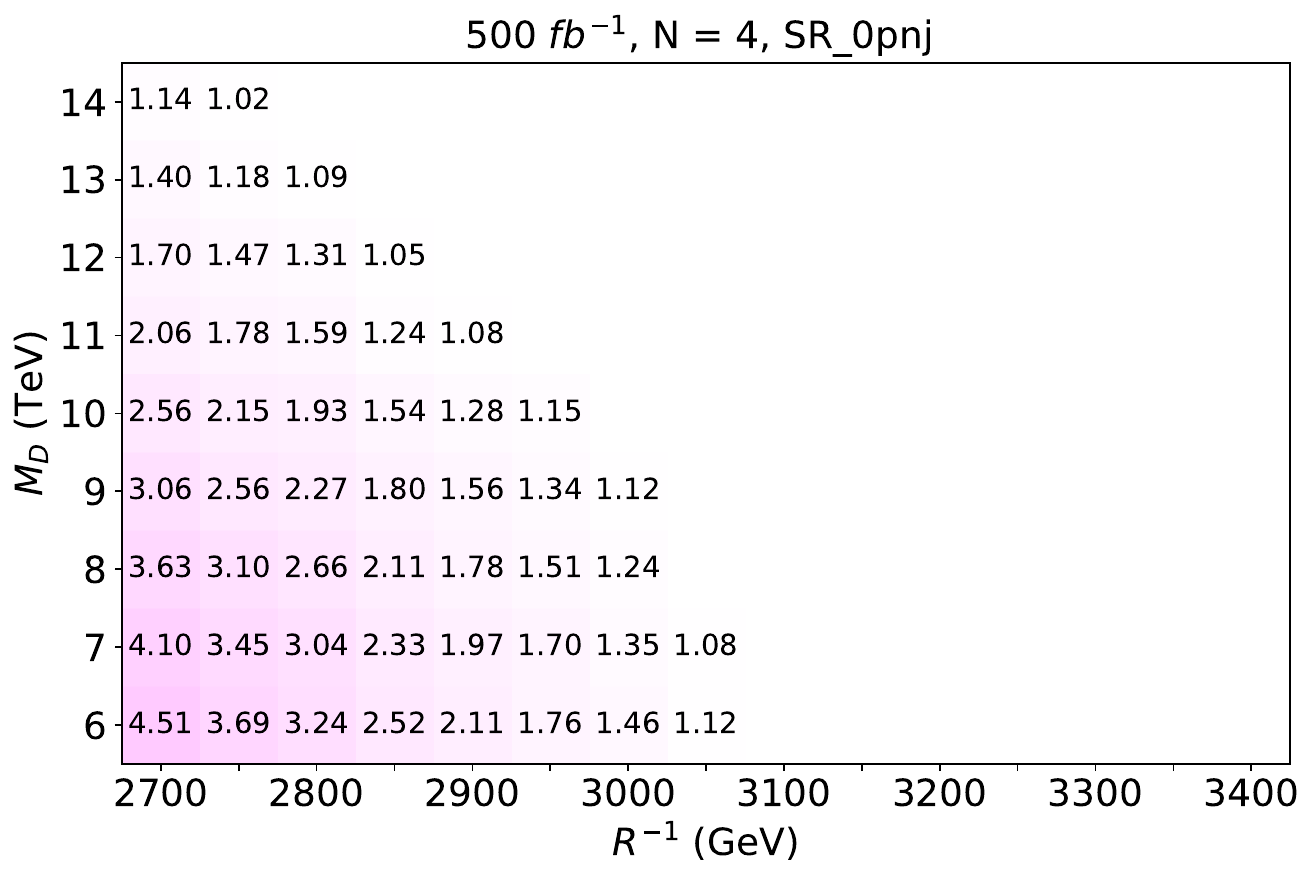}
	\caption{\label{fig: res21} Significance for different values of $R^{-1}$ and $M_D$ for the $N$ = 2 (upper left) and $N$ = 4 (other three) signal scenarios. We only present the results for the important signal regions (see text) for each case.  }
\end{figure}

\begin{figure}[htb!]
	\centering
	\includegraphics[width=0.48\textwidth]{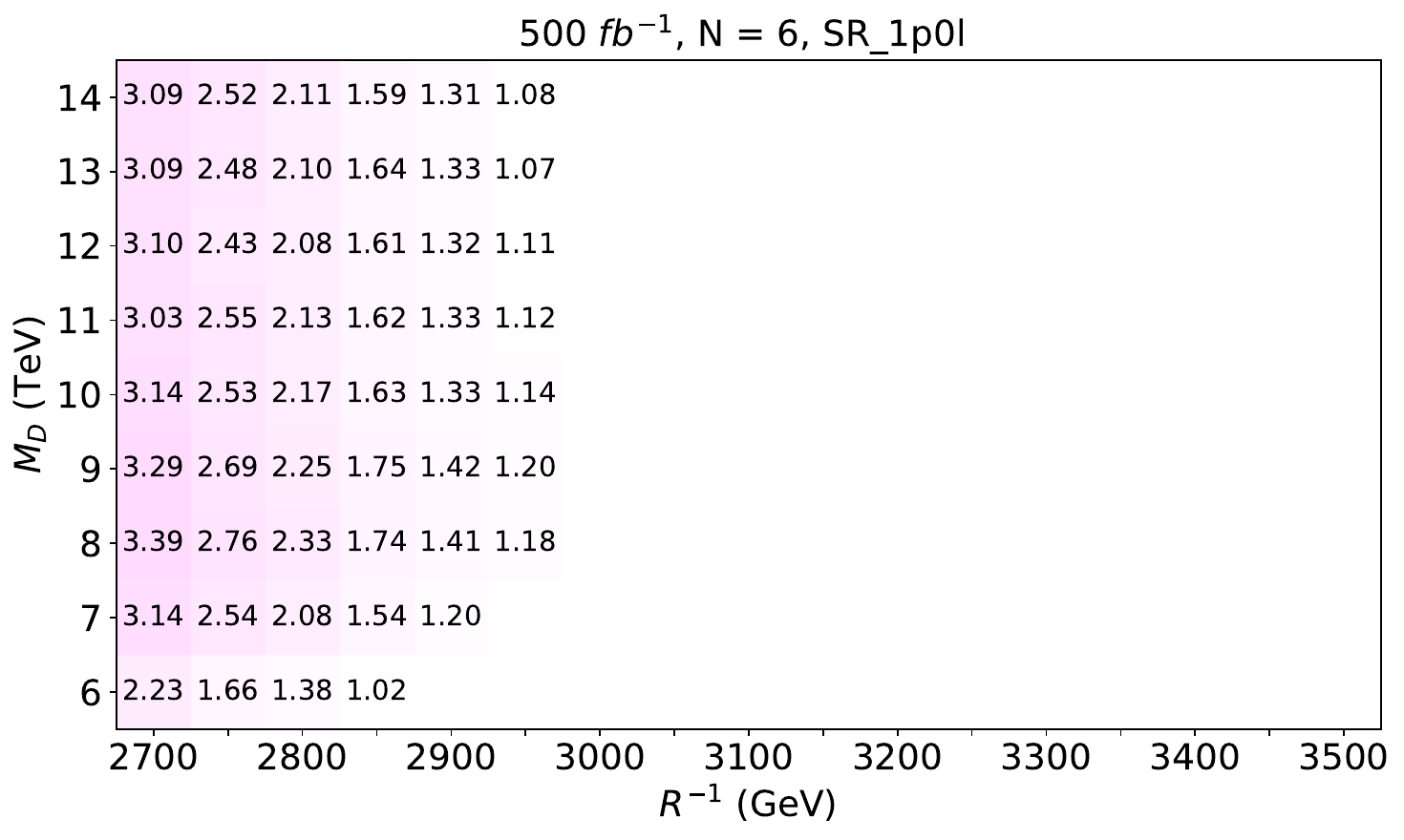} \quad
	\includegraphics[width=0.48\columnwidth]{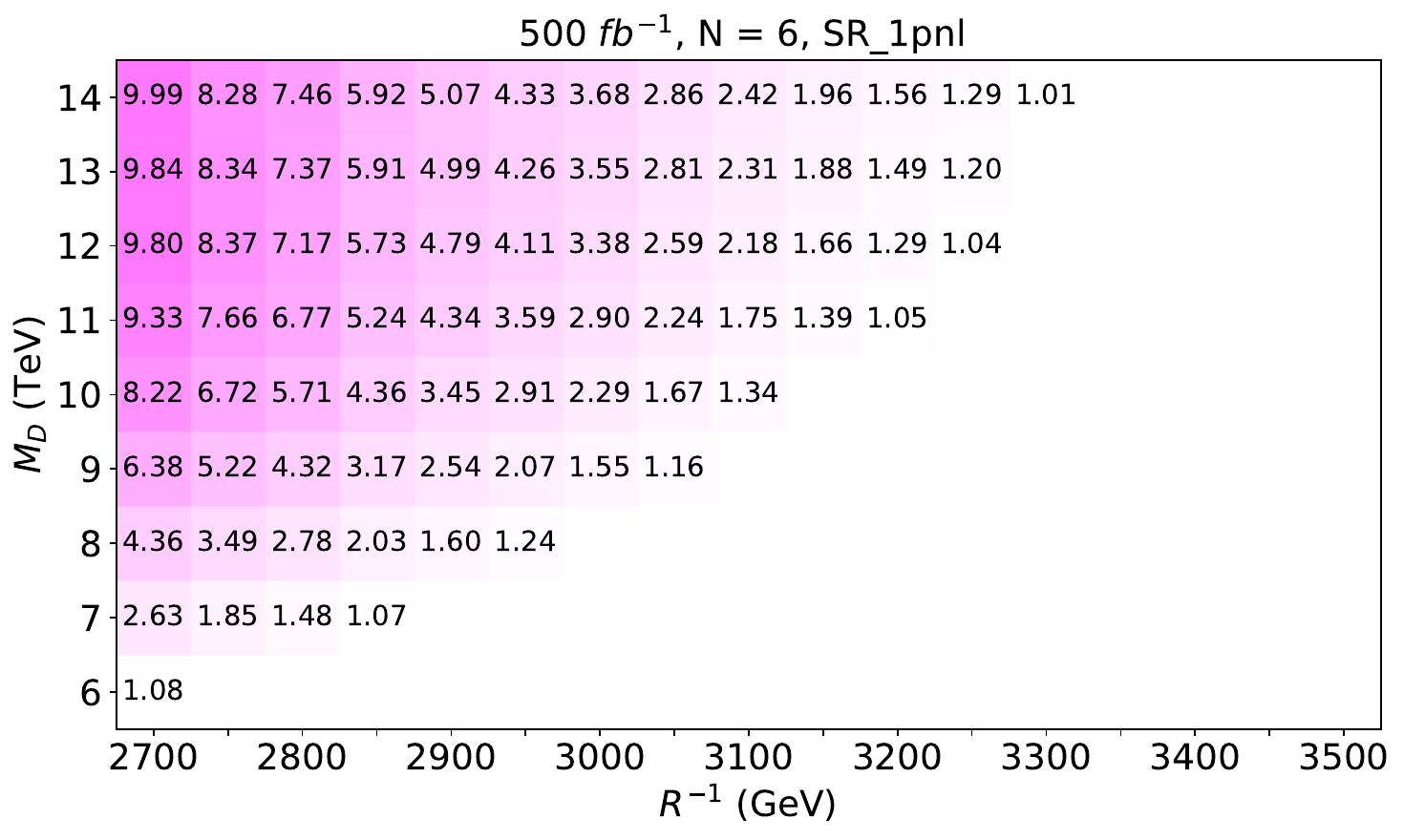}\\
    \includegraphics[width=0.48\columnwidth]{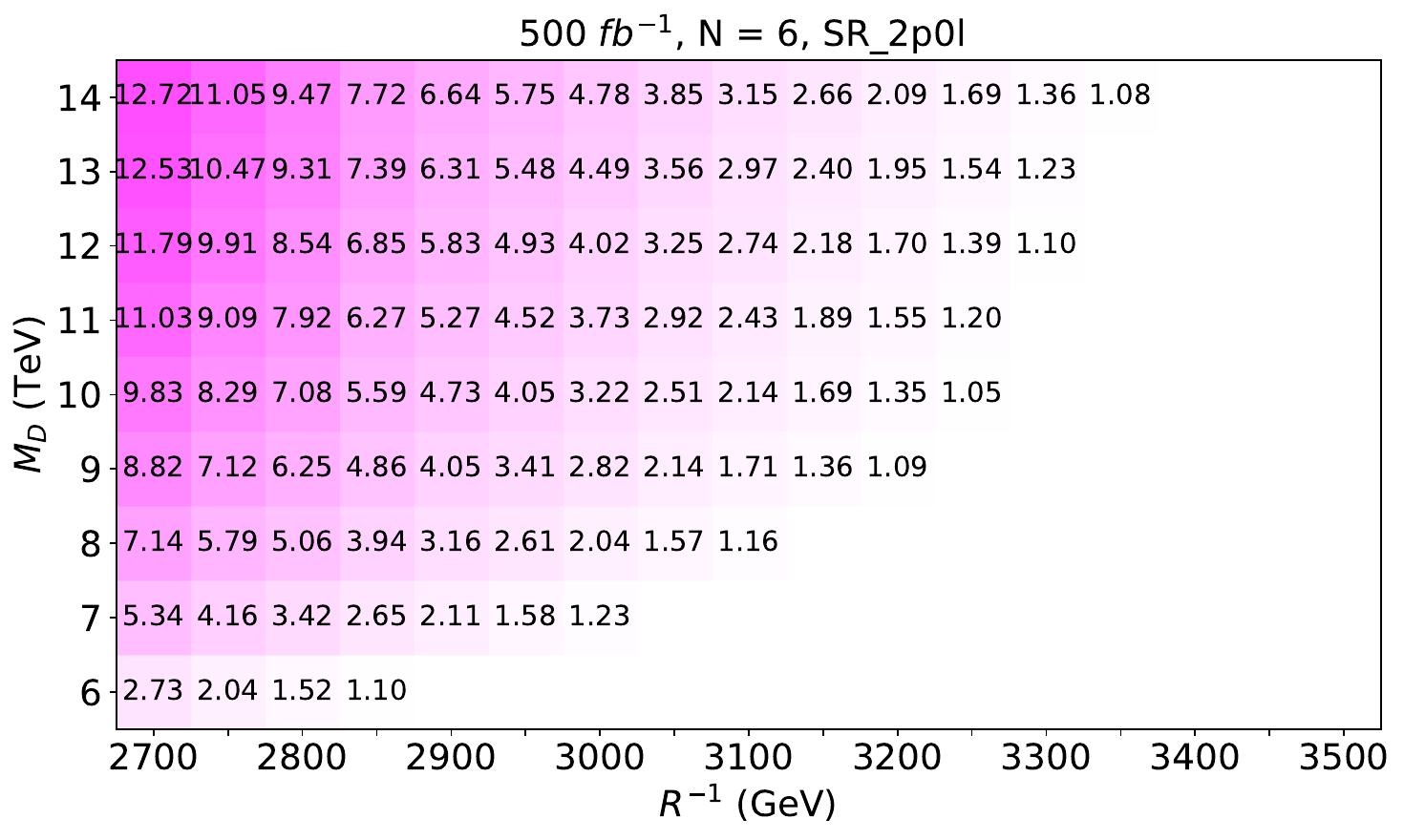}\quad
    \includegraphics[width=0.48\columnwidth]{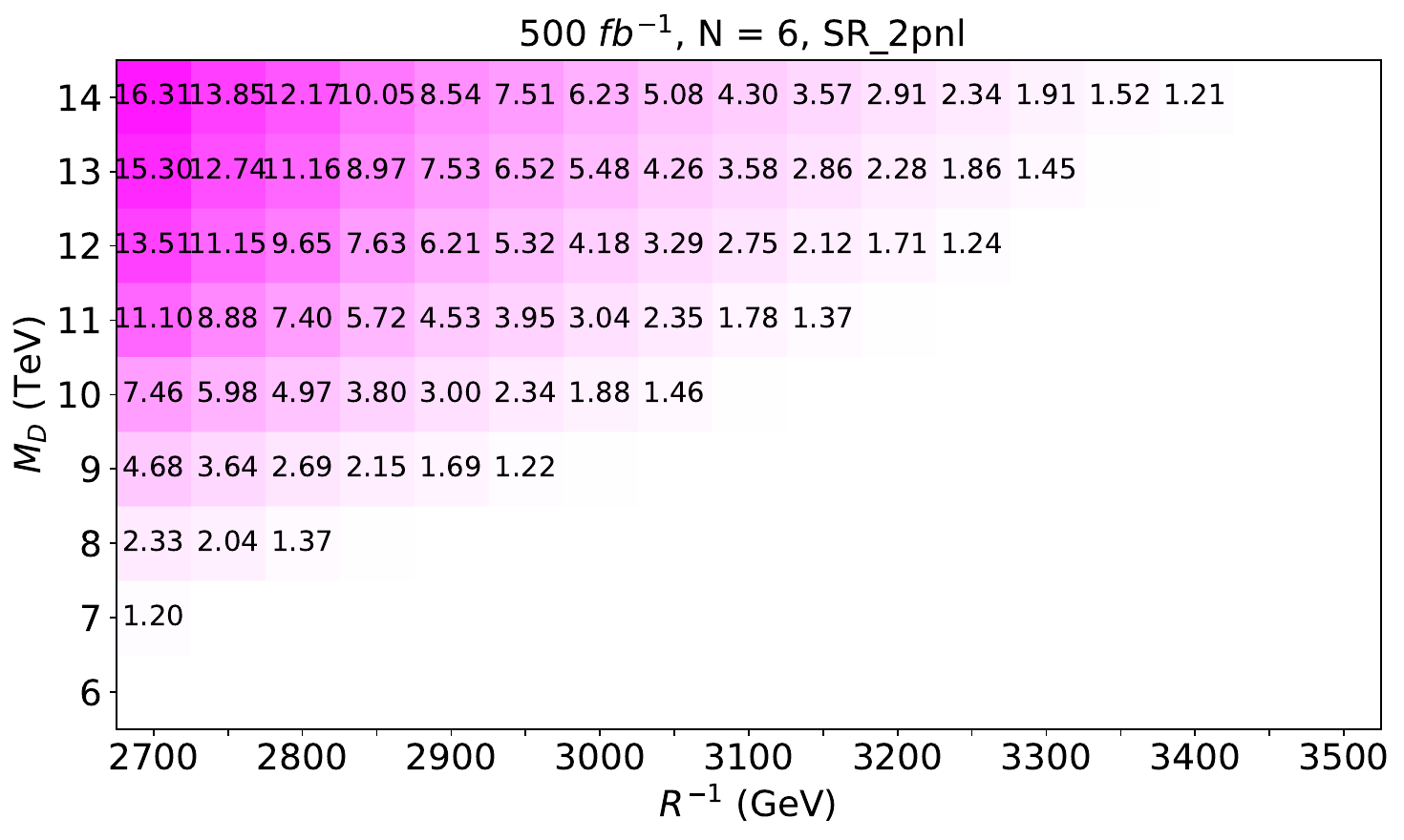}
	\caption{\label{fig: res22} Significance for different values of $R^{-1}$ and $M_D$ for the $N$ = 6 signal scenario.  }
\end{figure}

In Figure \ref{fig: res3}, we present the significance achieved by statistically combining the number of signal and background events across each signal region for different values of \( R^{-1} \) and \( M_D \). The results are based on an integrated luminosity of \( 500~\text{fb}^{-1} \). The upper left panel of Figure \ref{fig: res3} displays the results for the \( N = 2 \) scenario, showing that with \( 500~\text{fb}^{-1} \) of luminosity, the LHC can probe \( R^{-1} \) values up to 3090 GeV at 95\% C.L. significance across the full \( M_D \) range of 6 to 14 TeV. In the \( N = 4 \) scenario, shown in the upper right panel of Figure \ref{fig: res3}, a 95\% C.L. sensitivity extends up to \( R^{-1} = 3190 \) GeV for an \( M_D \) value around 14 TeV. Similarly, in the \( N = 6 \) scenario (bottom panel), sensitivity reaches approximately 3390 GeV for an \( M_D \) value near 14 TeV.
\\

\begin{figure}[htb!]
	\centering
	\includegraphics[width=0.48\columnwidth]{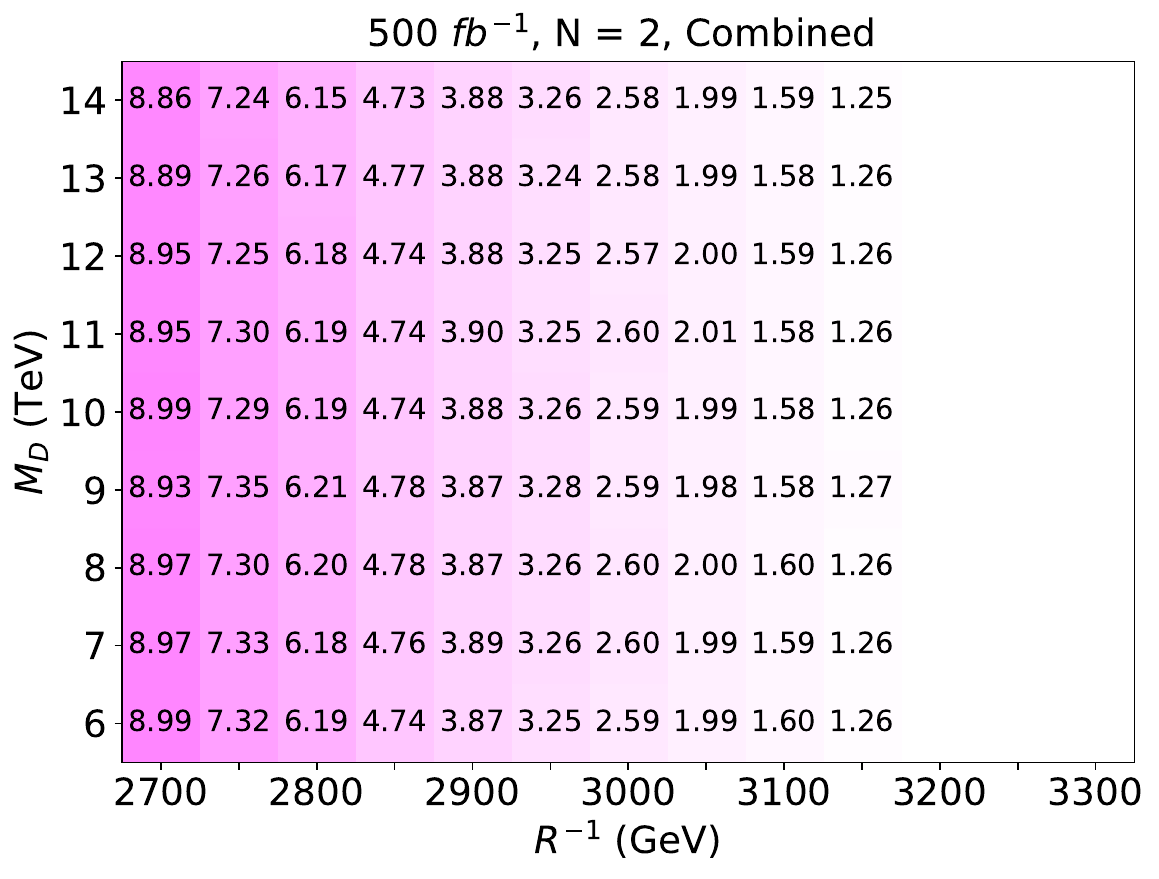} \quad
	\includegraphics[width=0.48\columnwidth]{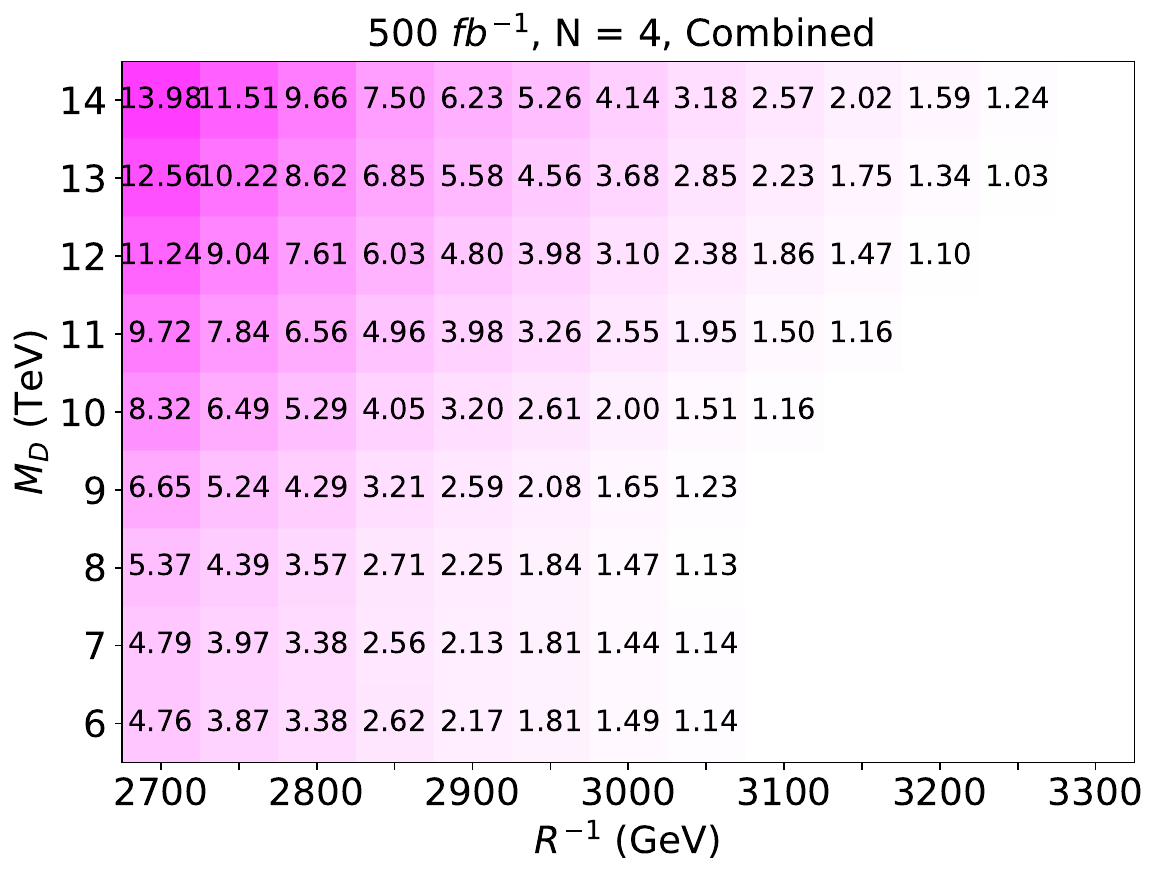}\\
    \includegraphics[width=0.48\columnwidth]{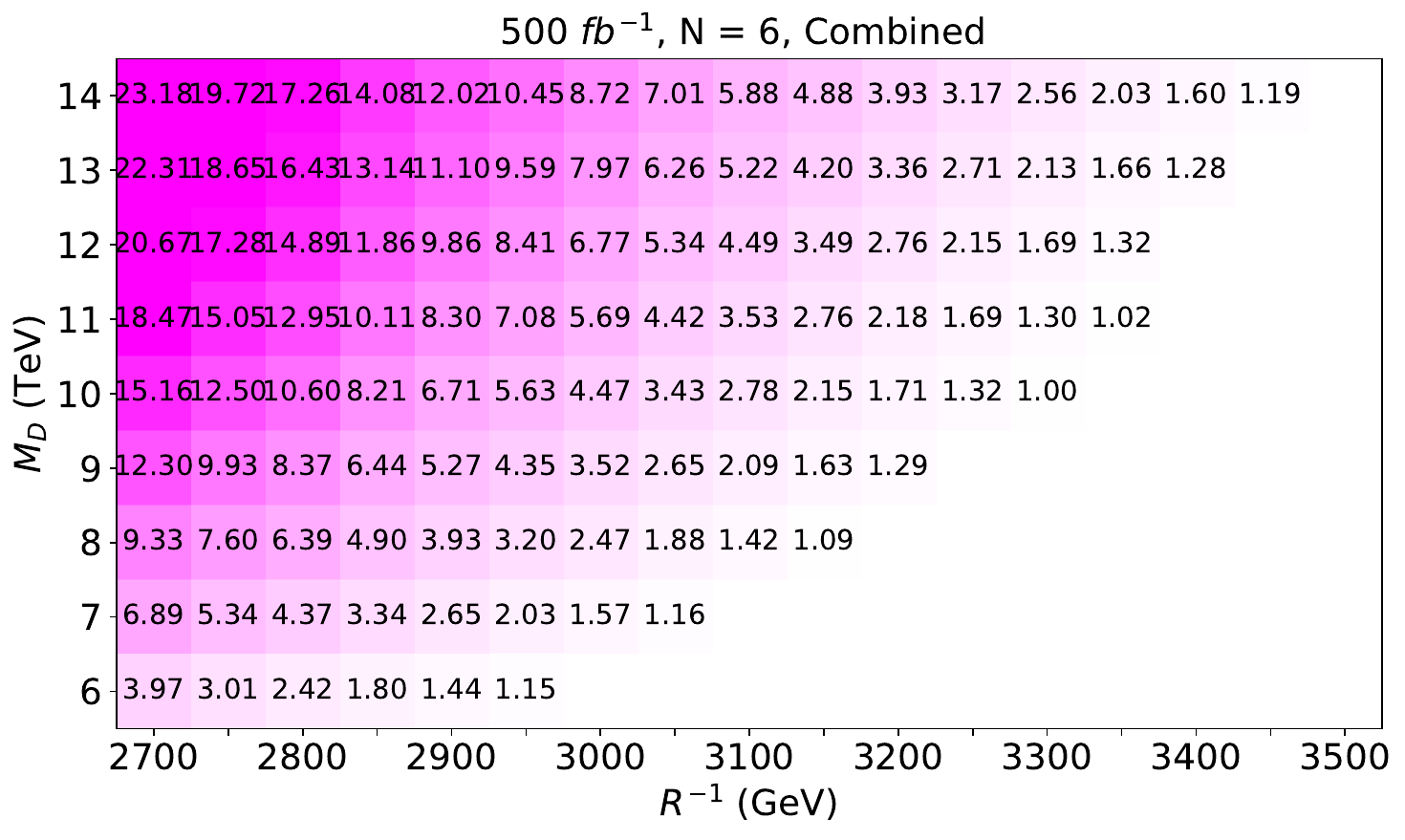}
	\caption{\label{fig: res3} Combined significance for different values of $R^{-1}$ and $M_D$ for the three signal scenarios: $N$ = 2 (upper left), $N$ = 2 (upper right), and $N$ = 6 (bottom).  }
\end{figure}

%In Figure \ref{fig: res4}, we present the required luminosity for achieving a median significance of 1.645 (corresponding to 95\% exclusion) for the three signal regions $N$ = 2 (upper left panel), $N$ = 4 (upper right panel), and $N$ = 6 (bottom panel). This result extends the results shown in Figure \ref{fig: res1}, considering all values of $M_D$ in the range of 6 TeV to 14 TeV, with a 1 TeV interval. To maintain consistency with the number of background events generated (See Table \ref{tab:back_cross1}), the results are shown up to an integrated luminosity of 500 $fb^{-1}$. The obtained results are consistent with the ones shown in figure \ref{fig: res3}.

%\begin{figure}[htb!]
%	\centering
%	\includegraphics[width=0.48\columnwidth]{Sections/final_lumi_N2.pdf} \quad
%	\includegraphics[width=0.48\columnwidth]{Sections/final_lumi_N4.pdf}\\
%    \includegraphics[width=0.48\columnwidth]{Sections/final_lumi_N6.pdf}
%	\caption{\label{fig: res4} Luminosity required for 95\% C.L. exclusion.  }
%\end{figure}
\section{Summary and Outlook}
\label{sec: sumandout}
We conducted a comprehensive study on the fat-brane MUED scenario, which involves expanding MUED to include additional dimensions ranging from $eV^{-1}$ to $KeV^{-1}$ that can only be accessed by gravity. By considering the decay mode of the level-1 photon, we identified three possible final state topologies at the LHC: the multijet, mono-photon, and diphoton final states. LHC has already performed searches for similar final states in the case of the MSSM and Gauge Mediated supersymmetry breaking models. In light of the observations being consistent with the predicted SM backgrounds, model-independent 95\% C.L. upper limits have been set on the visible cross-section. In our analysis, we have incorporated three such searches by the ATLAS collaboration, enabling us to impose constraints on the parameter space of the fat-brane MUED model. For the N=2 scenario, the multijet search has the best reach and excludes $R^{-1}$ below 2975 GeV independent of the value of $M_D$. For the N=4 case, the multi-jet and mono-photon searches probe complementary parts of the parameter space. The former(latter) excludes $R^{-1} <$ 2898(2958) for $M_D = $ 500(15000) GeV. For the case where N=6, the di-photon search is most effective and sets a lower limit of 3002 GeV on $R^{-1}$ for $M_D=$ 15000 GeV. The low-$M_D$ region of the parameter space is also sensitive to the multijet search, and for a $M_D$ of 5000 GeV, it excludes $R^{-1}$ below 2871 GeV.\\
Further, we forecast future limits at the 13 TeV LHC for 500 $fb^{-1}$ of integrated luminosity by proposing a novel search strategy that yields improved limits on the parameter space on interest. Our strategy advocates the implementation of fat jets and uses two BDT classifiers to discern top and W/Z fat jets from QCD fat jets. We also consider signal regions without fat jets, where we have modified the previous ATLAS analysis to improve their reach. Implementing our strategy, we project a lower limit of around 3320 GeV on $R^{-1}$ throughout the range of $M_D$ for the N=2 scenario, which is 345 GeV higher than the previous estimate. For the case where N=4 we can exclude $R^{-1}$ below 2942(3157) for $M_D = $ 5000(15000) GeV. As for the N=6 case, our proposed analysis has a reach of up to 3021(3225) GeV in $R^{-1}$ for $M_D = $ 5000(15000) GeV.\\
To conclude this section, it is worth mentioning that the fat jet signal regions hold promising potential for investigating the aforementioned parameter space provided the use of a classifier with superior performance, which we have deferred to future research endeavors.

\section*{ACKNOWLEDGMENTS}
KG acknowledges the financial support (MTR/2022/000989) provided by the MATRICS research grant, titled {\em "Exploring Next-to-Minimal Realizations of Universal Extra-Dimension Scenarios"}, funded by the Science and Engineering Research Board (SERB). KG also gratefully acknowledges the local hospitality provided by the Helsinki Institute of Physics, University of Helsinki, during his visit to Helsinki, where a significant portion of this project was conducted. The simulations were partly supported by the SAMKHYA: High-Performance Computing Facility provided by the Institute of Physics, Bhubaneswar.
\appendix
\section{Appendix}
\subsection{GMD widths of level-1 KK particles}
\label{sec:appendix-1}
 The gravity mediated decay width of KK-1 fermions is given by \cite{MACESANU_2006},
 \begin{equation}
     \begin{split}
         \Gamma(q^l \rightarrow q h^{\Vec{n}}) &= |F^c_{l|\Vec{n}}|^2 \frac{k^2}{2\times 384 \pi}\frac{M^3}{x^4}[(1-x^2)^4(2+3x^2)]\\
         \Gamma(q^l \rightarrow q A^{\Vec{n}}) &= |F^s_{l|\Vec{n}}|^2 \frac{k^2}{2\times 256 \pi} M^3 [(1-x^2)^2(2+x^2)] P_{55}\\
         \Gamma(q^l \rightarrow q \phi^{\Vec{n}}) &= |F^c_{l|\Vec{n}}|^2 \frac{k^2}{2\times 256 \pi} M^3 (1-x^2)^2 [c_{11}\frac{(1-x^2)^2}{x^4}+2c_{12} \frac{1-x^2}{x^2}+c_{22}]
     \end{split}
 \end{equation}
  Where M = l/R is the mass of the matter excitation, $x = m_{\Vec{n}}/M$ with $m_{\Vec{n}}$, the mass of the gravity mode. The form factor $F^{c/s}_{l|\Vec{n}}$ and the coefficients $P_{55}$,$c_{ij}$ appear because the gravity excitations are not all independent. They have the following forms: \\
  \begin{equation}
      \begin{split}
          |F^c_{l|n}|^2 &= \frac{4}{\pi^2}\frac{x_5^2}{(1-x_5^2)^2} [1+cos(\pi x_5)]\\
          |F^s_{l|n}|^2 &= \frac{|F^s_{l|n}|^2}{x_5^2}\\
          P_{55} &= 1-\frac{n_5^2}{n^2}\\
          c_{11} &= \omega^2(N-1)\\
          c_{12} &= -\frac{2}{N+2} P_{55}\\
          c_{22} &= \frac{2(N+1)}{(N+2)} P_{55}^2\\
      \end{split}
  \end{equation}
  where, $x_5 = 2\pi n_5 R/(lr)$. We can have similar expressions for the GMD decay width of KK-1 gauge bosons,\\ 
   \begin{equation}
     \begin{split}
         \Gamma(V^l \rightarrow V h^{\Vec{n}}) &= |F^c_{l|\Vec{n}}|^2 \frac{k^2}{3\times 96 \pi}\frac{M^3}{x^4}[(1-x^2)^3(1+3x^2+6x^4)]\\
         \Gamma(V^l \rightarrow V A^{\Vec{n}}) &= |F^s_{l|\Vec{n}}|^2 \frac{k^2}{3\times 32 \pi} \frac{M^3}{x^2} [(1-x^2)^3(1+x^2)] P_{55}\\
         \Gamma(V^l \rightarrow V \phi^{\Vec{n}}) &= |F^c_{l|\Vec{n}}|^2 \frac{k^2}{3\times 128 \pi} M^3 (1-x^2)^3 [c_{11}\frac{1}{x^4}+2c_{12} \frac{1}{x^2}+c_{22}]
     \end{split}
 \end{equation}
 
\subsection{Cut-flow chat for SRM and SRH signal regions}
\label{sec:appendix-2}
In Tables \ref{tab: SRM} and \ref{tab: SRH} we present the cut-flow tables for the SRM and SRH signal regions, respectively.
 \begin{table}[htb!]
	\begin{center}
		\begin{tabular}{l|c|c|c|c}
			\toprule 
            &\multicolumn{4}{r}{$m_{\Tilde{g}}=2000 GeV, m_{\Tilde{\chi_1^0}}=250 GeV$} \\ 
            \midrule
			  Cuts& \multicolumn{2}{c|}{$\gamma/Z$} & \multicolumn{2}{c}{$\gamma/h$} \\ 
			\midrule
             & Hepdata& Our Result & Hepdata  &  Our Result\\
            \midrule
			Trigger (one photon $p_T >$ 140 Gev) & $79.60$& $ 81.73 $ & $71.29$&  $ 75.56 $\\

             At least one photon                 & $79.51$& $ 81.73 $ & $71.24$&  $ 75.56 $\\

             Lepton Veto                         & $50.15$& $ 52.45 $ & $49.02$&  $ 52.78 $\\

             Leading photon $p_T >$ 300 Gev      & $42.33$& $ 40.26 $ & $41.87$&  $ 40.38 $\\

             $E_T^{miss} >$ 300 Gev             & $35.55$& $ 33.80 $ & $35.29$&  $ 33.74 $\\

             Number of Jets $\ge$ 5             & $34.52$& $ 33.54 $ & $34.22$&  $ 33.5 $\\

             $\Delta \phi(jet, E_T^{miss}) >$ 0.4 & $29.50$& $ 28.68 $ & $29.24$&  $ 28.49 $\\

             $\Delta \phi(\gamma, E_T^{miss}) >$ 0.4& $28.43$& $ 27.73 $ & $28.55$&  $ 27.65 $\\

            $H_T >$ 1600 GeV                        & $25.86$& $ 23.81 $ & $26.54$&  $ 24.65 $\\

            $R_T^4 <$ 0.9                         & $21.81$& $ 21.57 $ & $22.66$&  $ 22.25 $\\
		
			\bottomrule
		\end{tabular} 
		\caption{\label{tab: SRM} Cut-Flow table for the SRM signal region. The entries in the second and fourth columns are the results provided by the Atlas collaboration \cite{ATLAS:2022ckd} in the form of Hepdata. The entries in the third and fifth columns represent our simulated results.}
	\end{center}
\end{table}

 \begin{table}[htb!]
	\begin{center}
		\begin{tabular}{l|c|c|c|c}
			\toprule 
            &\multicolumn{4}{r}{$m_{\Tilde{g}}=2000 GeV, m_{\Tilde{\chi_1^0}}=250 GeV$} \\ 
            \midrule
			  Cuts& \multicolumn{2}{c|}{$\gamma/Z$} & \multicolumn{2}{c}{$\gamma/h$} \\ 
			\midrule
             & Hepdata& Our Result & Hepdata  &  Our Result\\
            \midrule
			Trigger (one photon $p_T >$ 140 Gev) & $92.73$& $ 92.3 $ & $76.03$&  $ 77.61 $\\

             At least one photon                 & $92.72$& $ 92.3 $ & $76.01$&  $ 77.61 $\\

             Lepton Veto                         & $88.15$& $ 88.79 $ & $73.73$&  $ 74.93 $\\

             Leading photon $p_T >$ 400 Gev      & $83.18$& $ 81.6 $ & $69.48$&  $ 68.39 $\\

             $E_T^{miss} >$ 600 Gev             & $65.41$& $ 64.03 $ & $54.66$&  $ 53.06 $\\

             Number of Jets $\ge$ 3             & $51.19$& $ 58.76 $ & $41.12$&  $ 41.03 $\\

             $\Delta \phi(jet, E_T^{miss}) >$ 0.4 & $42.84$& $ 48.53 $ & $35.02$&  $ 34.66 $\\

             $\Delta \phi(\gamma, E_T^{miss}) >$ 0.4& $42.70$& $ 48.33 $ & $34.92$&  $ 34.50 $\\

            $H_T >$ 1600 GeV                        & $26.28$& $ 27.95 $ & $23.60$&  $ 23.36 $\\
	
			\bottomrule
		\end{tabular} 
		\caption{\label{tab: SRH} Cut-Flow table for the SRH signal region. The entries in the second and fourth columns are the results provided by the Atlas collaboration \cite{ATLAS:2022ckd} in the form of Hepdata. The entries in the third and fifth columns represent our simulated results.}
	\end{center}
\end{table}

\subsection{BDT classifier}
\label{sec: appendix-3}
In this section, we will discuss the BDT classifier used in our analysis for boosted boson tagging. To train the classifier, we require both signal and background fat jet samples. To generate the signal fat jets, we have considered the processes $p p \rightarrow W+ W-$ and $p p \rightarrow Z Z$, and for the background fat jets, di-jet production is considered. In our analysis, we do not distinguish between the W and Z jets, and we consider them collectively to be V jets. For training the classifier, we have used the TMVA 4.3 toolkit \cite{Hocker:2007ht} integrated into ROOT 6.24\cite{Brun:2000es}. The training and evaluation process uses 1.5M fatjets of both V and QCD types. BDT, being a simple cut-based classifier, utilizes high-level features/variables (HLF). Before constructing these variables for our analysis, we pass the signal and background fat jets through a truth-level tagging stage. For signal fat jets, we demand that both the parent boson and all its decay products be present inside the cone of the fat jets. Similarly, for the QCD jets, we demand that the partonic quark/gluon be present inside the reconstruction radius. This truth level tagging enhances the purity of the sample, i.e., it ensures that the signal and background fat jets used for the training/evaluation of the classifier are properly reconstructed. These truth-tagged jets are then split in the ratio of 80:20 for training and testing the classifiers. \\

To train our classifier, we use the five kinematic features of the fat jets:  \\

\textbf{The invariant mass} of the fat jets is reconstructed from the four-momentum of the constituents. It has the following form\\
\begin{equation}
    M = \sqrt{\sum_i (E_i)^2 - \sum_i (p_i)^2}
\end{equation}
where the sum runs over all constituents of the fat jet.\\
\textbf{The Jet Charge \cite{Krohn:2012fg}} is reconstructed from the charge of individual tracks inside the fat jet, and we define it as,
\begin{equation}
    Q_k = \frac{\sum_i q_i (p_{Ti})^k}{\sum_i p_{T,i}}
\end{equation}
Where $p_{Ti}$ are the transverse momentum associated with the tracks of charge $q_i$ and k, the regularisation exponent has a value of 0.2 for our analysis.\\
\textbf{The N-subjettiness variable $\tau_N$ \cite{Thaler_2011}} gives a quantitative measure of the likeliness of a fat jet to have N substructures. We define it as,
\begin{equation}
    \tau_N = \frac{1}{\sum_k p_{Tk} R_0^{\beta}} \sum_k p_{Tk} min(\Delta R^{\beta}_{1,k}..\Delta R^{\beta}_{N,k})
\end{equation}
where the sum runs over all the jet constituents, $\beta$ is the thrust parameter, $R_0$ is the radius of the reconstruction cone, and $\Delta R{i,k}$ characterize the separation between the constituent k and the candidate sub jet i.\\
\textbf{b-tag} is a boolean observable that takes a value of 1 when at least one of the sub-jets inside a fat jet is a b-jet.\\
Table \ref{tab: bdtParametrs} summarizes the BDT hyperparameters used while training the classifiers. In Table \ref{tab: bdtSeparationRanking}, we present the method-unspecific and method-specific ranking of the kinematical variables used for training our classifiers. The table on the top corresponds to the top vs. QCD case, while that on the bottom corresponds to the W/Z vs. QCD classifier. Finally, we present the covariance matrix for the input features in Figure \ref{fig: covmat}.\\
\begin{table}[htb!]
\centering
\begin{tabular}{ll}
\toprule
BDT hyperparameter & Optimised choice \\
\midrule
NTrees & 1000 \\ 
MinNodeSize  & 5\% \\ 
MaxDepth & 4 \\ 
BoostType & AdaBoost\\ 
AdaBoostBeta & 0.1  \\ 
UseBaggedBoost & True \\ 
BaggedSampleFraction & 0.5 \\
SeparationType & GiniIndex \\ 
nCuts & -1 \\ 
\bottomrule
\end{tabular} 
\caption{\label{tab: bdtParametrs} Summary of optimised BDT hyperparameters.}
\end{table}

\begin{table}[!htb]
\centering
%\begin{tabular}{lll}
%\toprule
%Feature & Method-unspecific & Method-specific \\ 
%& separation & ranking \\ 
%\midrule 
%$m$ & 0.4336 & 0.49 \\ 
%$b$-tag & 0.3428 & 0.198 \\
%$\tau_{32}$ & 0.268 & 0.167 \\
%$\tau_{21}$ & 0.277 & 0.133 \\
%$Q_k$ & 0.00498 & 0.001 \\ 
%\bottomrule
%\end{tabular} 
\centering
\begin{tabular}{lll}
\toprule
Feature & Method-unspecific & Method-specific \\ 
& separation & ranking \\ 
\midrule 
$m$ & 0.0427 & 0.49 \\ 
$\tau_{21}$ & 0.387 & 0.326 \\ 
$\tau_{32}$ & 0.055 & 0.112 \\ 
$Q_k$ & 0.012 & 0.031 \\ 
$b$-tag & 0.019 & 0.03 \\
\bottomrule
\end{tabular}
\caption{\label{tab: bdtSeparationRanking} Method-unspecific separation and method-specific ranking of the input features.}
\end{table}

\begin{figure}[htb!]
	\centering
	\includegraphics[width=0.45\columnwidth]{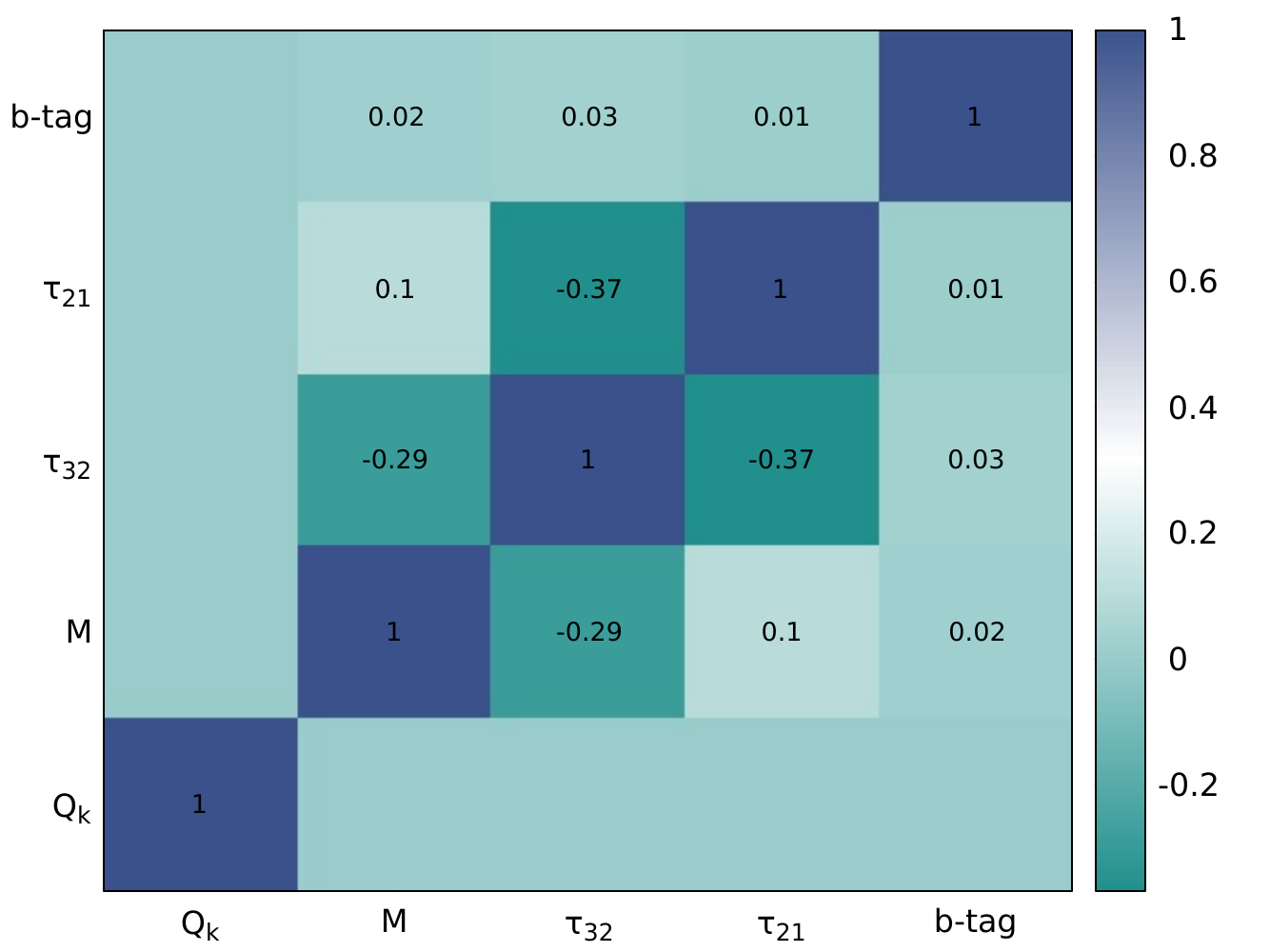}\qquad
    \includegraphics[width=0.45\columnwidth]{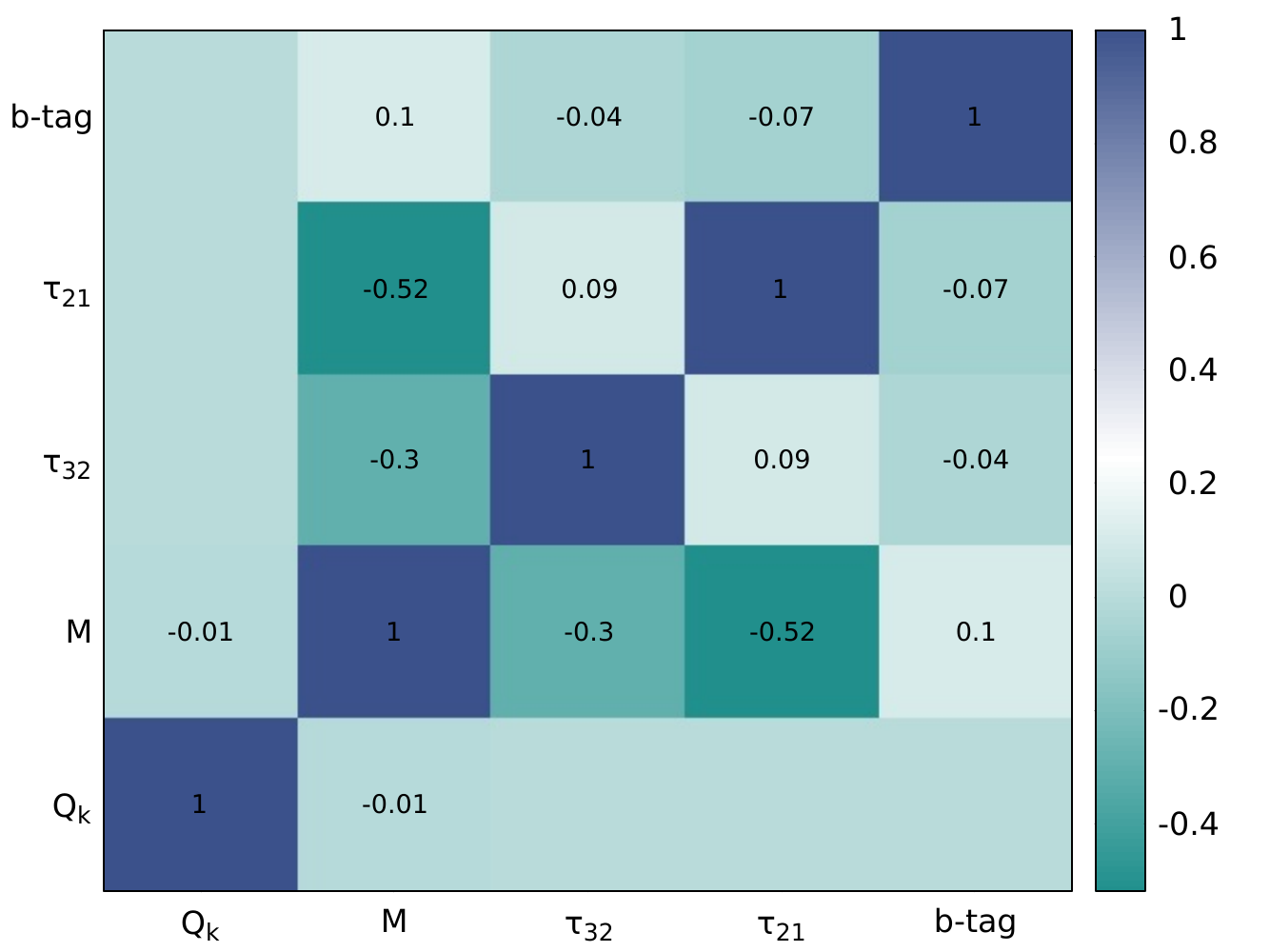}
	\caption{\label{fig: covmat} Correlations among the input features for the W/Z jets (left) and the QCD jets (right).  }
\end{figure}

We evaluate the performance of the classifier in terms of its receiver-operator-characteristic (ROC) curves and present our result in Fig. \ref{fig: rocplot}. The ROC curve also serves as a valuable tool in aiding the selection of the threshold BDT score for our final analysis. We find a threshold of 0.34, which corresponds to a signal efficiency of 60\%, suitable for our analysis. 

\begin{figure}[htb!]
	\centering
	\includegraphics[width=0.75\columnwidth]{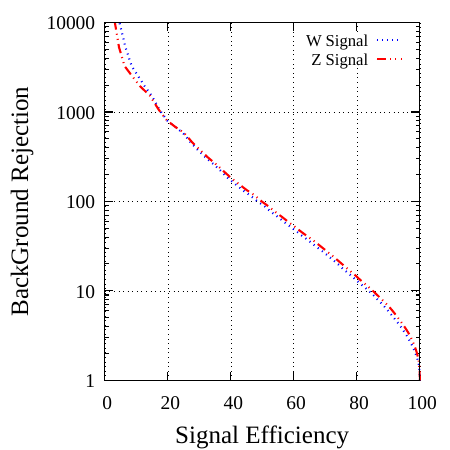}
	\caption{\label{fig: rocplot} The ROC curves for the W/Z vs. QCD classifier. We have defined background rejection as the reciprocal of background efficiency.  }
\end{figure}

\subsection{Variables constructed for the different signal regions}
\label{sec: appendix-4}
In Table \ref{tab: trainingvariables}, we present the list of variables constructed from events in the five signal regions and used in the training and evaluation of the BDT classifier. Only variables with checkmarks are used during the analysis. Most of these variables are self-explanatory. $\Delta \phi (i,j)$ stands for the separation between the particles i and j in the azimuthal plane. The azimuthal separation of different final state objects with the missing transverse momentum can help reduce background where a major contribution to the missing energy comes from objects that are not properly identified/reconstructed. The azimuthal separation between photons and leptons is expected to be higher for backgrounds like $W\gamma$ and $W\gamma \gamma$ where the lepton from the $W$ boson and the photon are more likely to get produced back to back. The variables $H_T$, defined as the scalar sum of the transverse momentum of all visible final state particles, $M_{eff} = (H_T + E_T^{miss})$, missing transverse momentum significance ($E_T^{miss}/\sqrt{H_T}$) are expected to have a higher value for the signal events considering the high mass of the level-1 KK particles involved. In Table \ref{tab: trainingvariables}, we have denoted with $j$ and $j^{1.0}$ final state jets with reconstruction radius 0.4 and 1.0, respectively, while the numbers in the subscript denote their $p_T$ ordering. For events with no fat jets or less than three $R=0.4$ jets, the corresponding transverse momentum variables and azimuthal separation with missing transverse momentum are zero-padded.

\begin{table}[!htb]
\centering
\begin{tabular}{l|c|c|c|c|c}
\toprule
Variable & SR\_1p0l & SR\_1pnl & SR\_2p0l & SR\_2pnl & SR\_0pnj \\  
\midrule 

$E_T^{miss}$ & \checkmark  & \checkmark & \checkmark  & \checkmark & \checkmark \\ 

$p_T (j_1)$ & \checkmark  & \checkmark & \checkmark  & \checkmark & \checkmark \\

$p_T (j_2)$ & \checkmark  & \checkmark & \checkmark  & \checkmark & \checkmark \\

$p_T (j_3)$ & \checkmark  & \checkmark & $\times$  & $\times$ & \checkmark \\

$p_T (\gamma_1)$ & \checkmark  & \checkmark & \checkmark  & \checkmark & $\times$ \\

$p_T (\gamma_2)$ & $\times$  & $\times$ & \checkmark  & \checkmark & $\times$ \\

$p_T (l_1)$ & $\times$  & \checkmark & $\times$  & \checkmark & $\times$ \\

$p_T (j^{1.0}_1) $ & \checkmark  & \checkmark & \checkmark  & \checkmark & \checkmark \\

$\Delta \phi (j_1,E_T^{miss})$ & \checkmark  & \checkmark & \checkmark  & \checkmark & \checkmark \\

$\Delta \phi (j_2,E_T^{miss})$ & \checkmark  & \checkmark & \checkmark  & \checkmark & \checkmark \\

$\Delta \phi (j_3,E_T^{miss})$ & \checkmark  & \checkmark & $\times$  & $\times$ & \checkmark \\

$\Delta \phi (\gamma_1,E_T^{miss})$ & \checkmark  & \checkmark & \checkmark  & \checkmark & $\times$ \\

$\Delta \phi (\gamma_2,E_T^{miss})$ & $\times$  & $\times$ & \checkmark  & \checkmark & $\times$ \\

$\Delta \phi (\gamma_1,l_1)$ & $\times$  & \checkmark & $\times$  & \checkmark & $\times$ \\

$\Delta \phi (\gamma_2,l_1)$ & $\times$  & $\times$ & $\times$  & \checkmark & $\times$ \\

$ \eta (j_1)$ & $\times$  & $\times$ & $\times$  &  $\times$ & \checkmark  \\

$ \eta (j_2)$ & $\times$  & $\times$ & $\times$  &  $\times$ & \checkmark  \\

$H_T$ & \checkmark  & \checkmark & \checkmark  & \checkmark & \checkmark \\

$E_T^{miss}/\sqrt{H_T}$ & \checkmark  & \checkmark & \checkmark  & \checkmark & \checkmark \\

$M_{eff}$ & \checkmark  & \checkmark & \checkmark  & \checkmark & \checkmark \\

$N(j \neq b)$ & \checkmark  & \checkmark & \checkmark  & \checkmark & \checkmark \\

$N (j=b)$ & \checkmark  & \checkmark & \checkmark  & \checkmark & \checkmark \\

$N (j^{1.0})$ & \checkmark  & \checkmark & \checkmark  & \checkmark & \checkmark \\

$N (l)$ & $\times$  & \checkmark & $\times$  & \checkmark & $\times$ \\

$N (\gamma)$ & $\times$  & $\times$ & \checkmark  & \checkmark & $\times$ \\

\bottomrule
\end{tabular}
\caption{\label{tab: trainingvariables} Variables used for the training of BDT classifiers.}
\end{table}

%=========================
\bibliographystyle{JHEP}
\bibliography{v0}
%==========================

\end{document}